%% file: vlqbsm.tex
\documentclass[a4paper,11pt]{article}
\pdfoutput=1 

\usepackage{jheppub}
\usepackage[T1]{fontenc} 
\usepackage{xspace}
\usepackage{slashed}
\usepackage{graphicx}
\usepackage{subcaption}
\usepackage{booktabs}
\usepackage{multirow}

\newcommand{\tprime}{\ensuremath{t^{\prime}}\xspace}

\newcommand{\antitprime}{\ensuremath{\bar t^{\prime}}\xspace}

\newcommand{\tpp}{\ensuremath{t^{\prime\prime}}\xspace}
\newcommand{\tppp}{\ensuremath{t^{\prime\prime\prime}}\xspace}

\newcommand{\antibar}[1]{\ensuremath{#1\bar{#1}}\xspace}

\newcommand{\tbar}{\ensuremath{\bar{t}}\xspace}
\newcommand{\ttbar}{\antibar{t}}

\def\mn{{\ensuremath{\mu\nu}}}

\newcommand{\Zgam}{\ensuremath{Z\, \gamma}\xspace}
\newcommand{\gamgam}{\ensuremath{\gamma\, \gamma}\xspace}
\newcommand{\Zt}{\ensuremath{Z\, t}\xspace}
\newcommand{\Wb}{\ensuremath{W^+ \, b}\xspace}
\newcommand{\higgst}{\ensuremath{h\, t}\xspace}
\newcommand{\St}{\ensuremath{S\, t}\xspace}
\newcommand{\Santit}{\ensuremath{S\, \bar t}\xspace}

\newcommand{\unit}[2]{\ensuremath{{#1\, \mathrm{#2}}}\xspace}
\newcommand{\ifb}{\ensuremath{\mathrm{fb}^{-1}}\xspace}
\newcommand{\BR}{\ensuremath{\mathrm{BR}}\xspace}
\newcommand{\BRs}{\ensuremath{\mathrm{BRs}}\xspace}

\newcommand{\Tr}{\ensuremath{\mathrm{Tr}\,}} 
\newcommand{\tr}{\ensuremath{\mathrm{tr}\,}} 
\let\Re\relax
\let\Im\relax
\newcommand{\Re}{\ensuremath{\mathrm{Re}\,}} 
\newcommand{\Im}{\ensuremath{\mathrm{Im}\,}} 

\newcommand{\uu}[1]{\ensuremath{\text{U}({#1})}}
\newcommand{\su}[1]{\ensuremath{\text{SU}(#1)}}
\newcommand{\so}[1]{\ensuremath{\text{SO}(#1)}}
\renewcommand{\sp}[1]{\ensuremath{\text{Sp}(#1)}}

\newcommand{\GeV}{\ensuremath{\text{Ge\kern-0.15ex V}}\xspace}
\newcommand{\TeV}{\ensuremath{\text{Te\kern-0.1ex V}}\xspace}

\newcommand{\SR}{{SR}\xspace}

\newcommand{\mtp}{\ensuremath{m_{\tprime}}\xspace}
\newcommand{\ms}{\ensuremath{m_S}\xspace}
\newcommand{\mz}{\ensuremath{m_Z}\xspace}

\newcommand*{\mgamgam}{\ensuremath{M_{\gamma\gamma}}\xspace}
\newcommand*{\mbkggamgam}{\ensuremath{M^{\mathrm{bkg}}_{\gamma\gamma}}\xspace}

\newcommand*{\mll}{\ensuremath{M_{\ell^+\ell^-}}\xspace}
\newcommand*{\mzgam}{\ensuremath{M_{Z \gamma}}\xspace}
\newcommand*{\mbkgzgam}{\ensuremath{M^{\mathrm{bkg}}_{Z \gamma}}\xspace}
\newcommand{\Mhat}{\ensuremath{\widehat M}\xspace}
\newcommand{\pt}{\ensuremath{p_{\text{T}}}\xspace}
\newcommand{\ET}{\ensuremath{E_{\text{T}}}\xspace}
\newcommand{\MET}{\ensuremath{E_{\text{T}}^{\text{miss}}}\xspace}
\newcommand{\Ht}{\ensuremath{H_{\text{T}}}\xspace}

\newcommand{\dR}{\ensuremath{\Delta R}\xspace}
\newcommand{\dRgamgam}{\ensuremath{\Delta R_{\gamma \gamma}}\xspace}

\newcommand{\sba}{s_{\beta-\alpha}}
\newcommand{\cba}{c_{\beta-\alpha}}

\newcommand*{\amcatnlo}{\textsc{aMC@NLO}\xspace}
\newcommand*{\delphes}{\textsc{Delphes}\xspace}
\newcommand*{\fastjet}{\textsc{FastJet}\xspace}
\newcommand*{\hathor}{\textsc{Hathor}\xspace}
\newcommand*{\higgsbounds}{\textsc{HiggsBounds}\xspace}

\newcommand*{\lhapdf}{\textsc{LHAPDF}\xspace}
\newcommand*{\lhpdf}{\textsc{LHPDF}\xspace}
\newcommand*{\madanalysis}{\textsc{MadAnalysis}\xspace}

\newcommand*{\madgraph}{\textsc{MadGraph}\xspace}
\newcommand*{\madgraphfive}{\textsc{MadGraph5}}
\newcommand*{\madspin}{\textsc{MadSpin}\xspace}
\newcommand*{\nnpdf}{\textsc{NNPDF}\xspace}
\newcommand*{\pythia}{\textsc{Pythia}\xspace}

\newcommand*{\runtwo}{Run~II\xspace}
\newcommand*{\runthree}{Run~III\xspace}

\newcommand{\nn}{\nonumber}

\newcommand{\be}{\begin{equation}}
\newcommand{\ee}{\end{equation}}
\newcommand{\bpm}{\begin{pmatrix}}
\newcommand{\epm}{\end{pmatrix}}

\newcommand{\mc}{\mathcal}

\newcommand{\al}{\alpha}
\newcommand{\gm}{\gamma}        \newcommand{\Gm}{\Gamma}
\newcommand{\kp}{\kappa}
\newcommand{\lm}{\lambda}       
\newcommand{\sg}{\sigma}

\newcommand{\lt}{\left}
\newcommand{\rt}{\right}

\def\eq#1{{eq.~\eqref{#1}}}
\def\fig#1{{figure~\ref{#1}}}
\def\sec#1{{section~\ref{#1}}}

\def\tab#1{{table~\ref{#1}}}
\def\tabs#1#2{{tables~\ref{#1}--\ref{#2}}}
\def\app#1{{appendix~\ref{#1}}}

\title{Signatures of vector-like top partners decaying into new neutral scalar or pseudoscalar bosons}

\author[a]{R.~Benbrik,}
\author[b,1]{E.~Bergeaas Kuutmann,\note{Corresponding author.}}
\author[c]{D.~Buarque Franzosi,}
\author[b]{V.~Ellajosyula,}
\author[b]{R.~Enberg,}
\author[c]{G.~Ferretti,}
\author[b]{M.~Isacson,}
\author[d,2]{Y.-B.~Liu\note{Also at University of Southampton.},}
\author[b]{T.~Mandal,}
\author[b]{T.~Mathisen,}
\author[e,3]{S.~Moretti\note{Also at Uppsala University.},}
\author[b,2]{L.~Panizzi.}

\affiliation[a]{Cadi Ayyad University, Marrakesh, Morocco}
\affiliation[b]{Uppsala University, Uppsala, Sweden}
\affiliation[c]{Chalmers University of Technology, G\"oteborg, Sweden}
\affiliation[d]{Henan Institute of Science and Technology, Xinxiang, P.R.China}
\affiliation[e]{University of Southampton, Southampton, UK}

\emailAdd{elin.bergeaas.kuutmann@physics.uu.se}

\abstract{We explore the phenomenology of models containing one Vector-Like Quark (VLQ), \tprime, which can decay into the Standard Model (SM)  top quark, $t$, and a new spin-0 neutral boson, $S$, the latter being either a scalar or pseudoscalar state. 
We parametrise the underlying interactions in terms of a simplified model which enables us to capture possible Beyond the SM (BSM) scenarios. 
We discuss in particular three such scenarios: one where the SM state is supplemented by an additional scalar, one which builds upon a 2-Higgs Doublet Model (2HDM) framework and another which realises a Composite Higgs Model (CHM) through partial compositeness. 
Such exotic decays of the \tprime can be competitive with decays into SM particles, leading to new possible discovery channels at the Large Hadron Collider (LHC). 
Assuming \tprime pair production via strong interactions, we design signal regions optimised for one $\tprime\to \St$ transition (while being inclusive on the other \antitprime decay, and vice versa), followed by the decay of $S$ into the two very clean experimental signatures $S\to\gamgam$ and $S\to Z(\to \ell^+\ell^-)\gamma$. 
We perform a dedicated signal-to-background analysis in both channels, by using Monte Carlo (MC) event simulations modelling the dynamics from the proton-proton to the detector level. 
Under the assumption of $\BR(\tprime\to \St) = 100\%$, we are therefore able to realistically quantify the sensitivity of the LHC to both the \tprime and $S$ masses, assuming both current and foreseen luminosities. 
This approach paves the way for the LHC experiments to surpass current VLQ search strategies based solely on \tprime decays into SM bosons ($W^\pm, Z$,  $h$).}

\begin{document}
\maketitle
\flushbottom

%%%%%%%%%%%%%%%%%%%%%%%%%%%%%%%%%%
\section{Introduction}
\label{sec:intro}
\input{introduction}

%%%%%%%%%%%%%%%%%%%%%%%%%%%%%%%%%%
\section{The simplified model}
\label{sec:model}
\input{model}

%%%%%%%%%%%%%%%%%%%%%%%%%%%%%%%%%%
\section{LHC constraints from $\gamma \gamma$ and $Z \gamma$ resonance searches}
\label{sec:LHCconstraints}
\input{LHCconstraints}

%%%%%%%%%%%%%%%%%%%%%%%%%%%%%%%%%
\section{Analysis}
\label{sec:analysis}
\input{analysis}

%%%%%%%%%%%%%%%%%%%%%%%%%%%%%%%%%
\section{Results}
\label{sec:interpretation}
\input{interpretation}

%%%%%%%%%%%%%%%%%%%%%%%%%%%%%%%%%%%%%%%%%%%
\section{Conclusions}
\label{sec:conclusions}
\input{conclusions}

%%%%%%%%%%%%%%%%%%%%%%%%%%%%%%%%%%%%%%%%%%%
\appendix

\section{Details of the models}
\label{sec:appendixmodels}
\input{appendix_models}

\section{Range of validity of the narrow-width approximation}
\label{sec:appendixrangeofvalidity}
\input{appendix_rangeofvalidity}

\section{Object definition}
\label{sec:object_definition_long}
\input{appendix_objects}

\section{Signal efficiencies}
\label{sec:appendixefficiencies}
\input{appendix_efficiencies}

%%%%%%%%%%%%%%%%%%%%%%%%%%%%%%%%%%%%%%%%%%%
\acknowledgments
This work would not have been possible without financing from the Knut and Alice Wallenberg foundation under the grant KAW 2017.0100, which supported EBK, DBF, VE, RE, GF, TMat and LP as a part of the SHIFT project. 
RB's work is supported by the Moroccan Ministry of Higher Education and Scientific Research MESRSFC and CNRST:\ Project PPR/2015/6.
RB, RE and SM are supported in part by the H2020-MSCA-RISE-2014  grant agreement No.\ 645722 (NonMinimalHiggs). 
TMan is grateful for support from The Royal Society of Arts and Sciences of Uppsala. 
Y-BL is supported by the Foundation of Henan Institute of Science and Technology (Grant no.\ 2016ZD01) and the China Scholarship Council (201708410324).
SM is supported in part by the NExT Institute and the STFC Consolidated Grant ST/L000296/1.
DBF would like to thank the  ITP of the G\"ottingen University and GWDG for providing computing resources.
MI wishes to thank the Swedish National Infrastructure for Computing (project SNIC 2019/3-68) for computing resources.
LP and Y-BL  acknowledge the use of the IRIDIS HPC facility at the University of Southampton.

\clearpage

%%%%%%%%%%%%%%%%%%%%%%%%%%%%%%%%%%%%%
\bibliographystyle{JHEP}
\bibliography{vlqbsm}

\end{document}

%% file: introduction.tex
During \runtwo at the LHC, the ATLAS and CMS experiments have collected almost \unit{150}{\ifb} and \unit{180}{\ifb} of data, respectively, at a centre-of-mass (CM) energy of \unit{13}{\TeV}. 
These data are now being analysed by the collaborations and, so far, no significant deviations from the SM have been recorded.  
This has significantly restricted the parameter space of the most common scenarios attempting to solve the hierarchy problem of the SM, such as supersymmetry and compositeness. 
Yet, it is important to find a viable solution to this flaw of the SM. 
This is inevitably connected to studying  both top quark and Higgs boson dynamics, as the hierarchy problem of the SM originates from their mutual  interactions. 
A pragmatic approach is to investigate BSM scenarios in which either of or both the top and Higgs sectors of the SM are enlarged through the presence of companions to the SM states ($t$ and $h$), by which we mean additional spin-1/2 and spin-0 states, respectively, with the same electromagnetic (EM) charge but different mass (naturally heavier) and possibly different quantum numbers as well. 

Some guidance in exploring the various BSM possibilities in this respect is afforded by experimental measurements of observables where both the top quark and the SM-like Higgs boson enter. 
On the one hand, a sequential fourth family of chiral SM quarks is strongly constrained indirectly from Higgs boson measurements due to their non-decoupling properties~\cite{Eberhardt:2012gv}, while VLQs (which transform as triplets under colour but whose left- and right-handed components have identical electroweak (EW) quantum numbers) can evade these bounds easily.
On the other hand, the possibility of the existence of additional Higgs bosons has not been excluded by experimental data and may well be theoretically motivated by the fact that neither the matter nor the gauge sectors are minimal. 
Moreover, the Higgs sector is extended in any supersymmetric model or in the 2HDM. 

Similarly, any model in which a  Higgs boson arises as a pseudo-Nambu-Goldstone Boson (pNGB), other than the minimal model based on the symmetry breaking pattern $\so{5}/\so{4}$, will include additional light (pseudo)scalars that might well have eluded direct searches due to their reduced couplings to the EW bosons and  top quark. 

Hence, it is of some relevance to assess the viability at the LHC of BSM models with both top quark partners (of VLQ nature) and companion scalar or pseudoscalar particles (both charged and neutral). 
In fact, it is particularly intriguing to investigate the possibility of isolating experimental signatures where the two particle species interact with each other, namely, when the \tprime decays into a new (pseudo)scalar.

So far, collider searches for a VLQ companion to the SM top quark \cite{Buchkremer:2013bha,Aguilar-Saavedra:2013qpa} have mostly been carried out under the assumption that it decays exclusively into SM particles, namely, a heavy quark ($b, t$) and a boson ($W^\pm, Z, h$), compatibly with the EM charge assignments. 
Specifically, for the case of a top-like VLQ, \tprime, the  decays considered are $\tprime \to \Zt$, $\tprime \to \higgst$ and $\tprime \to \Wb$, with varying branching ratios (\BRs) adding up to 100\%, see e.g.~\cite{Aaboud:2018pii,Aaboud:2018wxv,Aaboud:2018xpj,Aaboud:2018uek,Aaboud:2018saj,Aaboud:2018xuw,Sirunyan:2018qau,Sirunyan:2018yun,Sirunyan:2017pks,CERN-EP-2019-129}. 

It is thus important to ask how the presence of exotic decay channels of VLQs can affect the current bounds and whether these might actually be promising discovery channels on their own. 
This question has been asked in similar contexts in various preceding works~\cite{Serra:2015xfa,Anandakrishnan:2015yfa,Banerjee:2016wls,Dobrescu:2016pda,Aguilar-Saavedra:2017giu,Chala:2017xgc,Colucci:2018vxz,Banerjee:2018fsx,Han:2018hcu,Kim:2019oyh}, each concentrating on a specific BSM construction. 
Here, in contrast, we follow the approach of~\cite{Bizot:2018tds}, which adopts a set of simplified scenarios based on effective Lagrangians (motivated by compositeness).

In our paper, we build upon this last work, by adopting a simplified scenario which contains, above and beyond the SM particle spectrum, a top-like VLQ, \tprime, as well as an additional  scalar (or pseudoscalar) particle,  $S$, in turn leading to the new decay channel $\tprime \to \St$. 
As for the decay modes of $S$, we will concentrate on two of the experimentally cleanest channels accessible at the LHC, namely, $S \to \gamgam$ and  $S\to \Zgam$, with the $Z$ boson decaying in turn into electrons or muons. 
We will show in \sec{sec:model} that there exist well motivated phenomenological scenarios where these can indeed be decay modes with significant \BRs, for the case of both fundamental and composite Higgs states.  
In \sec{sec:LHCconstraints} we estimate LHC constraints using published ATLAS and CMS searches in $\gamgam$ and $\Zgam$ final states while in \sec{sec:analysis} we will describe our MC simulations, based on the pair production process $p\, p\to \tprime\,\antitprime$, followed by the decay chains $\tprime \to S(\to \gamgam)\, t$ or $\tprime \to S(\to \Zgam)\, t$, with the $\antitprime$ treated inclusively (and vice versa). 
Section~\ref{sec:interpretation} is then dedicated to interpreting the ensuing MC results in three theoretical scenarios embedding a \tprime alongside additional (pseudo)scalar states focusing on cases with $\BR({\tprime \to \St}) = 100\%$,  while in \sec{sec:conclusions} we conclude.

%% file: model.tex
The purpose of this section is to present the relevant details about the class of models whose phenomenology we aim to study. 
We begin with a general description of a simplified model that captures all relevant features. 
This is the model used for the analysis in \sec{sec:analysis}. 
We then justify the use of this simplified model by introducing three more specific models that can all be described with the same generic Lagrangian by a mapping of the fields and the couplings, provided that the processes considered in this paper are studied.

As discussed in the introduction, we are interested in exotic decays of a top partner \tprime (of mass $\mtp$) into the ordinary top quark $t$ and a scalar (or pseudoscalar) generically denoted by $S$ (of mass $\ms$) in the simplified model. 
We can thus augment the SM Lagrangian $\mc{L}_{\rm SM}$ by the following interaction Lagrangian with operators up to dimension five involving these two additional fields,
\begin{eqnarray}
\mc{L}_\text{BSM} &=& \kp^S_L~\antitprime_{R} t_{L} S + \kp^S_R~\antitprime_{L} t_{R} S + {\mathrm{ h.c.}} \nonumber \\ 
 && - \frac{S}{v} \sum_f m_f\left( \kp_f \bar{f}f + i\tilde{\kp}_f \bar{f}\gamma_5 f \right)  +
  \frac{S}{v}\left(2\lambda_W m_W^2 W^+_\mu W^{-\mu} +  \lambda_Z m_Z^2 Z_\mu Z^{\mu} \right)\nonumber \\
 && +  \frac{S}{16 \pi^ 2 v} \sum_V \left( 
   \kp_{V} g_V^2 \, V^a_{\mu\nu} V^{a\mu\nu}
   + \tilde{\kp}_V g_V^2 \,  V^a_{\mu\nu}\widetilde V^{a\mu\nu}  \right) \label{eq:LBSM} .
\end{eqnarray}
Here $\kp^S_L$ and $\kp^S_R$ are the Yukawa couplings of the $S$ to the $t$ and \tprime.  
In the second line, $f$ sums over all SM fermions (including the top $t$) and $\kp_f$ is the dimensionless reduced Yukawa coupling. 
In the last line $V_{\mu\nu}$ denotes the field strengths of the \uu{1}$_Y$, \su{2}$_L$ and \su{3}$_C$ gauge bosons $B_\mu,W_\mu,G_\mu$ in the gauge eigenbasis, $g_V$ is the associated gauge coupling ($g', g, g_s$ respectively) and $\widetilde V_{\mu\nu}= (1/2)\epsilon_{\mu\nu\rho\gamma}V^{\rho\gamma}$ is the dual field strength tensor. 
The coefficients $\tilde{\kp}_V$ and $\kp_V$ are couplings associated with dimension-five operators and are typically generated by loops of heavy particles or via anomalies. 
The couplings $\lambda_V$ for any gauge boson $V$ are only generated if $S$ is charged under some of the SM gauge groups and gets a vacuum expectation value (VEV) or if it mixes with such states, e.g., the Higgs boson. 
Since $\su{3}_{\mathrm{C}} $ and $\uu{1}_{\mathrm{EM}}$ are unbroken for the strong and EM interactions, $\lambda_V=0$ for the respective gauge bosons. 
We choose to normalise all terms with only one dimensionful parameter, the VEV $v=246\,\GeV$. 

In practice, we consider an $S$ state of either scalar or pseudoscalar nature, but not a mixture. 
We therefore do not consider CP-violation in this paper. 
This means that either $\tilde{\kp}_V$ or $\tilde{\kp}_f$ are zero, in the scalar case, or $\kp_V$, $\lambda_V$ and $\kp_f$ are zero, in the pseudoscalar case. 

The total widths of \tprime and $S$ are kept as free parameters in the simulation as an indication that other interactions and other states might be present. 
These interactions are not explicitly required to describe the process $p\, p\to \tprime \, \antitprime \to S S \ttbar$  apart from their contribution to the total widths. 
Here we only report the analytic expression for the partial width of the exotic \tprime decay, specifically. 
\begin{align}
\Gamma_{\tprime \to S t} &= \frac{1}{32\pi}\mtp
\left[\left(1 + x_{t}^2 - x_S^2\right) \left(|\kp^S_L|^2 + |\kp^S_R|^2\right)
  + 4x_{t}(\Re\kp^S_L\Re\kp^S_R +\Im\kp^S_L\Im\kp^S_R) \right]\nonumber \\ 
&\times \left(1+x_{t}^4+x_S^4-2x_{t}^2-2x_S^2 - 2x_{t}^2x_S^2\right)^{\frac{1}{2}} \ ,
\label{GamTtotS.EQ}
\end{align}
where $x_{t} \equiv m_{t}/\mtp$ and $x_S \equiv \ms/\mtp$. 
This formula is valid for decays into both scalar and pseudoscalar $S$. 

This defines the simplified model that will be used in the rest of this paper. 
Let us now briefly discuss three specific examples of models that motivate the use of the above simplified model and the mapping between the former and the latter. 
The results in this paper, given in terms of the simplified model above, can then easily be reinterpreted in terms of each model, if needed. 
In a forthcoming paper, we will specify these models in more detail and will discuss their specific phenomenology.

\subsection{Example 1: adding a VLQ and a scalar to the SM}
\label{sec:simpPC}

In order to illustrate how a particular model can be related to the phenomenological simplified model (\eq{eq:LBSM}), we will first present a simple model of top-quark partial compositeness (PC) in some detail. 
The model consists of the SM extended by a top partner VLQ and a scalar singlet. 
In this model the top quark acquires its mass via the mixing with the top partner. 
This model is not intended as a complete, realistic model, but provides an example of a model with an additional scalar $S$ that is neutral under the SM gauge group. 
We will only be concerned with the couplings between the top quarks and $S$, leaving the coupling inducing the decay of the $S$ to SM states as in \eq{eq:LBSM}. 

We denote the gauge eigenstates in the top sector by $\widetilde t_L$, $\widetilde t_R$ and $T$. 
The notation $\widetilde t_{L/R}$ is to prevent confusion with the mass eigenstates that are to be denoted by $t$ and $\tprime$. 
The Lagrangian for this model before EW symmetry breaking (EWSB) can be written as 
\begin{align}
\mc{L}_{\rm kin} \supset \bar{T}\lt(i\slashed{D} - M\rt)T + \frac{1}{2}\lt(\partial_\mu S\rt)\lt(\partial^\mu S\rt) - \frac{1}{2}m_{S}^2S^2,
\label{eq:LSkinetic}
\end{align}
\begin{align}
\mc{L}_{\rm int} &\supset - \lm_S^a S \bar{T}_L T_R - \lm_S^b S \bar{T}_L \widetilde t_R  - \widetilde y \lt(\bar{Q}_L\widetilde{H}\rt)\widetilde t_R - \lm_1\lt(\bar{Q}_L\widetilde{H}\rt)T_R - m_2 \bar{T}_L \widetilde t_R  \label{eq:LSetc}
 + \text{h.c.}\ ,
\end{align}
where the SM Higgs doublet is denoted by $H$ with $\widetilde{H}=i \sg_2 H^*$. 
The SM Yukawa coupling for the top quark is here denoted by $\widetilde y$ and $Q_L$ is the left-handed quark doublet of the third generation. 
The couplings $\lm_S^{a,b}$ are real if $S$ is a scalar and purely imaginary if $S$ is a pseudoscalar.
The mass $m_2$ is a non-diagonal entry in the mass matrix of \eq{eq:tpmassmat}. 
The remaining couplings are dimensionless. 
After EWSB,  we have a mass matrix
\be
\label{eq:tpmassmat}
\mc{L}_t \supset \bpm \bar{\widetilde t}_L & \bar{T}_L \epm \bpm m_{\widetilde t}  & m_1 \\ m_2 & M \epm \bpm \widetilde t_R \\ T_R \epm  + \text{h.c.},
\ee
where we defined $m_{\widetilde t} = \widetilde y v/\sqrt{2}$ and  $m_1=\lm_1v/\sqrt{2}$. 
The mass matrix can be diagonalised by bi-orthogonal rotations by the angles $\theta_{L,R}$, separately for left- and right-handed fermions, as follows  (where $s_X\equiv\sin \theta_X$ and $c_X \equiv \cos \theta_X$)
\be
\bpm {t}_{L,R} \\ \tprime_{L,R} \epm = \bpm c_{L,R} & -s_{L,R} \\ s_{L,R} & c_{L,R} \epm \bpm \widetilde t_{L,R} \\ T_{L,R} \epm \ ,
\label{tTrot.EQ}
\ee
where $\{t,\tprime \}$ are the mass eigenstates and the mixing angles are given by
\be 
\tan{(2\theta_L)} = \frac{2\lt(m_{\widetilde t} m_2+M m_1\rt)}{M^2-m_{\widetilde t}^2-m_1^2+m_2^2}  \ , \quad 
\tan{(2\theta_R)} = \frac{2\lt(m_{\widetilde t} m_1+M m_2\rt)}{M^2-m_{\widetilde t}^2+m_1^2-m_2^2} \ . \label{eq:rota}
\ee 
The mass eigenvalues $m_t$ and \mtp are found by computing the eigenvalues.
This model can be mapped to the simplified model Lagrangian in \eq{eq:LBSM} by performing the rotation in \eq{tTrot.EQ} inside \eq{eq:LSetc}. 
Focusing on the mixing terms yields 
\be
  \kp^S_L = \left(\lambda^a_S s_L c_R + \lambda^b_S s_L s_R\right)^* \, , \quad \kp^S_R = 
  \lambda^a_S c_L s_R - \lambda^b_S c_L c_R \,,
  \label{eq:kpSLkpSR}
\ee
while for the coupling to the top we have
\be
    \kp_t  =\Re \left( -\lambda^a_S s_L s_R + \lambda^b_S s_L c_R\right), \quad    \widetilde \kp_t  =\Im \left( -\lambda^a_S s_L s_R + \lambda^b_S s_L c_R\right).
\ee
There is also a diagonal term involving the \tprime, which is proportional to 
$\lambda^a_S c_L c_R + \lambda^b_S c_L s_R$. 
It is not included in the simplified model, but instead generates a contribution to the effective coefficients $\kp_V$ and $\widetilde \kp_V$ from loop diagrams.

Let us also briefly discuss the decays of the \tprime and $S$ in this model. 
The \tprime has both the standard and non-standard decay channels discussed above, where the width of the $\tprime \to \St$ channel is given by \eq{GamTtotS.EQ} with the couplings defined in \eq{eq:kpSLkpSR}. 
The scalar can, in general, decay into the final states $gg,\, \gm\gm,\, Z \gm,\, ZZ,\, WW$ and $\ttbar$. 
We always assume $\ms < \mtp$, which forbids the decay $S\to \tprime \antitprime$. 
Apart from the \ttbar channel, all the other decays are generated by loops of the $t$ and \tprime. 

\begin{figure}[tb]
\centering
\includegraphics[width=10cm]{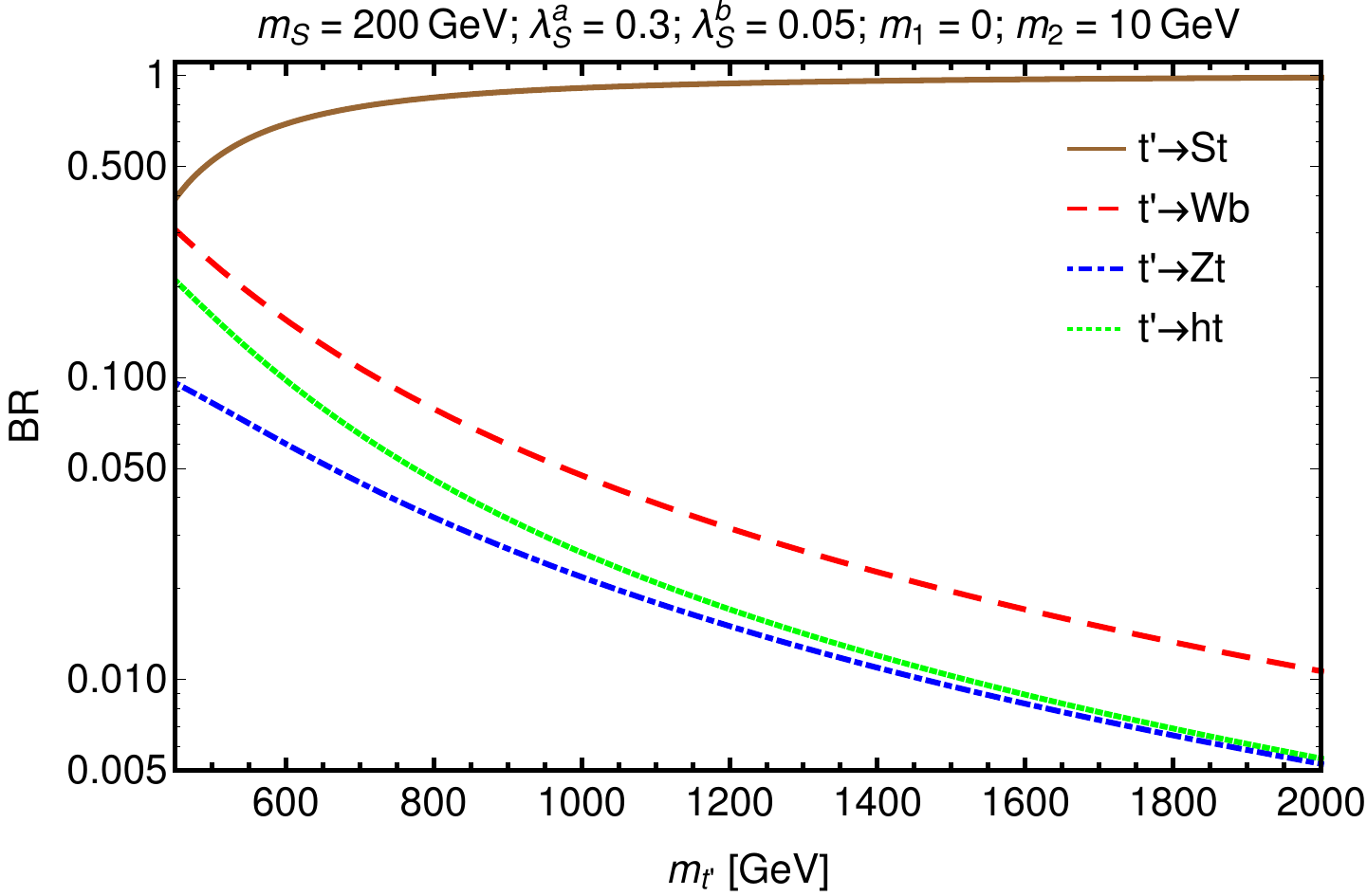}
\caption{\BRs of \tprime as a function of the mass for a specific parameter point.}
\label{fig:BR}
\end{figure}

We may now examine the decay of the \tprime and $S$ depending on the coupling of $T_L$ with $T_R$ and $\widetilde t_R$. The $\tprime \to \St$ decay is induced by the $\lm_S^a$ and $\lm_S^b$ couplings.  
If we are interested in a large $\BR({\tprime \to St})$, we may achieve that easily in a wide region of parameter space by considering suitable values of these couplings. 
For example, when the $T_L$ couples to $\widetilde t_R$ (i.e., $\lm_S^a=0,\lm_S^b\neq 0$), a small $\lm_S^b$ can induce large $\BR({\tprime \to St})$ as the $\lm_S^b$ part of $\kp^S_R$ in \eq{eq:kpSLkpSR} is proportional to $c_L c_R\sim 1$. 
If the $T_L$ only couples to $T_R$ (i.e., $\lm_S^a\neq 0,\lm_S^b=0$), a large $\BR({\tprime \to St})$ is realised when $\lm_S^a$ is sufficiently large, as the partial width is proportional to $(\lm_S^a/\lm_t)^2$. 
However, this will also increase the $s$-channel production of $S$ through $gg$ fusion, therefore, this scenario is heavily constrained by the $gg\to S\to\gm\gm$ resonance search data from the LHC. 
In \fig{fig:BR}, we show the \BRs of \tprime for a specific benchmark point where the  $\tprime \to S\,t$ channel has a \BR of almost 100\%.

As for the $S$ decay, the $S\to gg$ channel dominates if the $\ttbar$ decay is not kinematically allowed, $m_S<2m_t$. 
The total decay width is governed by $\Gm_{S\to gg}$, and hence the branching ratio in the $\gm\gm$ channel is approximately
\begin{align}
\BR({S\to\gm\gm})\simeq \frac{\Gm_{S\to\gm\gm}}{\Gm_{S\to gg}} = \frac{8\al_{EM}^2}{9\al_{S}^2}\approx 0.004 .
\label{eq:widthratio}
\end{align}
Despite the small BR, the $S\to\gm\gm$ decay  is a clean and  well motivated channel. 
For instance, in the search for  a VLQ decaying into a Higgs boson and a top, $\tprime \to  ht$, the $h\to \gamma\gamma$  decay channel (which has a BR of 0.23\%) is still sensitive~\cite{Khachatryan:2015oba}. 
We also note that there is no dedicated di-jet search, $\tprime \to S  t\to  gg t$, although it has been recently proposed in ref.~\cite{Cacciapaglia:2019zmj}. The current bounds estimated by a recast of R-parity violating (RPV) supersymmetry searches~\cite{Aad:2015lea} are not competitive.
Other loop induced channels are more suppressed than $S\to\gm\gm$. 
For example, the partial widths of $S\to Z\, \gamma$ and $S\to ZZ$, modulo negligible $m_Z$ corrections, are $2\tan^2\theta_W\Gm_{S\to\gm\gm}$ and $\tan^4\theta_W\Gm_{S\to\gm\gm}$, respectively. 

For $m_S > 2m_{t}$, the tree-level $S\to $ \ttbar channel usually dominates over the loop induced decays. 
However, in a region of parameter space, the \ttbar decay can be tuned down by suitable values of the off-diagonal entries in the mass matrix in \eq{eq:tpmassmat}. 
We find that, when $\sin\theta_L \ll \sin\theta_R$ (or equivalently $m_1 \ll m_2$), the effective $S t\bar t$ coupling, depending on the $\lm_S^a$ and $\lm_S^b$ couplings, is not sufficiently large to compete with the loop induced decays of $S$. 
The six tops final state via $\tprime \to \St \to t t\bar{t}$ has been discussed in ref.~\cite{Han:2018hcu} with both a recast from current searches and a dedicated analysis.

\subsection{Example 2: adding a VLQ to the 2HDM}
\label{sec:interpretation_2HDM}

The 2HDM (see~\cite{Branco:2011iw} for a review) is widely used as a minimal model for an extended Higgs sector that goes beyond additional singlet scalars. 
With additional vector-like top partners (see~\cite{Badziak:2015zez,Angelescu:2015uiz,Arhrib:2016rlj} for previous work), the 2HDM may be seen as the low-energy manifestation of a composite Higgs scenario, such as in~\cite{DeCurtis:2018zvh}. 
Specifically, we here consider a vector-like top partner $T$ with charge $+2/3$ in the singlet representation of the SM EW group. 
We further consider Yukawa couplings of the SM quarks of Type-II, i.e., that the up- and down-type quarks couple to different doublets.

The Higgs  sector of the 2HDM has an additional neutral scalar $H$, a pseudoscalar $A$ and a charged $H^\pm$ state. 
This enables us to obtain simple formulae where either $H$ or $A$ can play the role of $S$ in the simplified model Lagrangian in \eq{eq:LBSM}.
The details of the model and the involved parameters as well as the mapping onto the simplified model Lagrangian of \eq{eq:LBSM} are discussed in \app{sec:appendix2HDM}. 
Let us here only discuss the mixing of the physical top quark $t$ and top partner \tprime. 

The physical mass of the heavy top, $\mtp$, is different from the mass $M$ of the vector-like $T$ due to $t$--$T$ mixing. 
The mass matrix can be diagonalised in the same way as in \eq{tTrot.EQ} to obtain the physical states $(t_{L,R} , \tprime_{L,R})$ in terms of the gauge eigenstates $(\widetilde t_{L,R},T_{L,R})$. 
The mixing angles $\theta_L$ and $\theta_R$ are not independent parameters and we can derive similar relations to \eq{tTrot.EQ} (see \eq{eq:appmix2hdm}),
in terms of the Yukawa couplings $y_t$ and $\xi_T$ that couple the left-handed quark doublet $Q_L$ to the right-handed SM top $\widetilde t_R$ and the vector-like $T_R$, respectively (see \eq{eq:appLyuk} and \eq{equa:intr}). 
The two mixing angles in this case satisfy~\cite{Arhrib:2016rlj}
\begin{equation}
\tan\theta_L = \frac{\mtp}{m_t} \tan\theta_R , \quad\quad  
\quad\quad \frac{\xi_T}{y_t} = s_L c_L \frac{\mtp^2 - m^2_t}{m_t \mtp},
\end{equation}
while the mass of the \tprime is related to the Lagrangian parameters and the physical top quark mass via 
\begin{eqnarray}
\mtp^2 &=& M^2\left( 1 + \frac{\xi^2_T v^2}{2(M^2 - m_t^2)}\right).
\end{eqnarray} 
The $\tprime$--$t$ interaction can thus be described by three independent physical parameters: two quark masses, $m_t$ and $\mtp$, and a mixing angle, $s_L=\sin\theta_{L}$. 

In the 2HDM with a VLQ, the scalar $S$ is an additional Higgs boson. 
The dimension-five operators in \eq{eq:LBSM} are then generated through loops and in general $S$ can be produced through $gg\to S$. 
It can then decay in all the bosonic channels that we consider in this paper and, in addition, in fermionic ones. 
(The BRs in this model are discussed in \sec{sec:interpretation}.)
These channels give rise to constraints from all the usual collider observables. 
In addition,  the scalar sector of this model is subject to the same unitarity, perturbativity and vacuum stability constraints as the usual 2HDM \cite{Branco:2011iw,Kanemura:1993hm}. 
The Yukawa coupling $y_t$ is constrained from unitarity to be less than $4\pi$, while $\xi_T$  is a derived quantity. 
Since the new top partner will contribute to gauge boson self energies, the  mixing angle $\theta_L$ can be constrained from EW Precision Tests (EWPTs) such as the $S$ and $T$ parameters. 
Based on ref.~\cite{Arhrib:2016rlj}, such bounds require the mixing angle $\theta_L$ to be in the range $(-0.15, + 0.15)$. However, the constraints coming from BR$(b\to s \gamma)$ are  the most relevant ones, as the mixing angle is restricted to be in the range $(-0.1, +0.1)$ for large $m_{\tprime}$, i.e., around \unit{1}{\TeV}.

\subsection{Example 3: realisation in partial compositeness}
\label{sec:interpretation_composite}

Lastly, we present a Composite Higgs Model (CHM), which motivates the analysis in this paper by having a top partner with enhanced exotic decay mode and a pseudoscalar with dominant $\Zgam$ decay. 
The model is closely related to one of the earliest non-minimal models of composite Higgs with fermionic partial compositeness~\cite{Gripaios:2009pe}, based on the coset space $\su{4}/\sp{4}$, where Sp is the symplectic group. 
The usual Higgs field $\mathcal{H}$ is a bi-doublet of $\su{2}_L\times \su{2}_R$, which together with a singlet $S$ (usually denoted by $\eta$ in the CHM literature) forms the five dimensional anti-symmetric irreducible representation of $\sp{4}$,
\begin{equation}
   \mathcal{H} \oplus S \equiv \begin{pmatrix} H^{0*}&H^+\\ -H^{+*}&H^0\end{pmatrix} \oplus S \in (\mathbf{2},\mathbf{2}) \oplus (\mathbf{1},\mathbf{1}) = \mathbf{5}.
\end{equation}
This scenario has the further appeal of belonging to a class of models that can be obtained from an underlying gauge theory with fermionic matter~\cite{Barnard:2013zea,Ferretti:2013kya} and the additional features arising from this fact have been studied in, e.g.,~\cite{Belyaev:2016ftv}. 
Here, however, we want to focus on the bare bones of the model, namely the above-mentioned coset structure with the addition of one fermionic partner $\Psi$. 
(We only consider partial compositeness in the top sector). 

The fermionic sector also consists of a bi-doublet and a singlet in the $\mathbf{5}$ of $\sp{4}$. 
We will see that, as already anticipated in~\cite{Bizot:2018tds} (see also~\cite{Serra:2015xfa}), the possible decay patterns of the fermionic partners are richer than what is usually considered in current searches and, in particular, the lightest top-partner has an enhanced decay into the exotic channel $\tprime \to \St$.

To summarise, in addition to the SM fields the model has an additional pseudoscalar $S$, three top partners $T, T', \widetilde T$ (all of electric charge $+2/3$), a bottom partner $B$ (charge $-1/3$) and an additional coloured fermion $X$ of charge $+5/3$. 
Like in the previous example models, all of these fermions are vector-like Dirac spinors, to be thought of as in the gauge eigenbasis, i.e., before their mass matrices are diagonalised. 
The difference here is that there are more than one new fermion.

The mixing with the third family quarks of the SM depends on how they are embedded in a representation of $\su{4}$. 
We choose this embedding such that the custodial symmetry of~\cite{Agashe:2006at} is preserved, see \app{sec:appendixcomp} for details. 
In addition, the choice of having an elementary $\widetilde t_R$ distinguishes this model from similar ones studied in~\cite{DeSimone:2012fs}, where the $\widetilde t_R$ was taken to be fully composite. 
The elementary $\widetilde t_R$ seems more appealing, since chiral fermions are notoriously difficult to obtain from underlying strongly coupled theories. 
We do not address the origin of the bottom quark mass in this work, which would add additional model dependence that is not relevant for the experimental signatures of interest. 
See \app{sec:appendixcomp} for more details on the construction of the model and the singular value decomposition of the mass matrix.

We end up with four top quark mass eigenstates, which we denote, in increasing mass order, by
$t, \tprime, \tpp$ and $\tppp$. Here $t$ is the known SM top quark of mass $m_t=173\, \GeV$. 
We diagonalise the mass matrix numerically, but a perturbative expansion for the masses gives some insight into the mass spectrum. 
We find (see \app{sec:appendixcomp})
\begin{eqnarray}
& m_t = \displaystyle\frac{y_L y_R f v}{\sqrt{2}\Mhat}+{\mathcal{O}}\left({v^2}/{f^2}\right), \quad & \mtp = M, \\
\quad & m_{\tpp} = M + {\mathcal{O}}\left({v^2}/{f^2}\right), \quad & m_{\tppp} = \Mhat + {\mathcal{O}}\left({v^2}/{f^2}\right),
\end{eqnarray}
where $M$ is the mass parameter of the $\Psi$, $y_L$ and $y_R$ are the respective couplings of the $Q_L$ and $\widetilde t_R$ to the $\Psi$ and pNGBs while  $f$ is the ``pion decay constant'' of the strongly coupled theory. 
We also defined $\Mhat = \sqrt{M^2 + y_L^2 f^2}$. 
The mass of the bottom partner (mostly aligned with $B$) turns out to be of the same order as that of the heaviest top partner $m_{\tppp}$, while $X$ has mass equal to $M\equiv \mtp$ since it does not mix with anything. 

Substituting the mass eigenstates (see \app{sec:appendixcomp}) into the Lagrangian and considering the coupling that mixes the two lightest eigenstates $t$ and $\tprime$ with the pNGBs, we see that no mixing with the Higgs field $h$ arises, while the $S$ couples, up to terms of order ${\mathcal{O}}\left(v^2/f^2\right)$, as
\begin{equation}
    {\mathcal{L}} = - i y_R \, S\, \antitprime_{L} t_{R} - \frac{i y_L v M}{\sqrt{2} f \Mhat}\, S\, \antitprime_{R} t_{L} + \mbox{ h.c.},
\end{equation}
allowing us to match the models with the parameters of the phenomenological Lagrangian \eq{eq:LBSM}
\begin{equation}
\kp^S_R = - i y_R, \quad \kp^S_L = - \frac{i y_L v M}{\sqrt{2} f \Mhat}.
\end{equation}
From the analysis of the spectrum and of the couplings, we see that we can concentrate on a model with two mass degenerate VLQs $\tprime$ and $X$, with $\sim 100\%$ branching ratios $X\to W^+\, t$ and $\tprime \to \St$.  
The decay modes of \tprime to SM vector bosons are highly suppressed, \tprime being a singlet of $\su{2}_L\times \su{2}_R$. 
For this model, it is thus crucial to understand whether the BSM decay $\tprime \to \St$ can compete with the SM decay $X\to W^+\, t$ whose signatures have been looked for at the LHC~\cite{Aaboud:2018xpj} providing bounds to the model parameter $M>1.2\, \TeV$. 
We address this question in this work. 
Just above the \tprime  mass scale there is a further top partner, $\tpp$, with more diverse and model dependent decay modes, so it is  likely to be less relevant to experimental searches. The last top partner $\tppp$ and the $B$ are heavy and can be ignored altogether.

The coupling of the $S$ to gauge bosons can be motivated by the analysis of the underlying gauge theory~\cite{Barnard:2013zea,Ferretti:2013kya} and is given at leading order by the Lagrangian
\begin{equation}
{\mathcal{L}}_{S V V} = \frac{A \cos\theta}{16\pi^2 f}\, S\, \bigg(\frac{g^2 - g'^2}{2} Z_{\mu\nu}\widetilde Z^{\mu\nu} +
g g'  F_{\mu\nu}\widetilde Z^{\mu\nu} + g^2  W^+_{\mu\nu}\widetilde W^{-\mu\nu}\bigg), \label{etacouplings}
\end{equation}
where the ``Abelian'' field strength tensors are defined as $V_\mn = \partial_\mu V_\nu - \partial_\nu V_\mu$, thus omitting the ``non-Abelian'' part, which would contribute to interactions with three and four gauge bosons that we ignore here. 
$A$ is a model dependent dimensionless anomaly coefficient: $1\lesssim A \lesssim 10$. For instance, in the model analysed in~\cite{Bizot:2018tds} $A$ is given by the dimension of the representation of the hyper-fermions.
Note that there are no couplings of type $S S V$ since the $S$ does not acquire a VEV. 
Also, there is no anomalous coupling $S  F_{\mu\nu}\widetilde F^{\mu\nu}$ to the EM  field,  thus the decay $S\to\gamma\, \gamma$ is highly suppressed and for $m_S\lesssim 2m_W$ the decay $S\to Z\, \gamma$ has near 100\% branching ratio. 
Once again, we can match the current model with the remaining couplings of the phenomenological Lagrangian in \eq{eq:LBSM}: 
\begin{equation}
\widetilde \kp_W = -\widetilde \kp_B = \frac{A v}{2 f}\cos\theta.
\end{equation}
The mass of $S$ is expected to be small $m_S\lesssim m_h$ and thus in the region where the decay into $\Zgam$ is motivated. 
In this particular model, it is given by $m_h/(2\cos\theta)$ plus corrections proportional to explicit underlying fermions masses, which are disfavoured by fine tuning arguments. For $t_R$ symmetric for example, $m_\eta$ tends to vanish and should get its mass completely from underlying fermion masses. Other representations and other models give different expressions, but all agree on the approximate estimate that $m_S$ is light due to its pNGB nature.

As far as direct $S$ production goes, we observe that, choosing the spurion embeddings as above, no diagonal coupling of type $S \, \bar t^i t^i$ ($t^i = t,\, \tprime,\, \tpp,\, \tppp$) is directly generated~\cite{Gripaios:2009pe}. 
This means that the gluon fusion process is not present and the direct production proceeds mainly via EW vector bosons. 
Diagonal fermionic couplings for the top and for lighter fermions can be induced by further enlarging the model but we ignore them and consider the fermiophobic case. 
The coupling of $S$ to fermions is nevertheless generated via loop of gauge bosons and might be relevant for low $\ms$~\cite{Bauer:2017ris,Craig:2018kne}.

%% file: LHCconstraints.tex
To perform a phenomenological analysis of the $\gamma\gamma$ and $\Zgam$ final states it is necessary to estimate the allowed regions in the masses of the VLQ and (pseudo)scalar. 
This is done in this section by recasting one ATLAS and one CMS search at \unit{13}{\TeV} and providing the ensuing limits in the \mtp vs \ms plane. 

The searches used for the recast are briefly described in the following. 
\begin{itemize}
\item An ATLAS ``Search for new phenomena in high-mass diphoton final states''~\cite{Aaboud:2017yyg}, used to set constraints for the $\gamma\gamma$ final state. This search looks for resonances with spin 0 or 2 decaying into two photons. For the spin 0 resonances (of interest for our analysis) the explored diphoton invariant mass region ranges from \unit{200}{\GeV} to \unit{2700}{\GeV}. The search cuts on the transverse energy of the leading and subleading identified photons, $\ET>40\, \GeV$ and $\ET>30\, \GeV$, respectively, and requires \ET to be larger than a fraction of the diphoton invariant mass, $\ET>0.4 m_{\gamma\gamma}\, \GeV$ (leading photon) and $\ET>0.3 m_{\gamma\gamma}\, \GeV$ (subleading photon). 
\item A CMS ``Search for standard model production of four top quarks with same-sign and multilepton final states''~\cite{Sirunyan:2017roi}, used to set constraints for the $\Zgam$ final state. This search looks for final states with two (same-charge) or three leptons, and different numbers of jets and $b$-jets, depending on the signal region. No cuts are imposed on photons in the final state. The most relevant cuts are applied to the jet and $b$-jet multiplicity and differ depending on the signal region.
\end{itemize}

The recast simulations are done using \madgraphfive\_\amcatnlo~\cite{madgraph} with a dedicated UFO~\cite{Degrande:2011ua} model file corresponding to the simplified Lagrangian in \eq{eq:LBSM}. 
Events are generated at leading order and interfaced with \pythia8.2~\cite{pythia} and \delphes 3~\cite{delphes} for showering and fast detector simulation. 
As Parton Distribution Functions (PDFs), the \nnpdf 3.1 at NLO set~\cite{Ball:2017nwa} has been chosen, obtained through the \lhapdf~6 library~\cite{Buckley:2014ana} using PDF ID 303400. 
The recast and validation of the searches is then performed through \madanalysis~5~\cite{Conte:2012fm,Conte:2018vmg}. 

Simulations have been performed in a grid of $\tprime$ and $S$ masses: $\mtp$ has been varied in the range \unit{400}{\GeV} to \unit{1000}{\GeV} in steps of \unit{100}{\GeV}, while \ms starts from a minimum value of \unit{200}{\GeV} and increases in steps of \unit{100}{\GeV} until reaching the kinematical limit $\mtp - \ms - m_t = 0$. 
A point in the small mass gap region $\mtp - \ms - m_t = 10\, \GeV$ has been included as well. 

\begin{figure}[tbp]
\centering
\includegraphics[width=.48\textwidth]{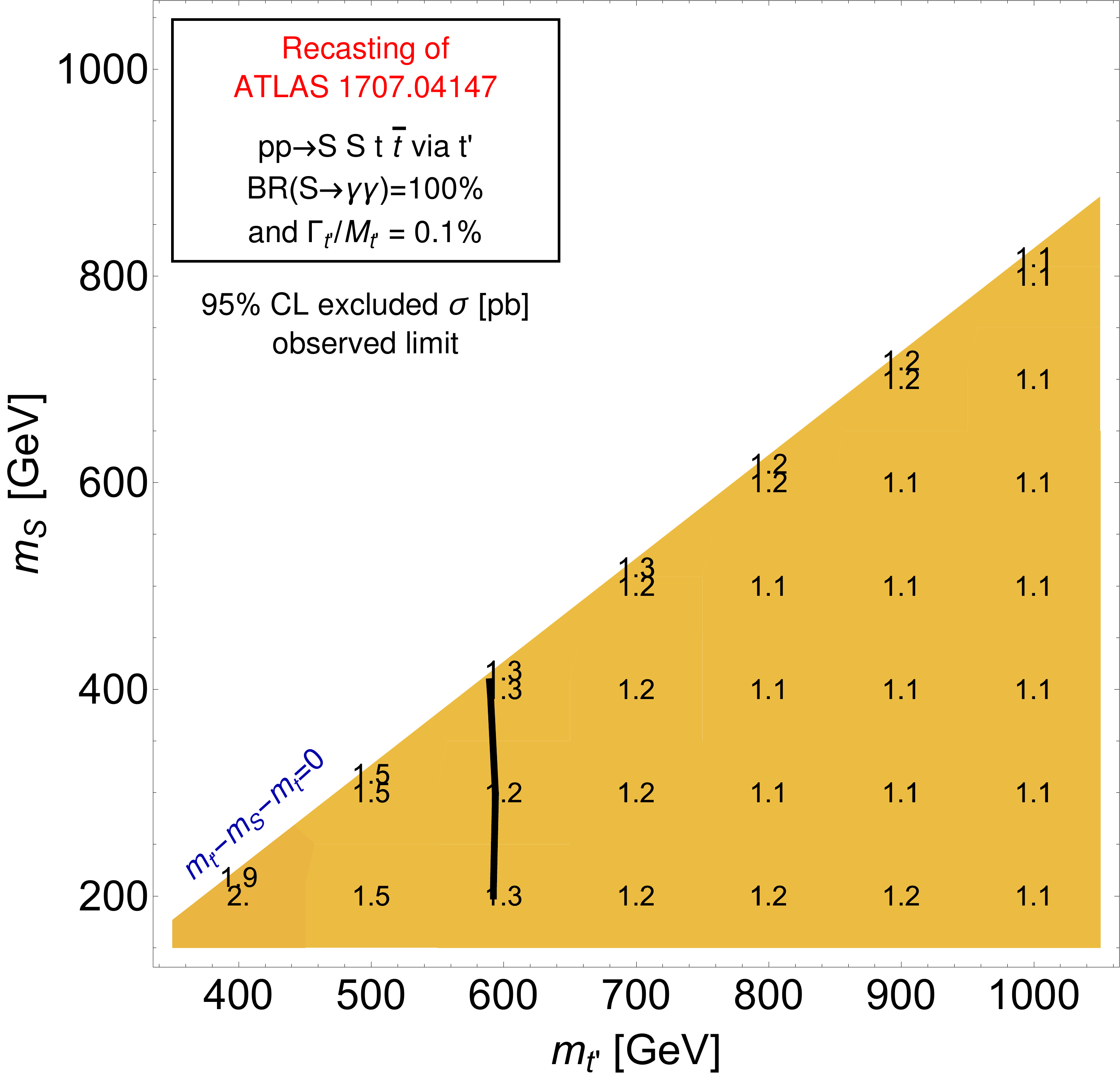}
\includegraphics[width=.48\textwidth]{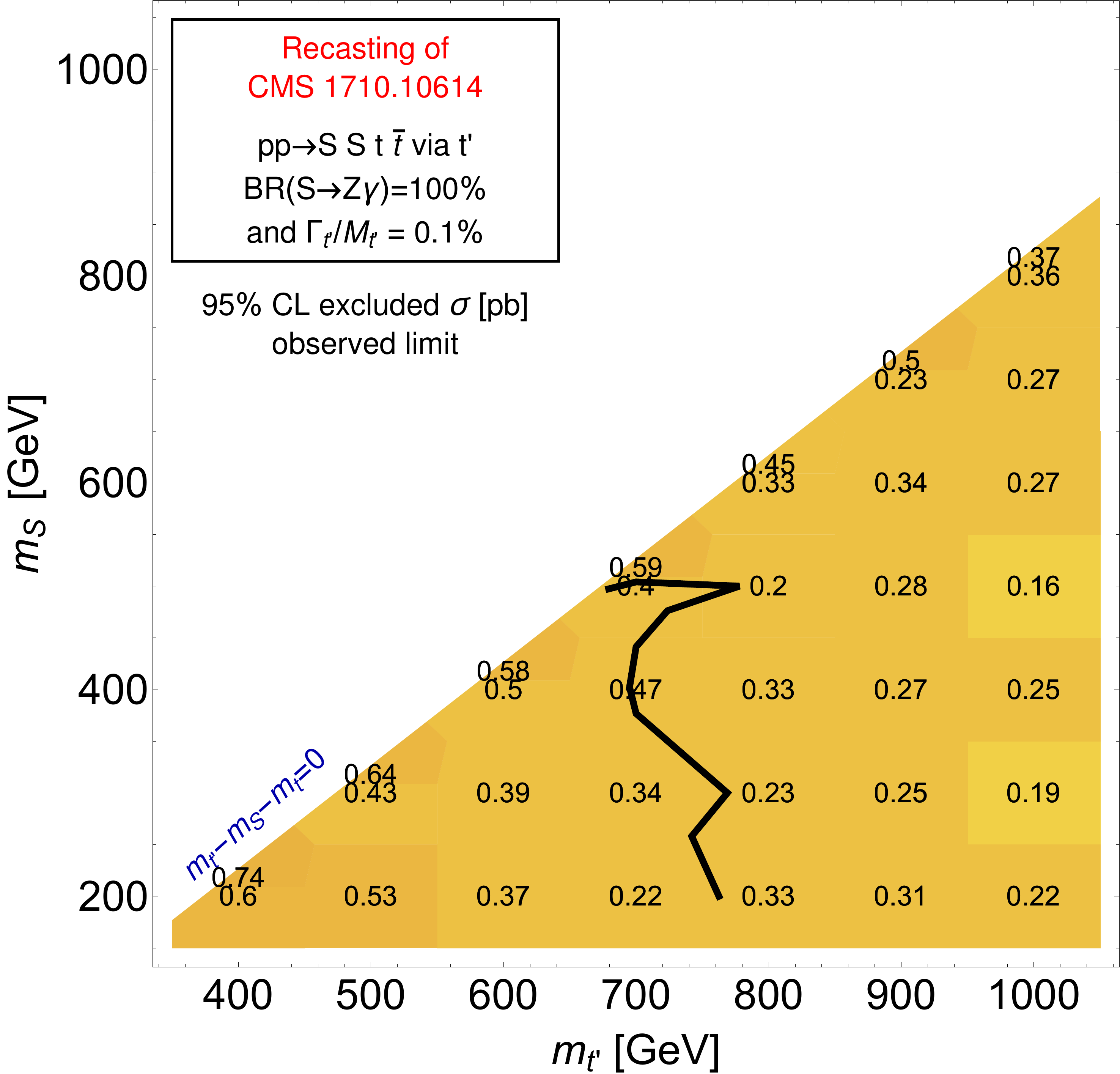}
\caption{\label{fig:recast_aa_aZ_resonance} Upper limits on the cross section in the \mtp vs \ms plane for the $\gamgam$ (left panel) and $\Zgam$ channels (right panel) from the recast of the ATLAS search~\protect\cite{Aaboud:2017yyg} and CMS search~\protect\cite{Sirunyan:2017roi}, respectively. The solid black lines represents the bounds on the two masses obtained by comparing the upper limits with the pair production cross section of \tprime at NLO+NNLL computed through \hathor~\protect\cite{Aliev:2010zk} under the assumption of 100\% \BRs for both \tprime and $S$ in the respective channels and in the narrow width approximation (NWA).}
\end{figure}

The results are shown in \fig{fig:recast_aa_aZ_resonance} as upper limits on the cross section (in pb). 
The observed bound on the \tprime and $S$ masses, represented as a solid black contour, has been obtained by comparing the upper bounds on the cross section with the cross section for pair production of \tprime obtained at NLO+NNLL through \hathor~\cite{Aliev:2010zk}, under the assumption of 100\% \BR for $\tprime \to \St$ and for $S \to \gamgam$ (\fig{fig:recast_aa_aZ_resonance} left panel) or $\Zgam$ (\fig{fig:recast_aa_aZ_resonance} right panel) in the narrow width approximation (NWA). 
The range of validity of the NWA in terms of the ratio between the total width and mass of \tprime is discussed in \app{sec:appendixnwa}. 
In the $\gamgam$ channels the allowed region for \mtp is above $\sim 600\, \GeV$ almost independently of \ms. 
In the $\Zgam$ channel the bounds are slightly more sensitive to the mass gap between the VLQ and the (pseudo)scalar, barring statistical fluctuations: the bound on \mtp is however between $\sim 700\, \GeV$ and $\sim 800\, \GeV$ for all the allowed \ms. 

The bounds obtained are typically weak compared to dedicated VLQ searches. We stress, however, that the bounds provided in this section are simply meant to give an idea about the optimal sensitivity of current searches for the final states considered above. In realistic scenarios the BRs of \tprime and $S$ into such final states will be likely smaller than 100\%, which trivially implies that the bounds will get weaker. In this case, other channels might be more sensitive depending on the BRs of the \tprime (and the recasting of different searches more sensitive to other final states has been performed, {e.g.} in \cite{Cacciapaglia:2019zmj}, after the appearance of this analysis). Indeed, only a \textit{combination} of bounds from different final states would give a full picture for any given benchmark point (defined in terms of masses and BRs of \tprime and $S$). The way bounds are provided in \fig{fig:recast_aa_aZ_resonance}, however, represents one of the elements of this picture. As a practical example, if a benchmark is considered in which the BRs of $\tprime \to \St$ or $S\to\gamgam$ or \Zgam are smaller than 100\%, the observed upper limits on the cross section represented by the grid of numbers in \fig{fig:recast_aa_aZ_resonance} can be directly compared with the $\sigma\times$ BRs of a given benchmark to determine the corresponding bound.

In the next section we propose a dedicated analysis to look for the signatures we are interested in leading to a much better sensitivity than the ones presented in \fig{fig:recast_aa_aZ_resonance}.

%% file: analysis.tex
\label{sec:analysis_strategy}

In its full generality, a top partner \tprime may decay into the usual three SM channels $\Wb$, $\Zt$, $\higgst$ or additional exotic channels. 
In this paper we are focusing our attention on the case of pair production $p\, p\to \tprime\, \antitprime$ and subsequent decay into the BSM channels $\tprime\to \St$, where $S$ is a neutral (pseudo)scalar decaying into SM EW diboson pairs.  
We have chosen the decays  $S\to \gamgam$ and $S\to \Zgam$ as our target signal, since they are experimentally very clean bosonic decay channels. 
In the case of the $\Zgam$ channel we only consider further leptonic decays of the $Z$.

The analyses are optimised to look for only one pair of photons or $Z\gamma$ final states originating from the same $S$.
When limits from these analyses are reinterpreted in specific models, the BRs of the $S$ can significantly affect the limits therein. In order to reinterpret the results in the models described in \sec{sec:model}, we need to evaluate the efficiencies of the signal region cuts while taking into consideration all possible decays of $S$.
We assume $\tprime$ decays at 100\% rate as $\tprime\to \St$. 
For $S$, we consider all the possible bosonic decay channels necessary to ensure gauge invariance in the CHM,\footnote{Note that additional \emph{sizable} decays are  present for the 2HDM+VLQ case, specifically, $gg$, $t\bar t^{(*)}$  and $hh$ (as appropriate for $S=H$ and $A$) decays, which are then simulated or estimated for the corresponding signal.}
\begin{equation}
 S \to \{\gamgam, \Zgam, WW, ZZ\}.
 \label{eq:Sdecays}
\end{equation}

In this section we briefly define the objects used in the analyses (with a longer discussion for reproducibility in \app{sec:object_definition_long}), then describe the tools and processes for the simulation of events to model signal and background (\sec{sec:simulations}), and finally we present event selections to extract the signal in the two considered signal regions (\SR): the {$\gamgam$ \SR} in \sec{sec:analysis_aa} and the {$\Zgam$ \SR} in \sec{sec:analysis_az}. 

\begin{figure}[tp]
\centering\includegraphics{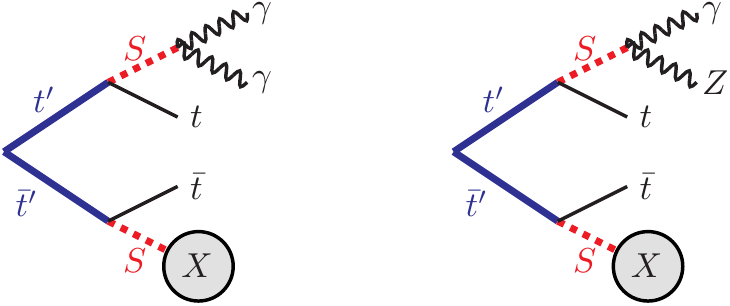}
\caption{\label{fig:semiinclusivetopology} Pair production of \tprime with decay of the \tprime into (anti-)top and $S$ in both branches. $S$ is then decayed in one branch into $\gamgam$ or $\Zgam$, depending on the signal pursued, and inclusively in the other branch.}
\end{figure}

\subsection{Object definition}
\label{sec:object_definition}

In the following the definition and selection of objects at reconstructed level are briefly outlined. 
A more detailed account can be found in \app{sec:object_definition_long}. 
The default ATLAS \delphes card~\cite{delphes} is used, with minor modifications and calorimeter objects that fall in the calorimeter transition region $1.37 < |\eta| < 1.53$ are excluded. 
Isolation and overlap removal is done in the \delphes card for most of the objects. 

The basic objects used are {photons} ($\gamma$), {leptons} ($\ell$), {jets} ($j$) and {$b$-jets} ($j_b$). 
Photons are required to have a $\pt > 30\,\GeV$ and $ |\eta|  < 2.37$. 
Leptons in this paper are understood to mean electrons or muons only, and not $\tau$-leptons. Leptons must fulfil $\pt > 25\, \GeV$ and $| \eta | < 2.47$.
Jets are reconstructed by using the \fastjet~\cite{fastjet} package and \delphes with the anti-$k_t$ algorithm~\cite{Cacciari:2008gp} using $R=0.4$. Jets are required to pass $\pt > 25\, \GeV$ and $| \eta | < 2.47$. 
In \delphes, a $b$-jets is a jet which contains a truth $b$-quark. 

The compound objects used are $Z$ bosons, missing transverse energy (\MET) and the scalar transverse energy (\Ht).
$Z$ bosons are identified as two opposite-sign same-flavour leptons with $|\mll - \mz|<10\, \GeV$, where \mll is the invariant mass of the reconstructed leptons. 
\MET is defined as $\vec{E}_{\text{T}}^{\text{miss}} = - \sum_i \vec{\pt}(i)$~\cite{delphes}, where $i$ runs over the energy deposits in the calorimeter. 
\Ht is the scalar sum of the \pt of all reconstructed basic objects used in the analysis (jets, muons, electrons and photons).

\subsection{Simulations}
\label{sec:simulations}

All simulations in this study have been performed using the following framework: \madgraph5\_\amcatnlo~\cite{madgraph} was used to generate events at leading order accuracy. 
\pythia8.2~\cite{pythia} and \delphes3~\cite{delphes} have been used for showering and fast detector simulation, respectively. 
For the signal simulations, the parton distribution function (PDF) \nnpdf 3.1 at NLO set~\cite{Ball:2017nwa} set has been chosen, obtained through the \lhpdf~6 library~\cite{Buckley:2014ana} using PDF ID 303400. 
For the background simulations instead the \madgraph default \nnpdf 2.3 LO with PDF ID 230000 has been used. 

\begin{figure}[tp]
\centering
\includegraphics[width=0.69\textwidth]{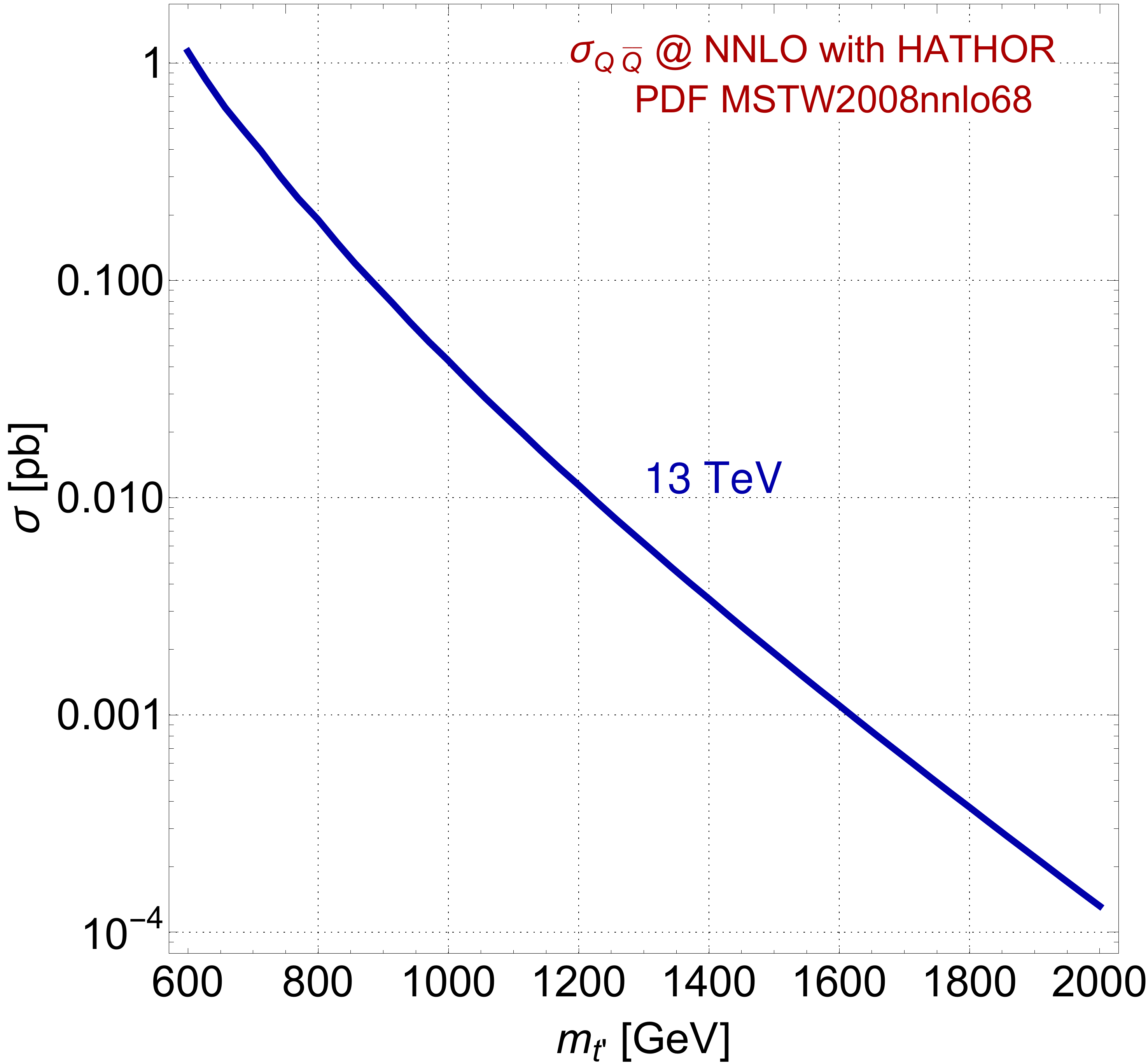}
\caption{\label{fig:pairsigma} Pair production cross section of \tprime at NLO+NNLL computed through \hathor~\protect\cite{Aliev:2010zk}, with NNLO MSTW2008~\protect\cite{Martin:2009iq} PDFs. }
\end{figure}

The numerical values of the pair production cross-sections, which only depend on \mtp, are shown in \fig{fig:pairsigma}.
They were computed through \hathor~\cite{Aliev:2010zk}, with NNLO MSTW2008~\cite{Martin:2009iq} PDFs. 

\begin{figure}[t]
  \begin{center}
    \includegraphics[width=0.95\textwidth]{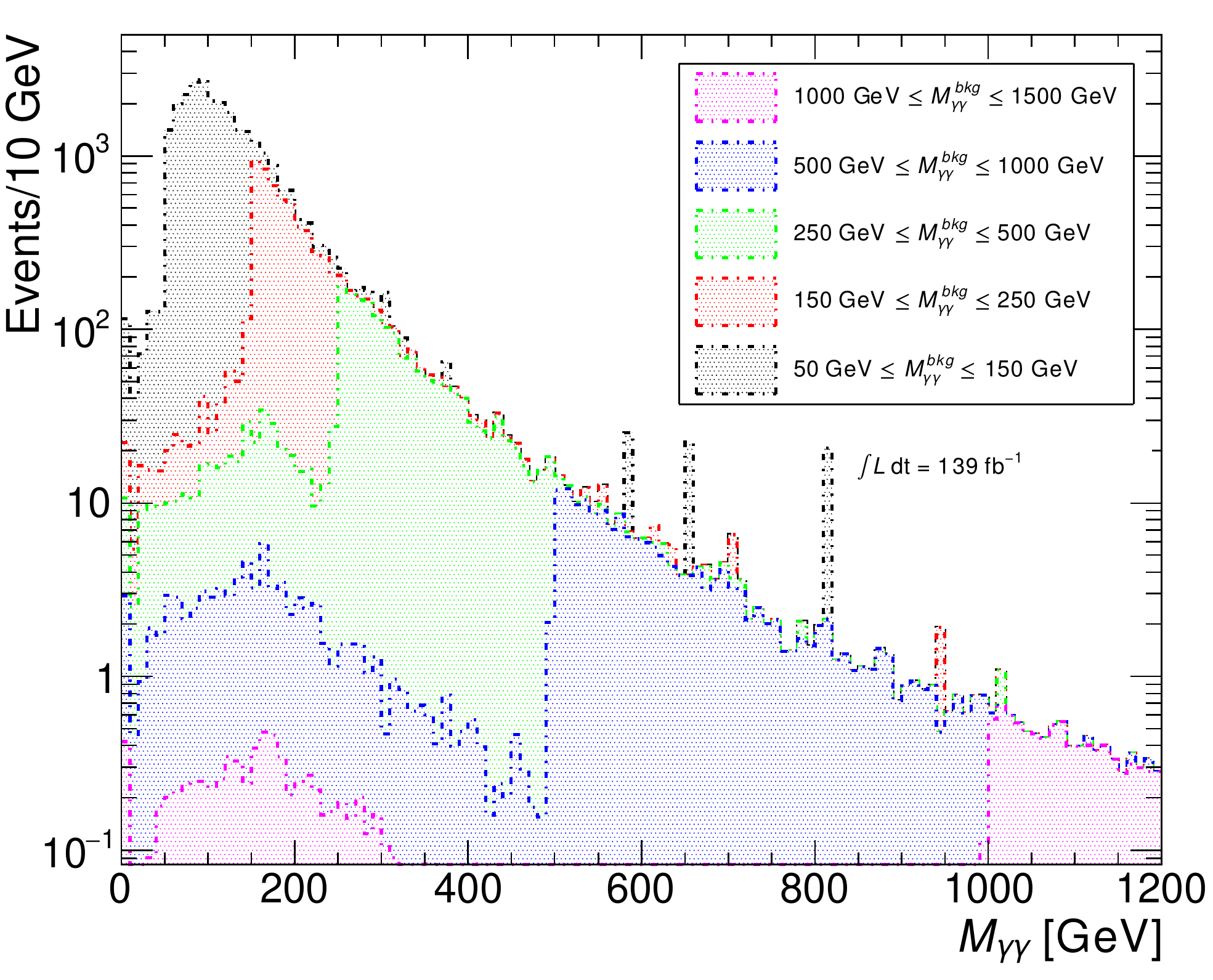}
        \caption{Invariant mass \mgamgam of the photon pair at reconstructed level for each \mbkggamgam slice in the $\gamgam + \text{jets}$ background. At least two photons and one $b$-jet, as defined in \sec{sec:object_definition}, are required. 
The contributions from the slices are stacked. 
\label{fig:bkg_gammagamma}}
  \end{center}
\end{figure}

The background  of the $\gamgam$ \SR  is dominated by $pp \to \gamgam + \text{jets}$ mediated by QCD interactions. 
The backgrounds $\gamgam+t+\text{jets}$ and $\gamgam+\ttbar$ were found to be negligible and hence are not considered for the diphoton analysis.
Events from the $pp \to \gamgam + \text{jets}$ process are generated with up to three jets, including jets initiated by $b$-quarks, in the matrix element. 
The final jets after showering and jet clustering are matched to the original partons with the MLM method~\cite{Mangano:2006rw} as implemented in \pythia. 
In the simulation of the initial state $b$-quarks are explicitly considered as part of the incoming protons. 
This accounts for processes with an odd number of $b$-jets in the final state, such as those initiated by $gb \to \gamgam + u \bar{u} b$. 
To ensure enough statistics in the high mass tail the events are generated in slices of the diphoton invariant mass \mbkggamgam with $\sim1$~M events per slice, where \mbkggamgam refers to the invariant mass of the generated (not reconstructed) photons. 
Table~\ref{tab:bkg_slices} lists the slices along with the fiducial cross section for each slice. 
The invariant mass of the two photons for all slices is shown in \fig{fig:bkg_gammagamma}. 
If there are more than two photons in the event, the pair with invariant mass closer to $160\, \GeV$ is shown in this figure. 
The high-mass slices have small tails towards lower masses, which occurs when one or both of the hard photons is lost in the reconstruction and the selected photons originate from e.g. the hadronisation process. 
The contribution from these mis-reconstructions is typically small and can be mitigated further with \dR cuts on the photons. 
The small peak at $160\, \GeV$ is due to the selection requirement that the invariant mass of the photons is close to $160\, \GeV$.
The total fiducial cross section in the $\mbkggamgam > 50\, \GeV$ region is calculated by generating $25$K events in the allowed range using the same setup as in the full event generation, resulting in \unit{74.0}{pb}, in good agreement with the sum of the fiducial cross sections for the individual slices.

\begin{table}[t]
   \centering
   \begin{tabular}{r@{$~-~$}l|r@{$~\pm~$}l|r@{$~\pm~$}l}
   \multicolumn{2}{c|}{Background process} & \multicolumn{2}{c|}{$\sigma_\text{fid.}(\gamgam +\text{jets})$~[pb]} & \multicolumn{2}{c}{$\sigma_\text{fid.}(Z\gamma +\text{jets})$~[pb]} \\
   \hline
     50 &  150~\GeV & $69.0$ & $0.2$    & $3.223    $ & $ 0.003$    \\
    150 &  250~\GeV & $3.577   $ & $ 0.006$   & $1.010   $ & $ 0.001$   \\
    250 &  500~\GeV & $(91.3   $ & $ 0.2)\times10^{-2}$   & $(22.56   $ & $ 0.02)\times10^{-2}$   \\
    500 & 1000~\GeV & $(99.2$ & $ 0.2)\times10^{-3}  $  & $(25.43 $ & $ 0.03)\times10^{-3}$  \\
   1000 & 1500~\GeV & $(63.6 $ & $ 0.2)\times10^{-4}$ & $(1.764 $ & $ 0.002)\times10^{-3}$ \\
   \hline
   \hline
   \multicolumn{2}{c|}{Sum} & $73.6 $ & $ 0.1$ & $4.486 $ & $ 0.003$ \\
   \hline
   \multicolumn{2}{c|}{Estimated total} & $74.0  $ & $ 0.6$  & $4.45 $ & $ 0.03$ \\
   \end{tabular}
\caption{Fiducial cross section for each mass slice of the two major background processes. For the $\gamgam + \text{jets}$ background the slices refer to $\mbkggamgam$ while for the $\Zgam + \text{jets}$ background the slices refer to \mbkgzgam at the generator level. The sums of the fiducial cross sections over all slices for each process are also listed together with their estimated value.\label{tab:bkg_slices}}
\end{table}

\begin{figure}[t]
  \begin{center}
    \includegraphics[width=0.95\textwidth]{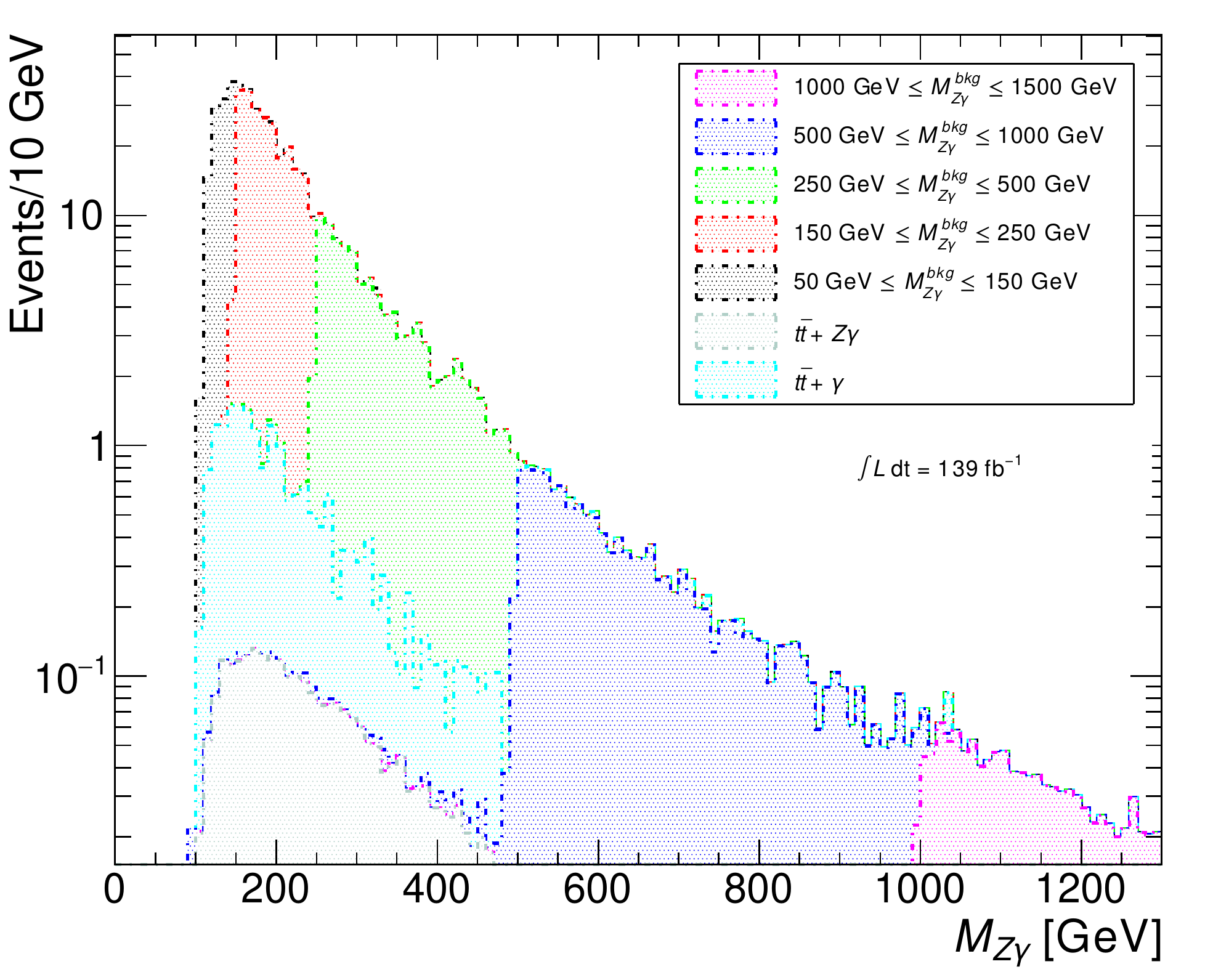}
       \caption{Invariant mass \mzgam of the reconstructed $Z$ boson and the photon for each mass slice in the $\Zgam + \text{jets}$ background, as well as for the $\ttbar+\gamma$ and $\ttbar+Z+\gamma$ backgrounds. At least one $Z$ boson, one photon and one $b$-jet, as defined in \sec{sec:object_definition}, are required. \label{fig:bkg_Zgamma}}
  \end{center}
\end{figure}

The dominant background in the $S \to \Zgam$ final state is $pp \to \Zgam + \text{jets}$, with $Z \to \ell^+\ell^-$. 
Events from this process are generated using the same setup as for the $\gamgam + \text{jets}$ background, with up to two hard jets in the matrix elements. 
For the same reason as for $\gamgam + \text{jets}$ the event generation for the $Z\gamma + \text{jets}$ background is performed in slices of the invariant mass of the generator-level $Z$ and $\gamma$, \mbkgzgam,  with $\sim 2$M events each, listed in \tab{tab:bkg_slices} together with their fiducial cross section. 
The latter at $\mbkgzgam > 50\, \GeV$ is estimated to be \unit{4.451}{pb} by generating $25$K events in the allowed kinematic range, which, again, is in good agreement with the sum of the fiducial cross sections of the slices. 
SM top-quark pair production associated to a photon and to a $Z$ and a photon can also give relevant contributions to the background. 
We generated 150K events of the process $\ttbar+Z\gamma$ and let the top decay inclusively and the $Z$ leptonically via \madspin. 
For  $\ttbar+\gamma$ we generated 300K events and required the top quarks to decay leptonically to either electrons or muons. 
We use the LO cross sections \unit{0.315}{fb} for decayed $\ttbar+Z+\gamma$ and \unit{94}{fb} for decayed $\ttbar+\gamma$ events. 
The invariant mass of the  $\Zgam$ system, for each of the mass slices of $\Zgam$ + jets, together with $\ttbar+\gamma$ and $\ttbar+Z+\gamma$, is shown in \fig{fig:bkg_Zgamma}. 
In that figure, at least one $Z$ boson, one photon and one $b$-jet, according to the definitions in \sec{sec:object_definition}, are required. 
If there are more than one $Z$ and/or $\gamma$ candidate we choose the system with invariant mass closer to $160\, \GeV$ to present in this specific plot. 

In both final states, non-prompt backgrounds are also possible. 
These are expected to be reduced significantly since we use tight identification requirements for leptons and photons. 
Furthermore, in analyses with similar final states, the backgrounds with one or more jets mis-identified as photons was found to be significantly smaller than those with prompt photons~\cite{Aaboud:2017yyg_aux}.
Thus, we do not consider non-prompt background sources in either of the final states.

For the signal simulation and definition, we generated the process $pp\to \tprime \bar{\tprime}$ with $\tprime \to \St$ and  $S$ decaying into EW bosons, \eq{eq:Sdecays}. 
We define our signal samples as any possible decay combination, $(S\to X)(S\to Y)$ where $X,Y\in\{\gamgam, \Zgam, WW, ZZ\}$. 
Both the $Z$ and $W$  decay inclusively in our signal definition. 

The UFO model for signal simulations is the same one used for recasting LHC bounds, corresponding to the simplified Lagrangian of \eq{eq:LBSM}. 
Decays of interest are thus turned on or off by setting the corresponding couplings. 
In the following analysis, couplings are set such that the widths for the top partner \tprime and scalar $S$ are $0.1\%$ of their mass, to allow the use of the NWA. A quantitative determination of this parameter, performed in \app{sec:appendixnwa}, is essential to determine the range of validity of signal simulations in experimental analyses and also for the subsequent reinterpretation of results in terms of theoretical models. 

For the simulations, we use $\kappa_S^R=0$, keeping only the $\kappa_S^L$ coupling. 
This is an important assumption, as fixing a different chirality of the top coupling can lead to observable differences. Indeed, it is known that the dominant chirality of the couplings of a VLQ interacting with the SM top quark can be probed by looking at the transverse momentum of the decay products of the $W$ boson emerging from the top quark~\cite{Belanger:2012tm,Barducci:2017xtw}. Differently from the SM case, however, here the kinematics of the decay products of $\tprime$ is not only affected by its mass, but also by the $S$ mass.

Similarly we turn off the scalar $S$ couplings, $\kappa_W=\kappa_B=\lambda_W=\lambda_Z=0$, when we assume a pseudoscalar nature of the $S$ state. 
The scalar or pseudoscalar nature of $S$ can also in principle affect the kinematical distributions of its decay products. We have therefore performed simulations imposing specific decay channels, to check, at reconstruction level but without including detector effects, how large differences can be between the above scenarios in differential distributions. We found that there is no observable difference in our predictions with respect to a scalar $S$ in terms of kinematical distributions.
In view of this indistinguishability, in the 2HDM+VLQ case, we will assume the $S$ state to represent alternatively a CP-even and CP-odd neutral Higgs states entering the \tprime decay.

\subsection{$S\to \gamma \gamma$ signal region}
\label{sec:analysis_aa}

In this section, the diphoton final state is presented. 
From an experimental point of view, the diphoton final state gives a very clean signature in the detector, which makes it attractive to study. 

We considered  \tprime masses $\mtp= 600$ to $1800\, \GeV$ in steps of $200\, \GeV$, every kinematically allowed $S$ mass is investigated, via the discrete values of $\ms = 100\, \GeV$,  $200\, \GeV$, $400\, \GeV$, and then in steps of $200\, \GeV$ up to the highest kinematically available mass, $\ms = \mtp - 200\, \GeV$.    
The wide selection of $S$ and $\tprime$ masses enables the possibility to study both threshold effects and highly boosted decay products. 

To select the signal we demand the presence of 2 photons and 1 $b$-jet defined according to \sec{sec:object_definition}. 
If more than one pair of photons is present we choose the pair whose invariant mass is closer to \ms and define these photons as ``best'' photon candidates, $\gamma_1$, $\gamma_2$. 
Unless otherwise specified, a pair of photons is assumed to be the ``best'' pair. 
The invariant mass of the system with the two ``best'' photon candidates  is required to be within $20\, \GeV$ from the nominal $S$ mass, $|\mgamgam - \ms| < 20\, \GeV$. 

\begin{figure}[t]
  \begin{center}
      \includegraphics[width=0.95\textwidth]{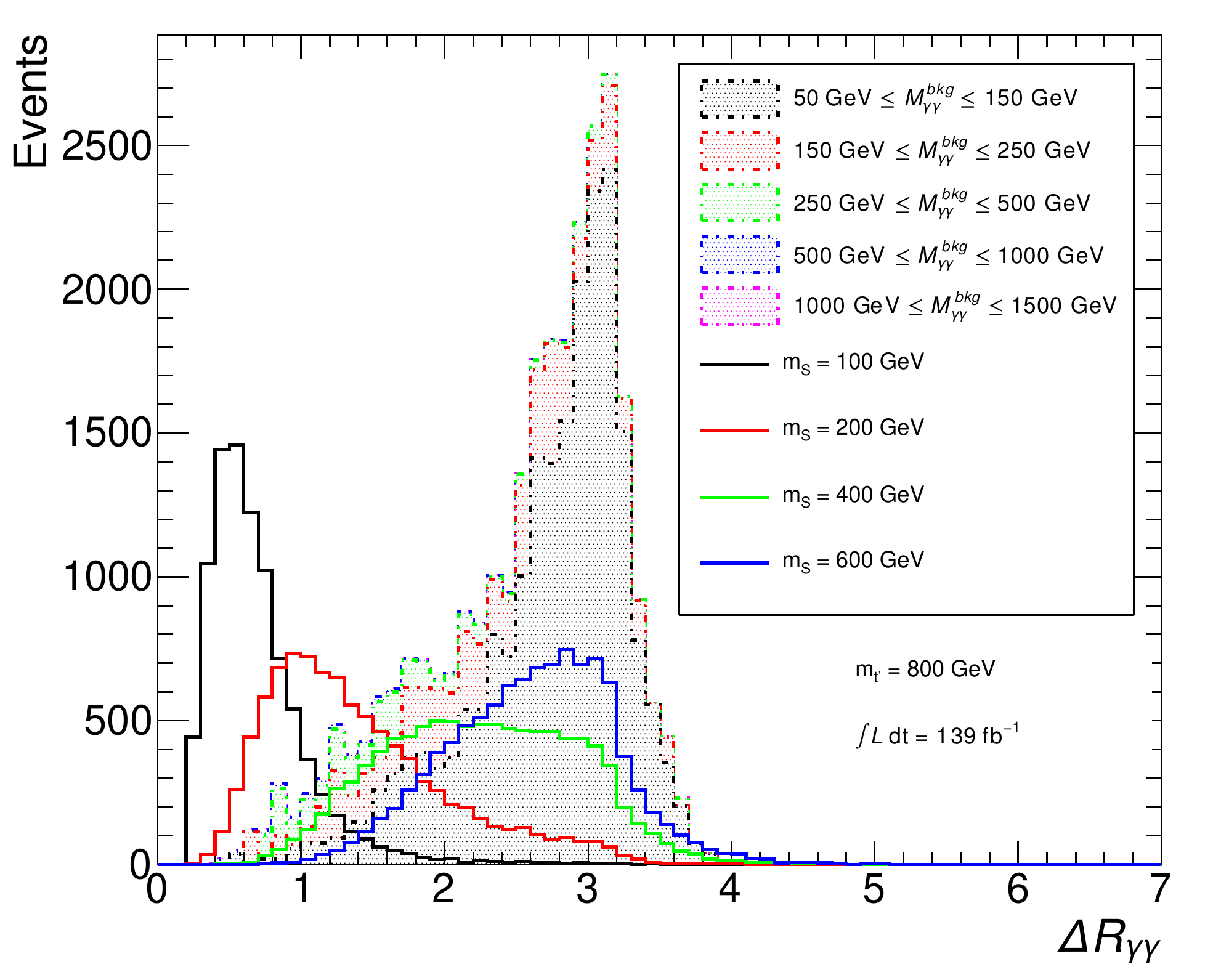}
      \caption{Distributions of \dRgamgam of the two photons with  invariant mass closest to \ms, at the reconstructed level, with $\mtp = 800\, \GeV$ and various \ms values. In the plot, cuts 1 and 2, as defined in \tab{tab:cuts_gammagamma}, have been applied. \label{F:dRgamgam_gammagamma}}
  \end{center}
\end{figure}

In order to further enhance the signal discrimination with respect to  the background for low \ms values we use the fact that the $S$ is produced in a \emph{boosted} regime. 
The top partners \tprime and $\bar{\tprime}$ will be produced  nearly at rest and the pair will be back-to-back. 
The large difference in mass between \tprime and $S$ will make $S$  boosted and thus also the photon pair from $S$ will be  collimated. 
In \fig{F:dRgamgam_gammagamma} we show the $\dRgamgam$ distributions for different \ms and for $\mtp=800\,\GeV$ fixed. 
We take advantage of this characteristic signal profile and require  $\dRgamgam<2.3$ from $\ms= 100\, \GeV$ to $\ms = 200\, \GeV$. 

The selection cuts are summarised in \tab{tab:cuts_gammagamma}. 
Note that, due to limitations in statistics, the cuts are sub-optimal. 
The discrimination between signal and background could be improved significantly by tightening the cuts in a real experimental analysis. 
 
\begin{table}[t]
  \centering
  \begin{tabular}{c|l}
Cut no. & Description \\
\hline
1  &  $N_\gamma \geq 2$ \\
2  &  $N_{\text{b-jets}} \geq 1$\\ 
3  &  $|\mgamgam - \ms| < 20\, \GeV$ \\
4  &  $\dRgamgam< 2.3$ ($\ms\leq 200\,\GeV$)\\
  \end{tabular}
  \caption{Selection cuts applied to the $S\to \gamgam$  signal region. The cuts are described in detail in the text. Refer to \sec{sec:object_definition} for the definition of the objects. \label{tab:cuts_gammagamma}}
\end{table}

In \tab{tab:SB_efficiencies_gammagamma} we show the efficiencies (number of events left after the cut divided by the number before the cut) of the selection cuts numbered in \tab{tab:cuts_gammagamma} for different \ms values.
In the upper part of the table, the signal process is defined with both $S$ decaying into diphotons, i.e., $\ttbar S(\to \gamgam)S (\to\gamgam)$ in the final state. 
This is the process we use to optimise the selection cuts. 
We display only the $\mtp = 1\,\TeV$ case in the table. 
In the lower part of the table, the efficiencies for the background sample are displayed.
It can be noticed that the last two cuts are the most efficient ones in removing the background and keeping signal events.

\begin{table}[htb]
  \centering
  \begin{tabular}{c|r|r|r|r|r}
 \ms & $100\,\GeV$ &  $200\,\GeV$ &  $400\,\GeV$ &  $600\,\GeV$  & $800\,\GeV$ \\
\hline
Cut no. & \multicolumn{5}{c}{Signal $\ttbar(S\to \gamgam) (S\to \gamgam)$ efficiency (\%)} \\
\hline
1 &  98.1 &  98.8 &  99.1 &  99.0 &  98.8 \\ 
2 &  48.8 &  47.9 &  51.0 &  54.8 &  60.4 \\ 
3 &  35.9 &  35.9 &  39.4 &  42.9 &  46.4 \\ 
4 &  35.8 &  34.0 &  39.4 &  42.9 &  46.4 \\ 
\hline \hline
Cut no. & \multicolumn{5}{c}{Background efficiency (\%)} \\
\hline
1 &  15.4 &  15.4 &  15.4 &  15.4 &  15.4  \\
2 &$2.8\times 10^{-1}$ &$2.8\times 10^{-1}$ & $2.8\times 10^{-1}$ & $2.8\times 10^{-1}$ & $2.8\times 10^{-1}$ \\      
3 &$5.7\times 10^{-2}$ &$1.2\times 10^{-2}$ & $1.1\times 10^{-3}$ & $2.5\times 10^{-4}$ & $7.1\times 10^{-5}$ \\
4 &$2.0\times 10^{-2}$ &$1.9\times 10^{-3}$ & $1.1\times 10^{-3}$ & $2.5\times 10^{-4}$ & $7.1\times 10^{-5}$ \\

  \end{tabular}
  \caption{Signal and background efficiencies in percent following the cuts listed in \tab{tab:cuts_gammagamma}, for the \gamgam SR and $\mtp=1000\, \GeV$. \label{tab:SB_efficiencies_gammagamma}}
\end{table}

The final efficiencies for the signal decay channel $S(\to \gamgam)S(\to\gamgam)$ are discussed in \sec{sec:efficiencies}. 
The efficiencies for the other signal decay channels with at least one branch decaying into $\gamgam$ are presented in \app{sec:appendixefficiencies}.

\subsection{$S\to \Zgam$ signal region}
\label{sec:analysis_az}

In the $S \to \Zgam$ final state we require at least one $Z$ boson candidate reconstructed according to the definitions in \sec{sec:object_definition}. 
In addition to the $Z$ candidate we require the presence of at least one isolated photon. 
The system of one isolated photon and one $Z$ candidate whose invariant mass is closest to the nominal $S$ mass is called the ``best $S$ candidate''. 
To efficiently distinguish the signal from the background we exploit the high multiplicity of objects and high total energy of a  typical signal event.  
We require  $\Ht+\MET>0.3 \mtp$, where $\Ht$ is the scalar sum of the \pt of all reconstructed basic objects and $\MET$ is the missing transverse energy of the event as described in \sec{sec:object_definition}. 
We finally require the invariant mass of the $S$ candidate to be within $15\, \GeV$ of the nominal $S$ mass, i.e., $|\mzgam - \ms|<15\GeV$. 
A summary of these selection cuts is presented in \tab{tab:Zacuts}, with some information on the object definitions for convenience.

\begin{figure}
\centering
\includegraphics[width=0.48\textwidth]{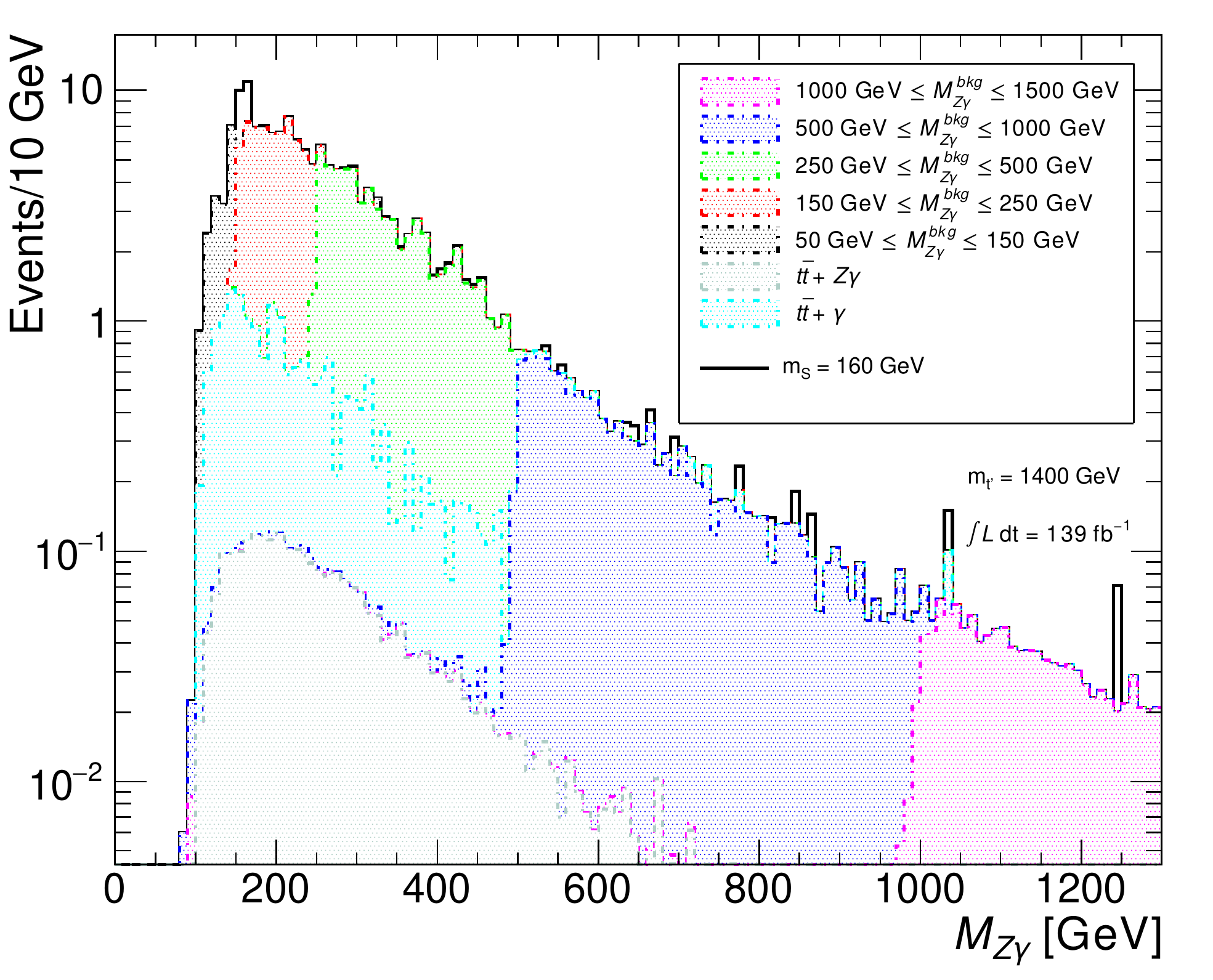}
\includegraphics[width=0.48\textwidth]{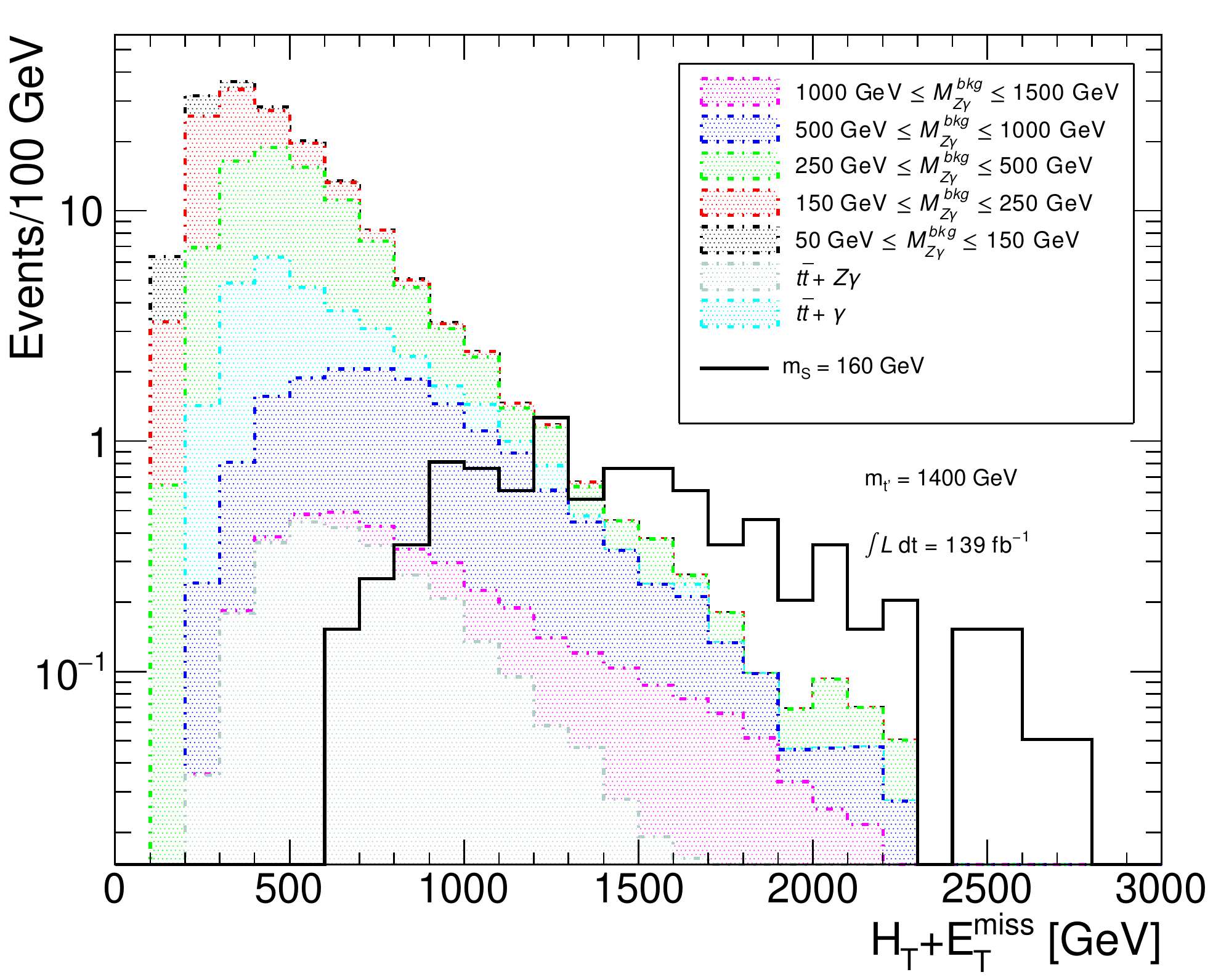}\\
\caption{Distributions of \mzgam and $\Ht+\MET$ for $\ms=160\,\GeV$ and $\mtp=1400\,\GeV$. \label{fig:Zadist}}
\end{figure}

The  distributions of \mzgam before cut 5 and $\Ht+\MET$ before cut 4 and 5 are shown in \fig{fig:Zadist}, for the masses $\ms=160\,\GeV$ and $\mtp=1400\,\GeV$. 
There is a great discriminating power in the $\Ht+\MET$ observable due to the large multiplicity and energy of a typical signal event. 
We note that the used cut is not optimised to suppress the background due to lack of MC statistics. 
A realistic experimental analysis could harden this cut to further reduce the background and use data-driven methods to estimate it without relying too much on MC estimates. 
\begin{table}[t]
  \centering
  \begin{tabular}{c|l}
Cut no. & Description \\
\hline
1  &  $N_Z \geq 1$\\
2  &  $N_\gamma \geq 1$\\
3  &  $N_{\text{b-jets}} \geq 1$\\ 
4  &  $\Ht+\MET>0.3\mtp$  \\
5  &  $|\mzgam-\ms|<15\,\GeV$\\
  \end{tabular}
 \caption{Selection cuts applied to the $\Zgam$ signal region. Details on the cuts are given in the text. Refer to \sec{sec:object_definition} for the definition of the objects. \label{tab:Zacuts}} 
\end{table}

For illustrative purposes, in \tab{tab:SB_efficiencies_Zgamma}, we show the efficiencies of the selection cuts numbered in \tab{tab:Zacuts} for different \ms values. 
We display only the case $\mtp = 1400\,\GeV$ in the table. 
In the upper subtable, the signal process is defined with both $S$ decaying into $\Zgam$,  $S(\to \Zgam )S(\to \Zgam)$ in the final state. 
This is the process we use to optimise the selection cuts. 
In the lower subtable, the efficiencies for the background sample are displayed. 
Except the mass-window cut for the $S$ candidates, all cuts depend on \mtp.
\begin{table}[htb]
  \centering
  \begin{tabular}{c|r|r|r|r}
 \ms & $130\,\GeV$ &  $160\,\GeV$ &  $400\,\GeV$  & $800\,\GeV$ \\
\hline
Cut no. & \multicolumn{4}{c}{ Signal $(S\to \Zgam)(S\to \Zgam)$ efficiency (\%)} \\
\hline
1 & 4.11  & 4.81   & 5.80  &  5.68 \\      
2 & 2.74  & 3.97   & 5.39  &  5.13\\ 
3 & 1.51  & 2.31   & 3.27  &  3.61 \\ 
4 & 1.51  & 2.31   & 3.27  &  3.61 \\ 
5 & 1.19  & 1.77   & 2.43   & 2.36  \\ 
\hline \hline
Cut no. & \multicolumn{4}{c}{Background efficiency (\%)} \\
\hline
1 & 21.7   & 21.7      & 21.7   & 21.7     \\	
2 & 4.13   & 4.14      & 4.14  & 4.14  \\ 
3 & 0.0731 & 0.0731    & 0.0731 & 0.0731 \\ 
4 & 0.0461 & 0.0463    & 0.0463 & 0.0463 \\   
5 & 5.22$\times 10^{-3}$ & 8.09$\times 10^{-3}$ & 9.39$\times 10^{-4}$  & 6.81$\times 10^{-5}$ \\   
\end{tabular}
  \caption{Signal and background efficiencies in percent following the cuts listed in \tab{tab:Zacuts}, for the \Zgam SR and $\mtp=1000\, \GeV$. \label{tab:SB_efficiencies_Zgamma}}
\end{table}

\subsection{Efficiencies}
\label{sec:efficiencies} 

\begin{figure}[t]
  \begin{center}
      \includegraphics[width=0.49\textwidth]{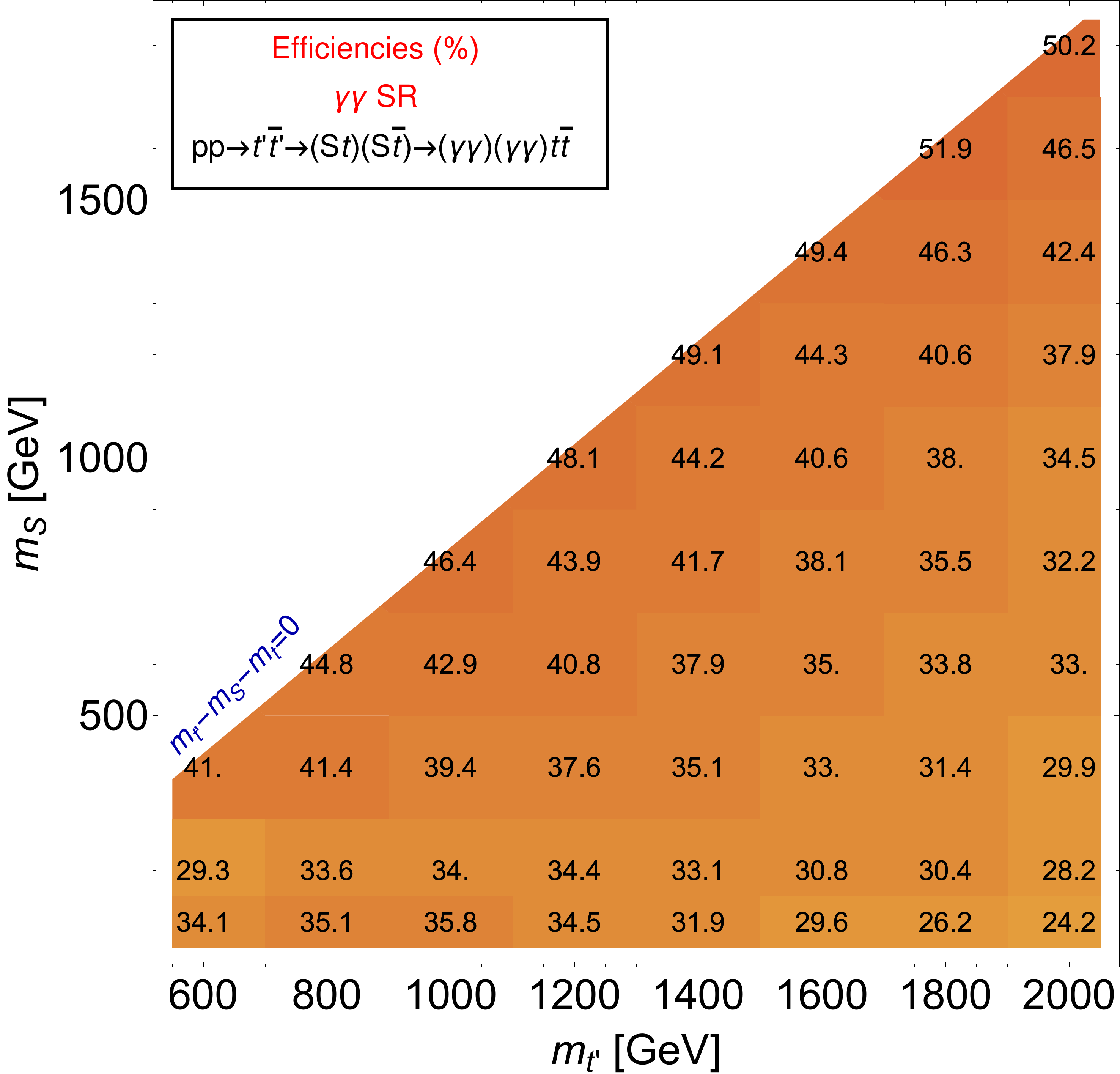}
      \includegraphics[width=0.49\textwidth]{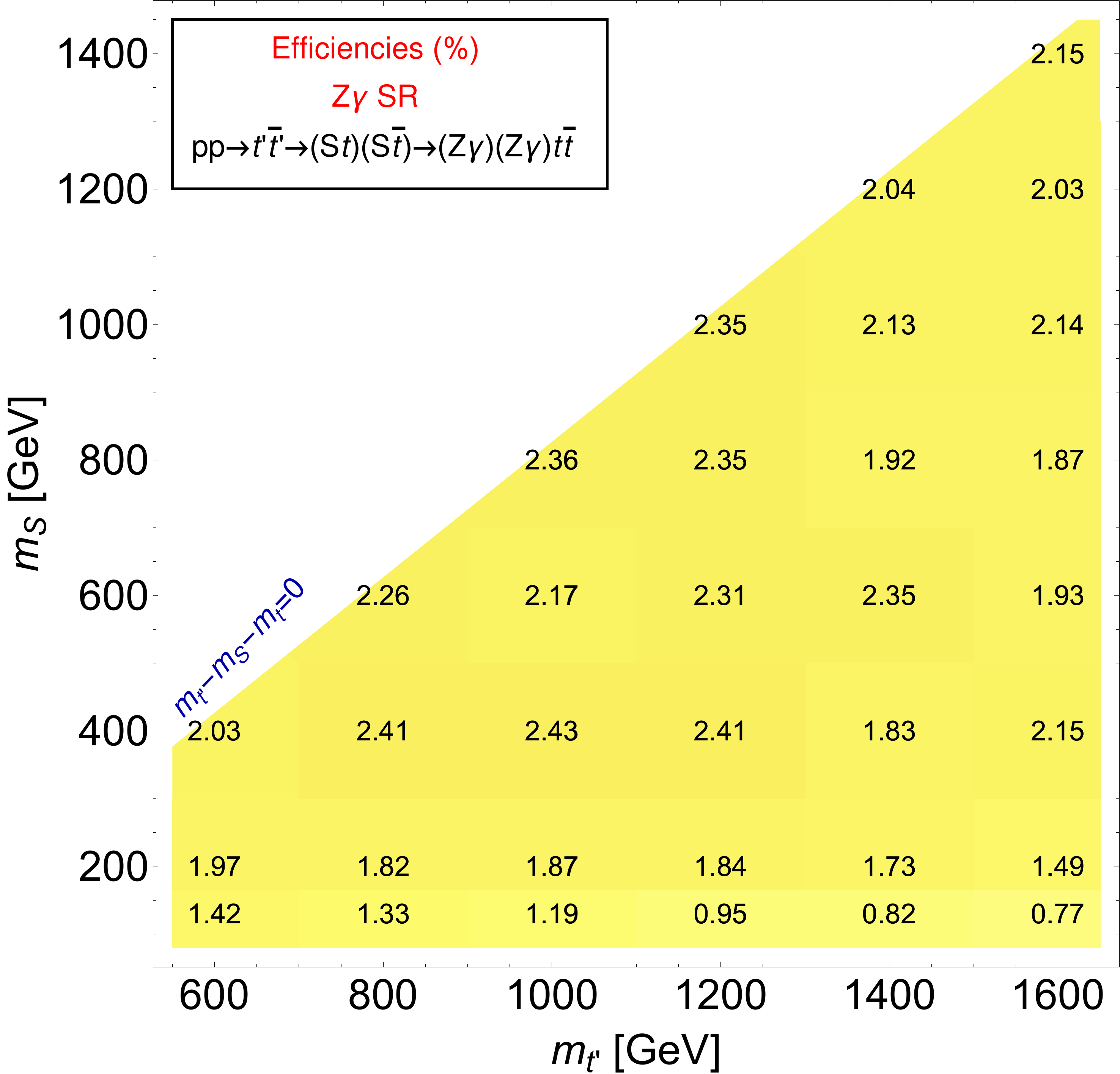}
      \caption{{Left:} Efficiencies for the $\gamgam$ SR for the signal decay channel $S(\to \gamgam)S(\to\gamgam)$. {Right:} Efficiencies for the $\Zgam$ SR for the signal decay channel $S(\to \Zgam)S(\to \Zgam)$.\label{fig:efficiencies}}
  \end{center}
\end{figure}

The signal efficiencies for the two different signal regions are the last piece of information necessary for reconstructing the number of signal events. 
In \fig{fig:efficiencies} we provide, as illustrative examples, the efficiencies for the $(\gamgam)(\gamgam)$ channel in the $\gamgam$ SR and for the $(\Zgam)(\Zgam)$ channel in the $\Zgam$ SR, for which the selections have been optimised. 
Further efficiency plots for different channels are provided in \app{sec:appendixefficiencies}. All efficiencies have been computed considering signal samples of $10^4$ MC events, corresponding to a statistical uncertainty of the order of 10\% which can affect the evaluation of efficiencies especially when they are small.
The whole set of efficiencies, combined with the \BRs chosen in \sec{sec:analysis_strategy}, allows one to compute the expected total number of events via \eq{Stot} in the following section, where the results of the study are discussed. 

In the next section we will show how to estimate the number of events for both signal and backgrounds for different model assumptions and devise a simple statistical framework for model interpretation.

%% file: interpretation.tex
In this section we discuss the discovery potential of LHC for the models introduced previously. 
Essentially, we propose a counting experiment comparing the number of expected background events with the number of signal events.

The expected number of background events in one of the signal regions SR $\in \{\gamgam,\Zgam\}$, $B_{\rm SR}$, is given by 
\begin{equation}
B_{\rm SR}(\ms,\mtp) = L~\sigma_{B_{\rm SR}} \epsilon_{B_{\rm SR}}(\ms,\mtp)
\end{equation}
with $L$ the integrated luminosity, and $\sigma_{B_{\gamgam}}=74.0\, \text{pb}$ and  $\sigma_{B_{\Zgam}}=4.58\, \text{pb}$ our best estimate of the total background cross section for the $\gamgam$ and $\Zgam$ signal regions, respectively, and $\epsilon_{B_{\rm SR}}$ the efficiency after all cuts in the corresponding SR. 

The number of background events can be extracted for arbitrary values of \ms and \mtp by interpolating the data presented in \tabs{tab:BGgammagamma}{tab:BGZgamma}. 
\begin{table}[htb]
  \centering
\begin{tabular}{c|c}
\ms [\GeV{}]  & $\sigma_{B_{\gamgam}}\epsilon_{B_{\rm SR}}(\ms)$ [pb]\\
\hline
$100$ &  $0.0146$ \\
$200$ &  $0.00144$ \\
$400$ &  $8.41\times 10^{-4  }$ \\
$600$ &  $1.82\times 10^{-4  }$ \\
$800$ &  $5.23\times 10^{-5}$ \\
$1000$ & $2.14\times 10^{-5}$ \\
$1200$ & $7.64\times 10^{-6}$ \\
$1400$ & $3.10\times 10^{-6}$ \\
\end{tabular}
  \caption{The background cross section times efficiency $\sigma_{B_{\gamgam}} \epsilon_{B_{\gamgam}}(\ms)$ (in pb)  relevant for the $\gamgam$ signal region. For this signal region the efficiency is independent of \mtp. \label{tab:BGgammagamma}}
\end{table}

\begin{table}[htb]
  \centering
\begin{tabular}{c | c c c c c c}
\multirow{2}{*}{\ms [\GeV{}]} & \multicolumn{6}{c}{\mtp [\GeV{}]} \\
        & 600		      & 800 		     & 1000		    & 1200 		   & 1400 		& 1600 \\ 
\hline
130     & 5.87$\times 10^{-4}$ &  3.94$\times 10^{-4}$ &  2.39$\times 10^{-4}$ & 1.39$\times 10^{-4}$ & 7.32$\times 10^{-5}$ & 5.15$\times 10^{-5}$ \\
160     & 7.61$\times 10^{-4}$ &  5.90$\times 10^{-4}$ &  3.70$\times 10^{-4}$ & 2.34$\times 10^{-4}$ & 1.54$\times 10^{-4}$ & 9.65$\times 10^{-5}$ \\
200     & 4.79$\times 10^{-4}$ &  4.37$\times 10^{-4}$ &  3.47$\times 10^{-4}$ & 2.47$\times 10^{-4}$ & 1.47$\times 10^{-4}$ & 9.48$\times 10^{-5}$ \\
400     & 4.55$\times 10^{-5}$ &  4.42$\times 10^{-5}$ &  4.30$\times 10^{-5}$ & 4.10$\times 10^{-5}$ & 3.75$\times 10^{-5}$ & 3.24$\times 10^{-5}$ \\
600     &                     &  9.98$\times 10^{-6}$ &  9.88$\times 10^{-6}$ & 9.72$\times 10^{-6}$ & 9.39$\times 10^{-6}$ & 8.96$\times 10^{-6}$ \\
800     &                     &                      &  3.12$\times 10^{-6}$ & 3.12$\times 10^{-6}$ & 3.05$\times 10^{-6}$ & 3.02$\times 10^{-6}$ \\
1000    &                     &                      &                      & 1.16$\times 10^{-6}$ & 1.11$\times 10^{-6}$ & 1.11$\times 10^{-6}$ \\
1200    &                     &                      &                      &                     & 5.01$\times 10^{-7}$ & 4.94$\times 10^{-7}$ \\
1400    &                     &                      &                      &                     &                     & 2.17$\times 10^{-7}$ \\ 
\end{tabular}
  \caption{The background cross section times efficiency $\sigma_{B_{\Zgam}} \epsilon_{B_{\Zgam}}(\ms,\mtp)$ (in pb) relevant for the $\Zgam$ signal region. \label{tab:BGZgamma}}
\end{table}
It should be noted that we only present the estimates for the irreducible background. 
This turns out to be negligible in the high mass region and its values are presented only to show this fact and for completeness. 
Fake rates are also expected to be negligible in the high-mass region~\cite{ATL-PHYS-PUB-2013-009}.

The number of expected signal events for each SR is given by
\begin{align}
S_{\rm SR}=L~\sigma(\mtp)~&\left( \sum_{ X, Y} \epsilon_{\rm SR}^{Y,X} \BR({S\to  X}) \,  \BR({S\to Y}) \right), 
 \label{Stot}
\end{align}
where $\epsilon_{\rm SR}^{Y,X}$ is the final efficiency in appropriate signal region SR for the signal sample with decay $(S\to X)(S\to Y)$ with $X,Y\in \{ \gamgam, \Zgam, WW,ZZ\}$. 
(In these expressions we assume the validity of the NWA and assume 100\% \BR $\tprime\to \St$ and  $\antitprime\to \Santit$.)

In \app{sec:appendixefficiencies} we tabulate the above efficiencies,  allowing one to estimate the signal in any of the theoretical models discussed here  by simply computing the corresponding \BR.  
The discovery potential for a more generic model can be also estimated using the numbers provided as long as the efficiency times BR of any extra decay channel is known to be small.

Having computed the number of signal ($S$) and background ($B$) events, we estimate the significance by employing the formula~\cite{Li:1983fv,Cousins:2007bmb,Cowan:2010js}
\begin{equation}
z=\sqrt{2}\left\{(S+B)\ln\left[\frac{(S+B)(B+\sigma_b^2)}{B^2+(S+B)\sigma_b^2}\right]
	-\frac{B^2}{\sigma_b^2}\ln\left[1+ \frac{\sigma_b^2 S}{B(B+\sigma_b^2)} \right] \right\}^{1/2} \,,\label{CowanZ}
\end{equation}
that is obtained by using the ``Asimov'' data-set into the profile likelihood ratio. 
The explicit expression above, containing the uncertainty $\sigma_b$ on the background, is found in ref.~\cite{CowanMPI}.

We consider an overall $\sigma_b=10\% B$ systematic uncertainty on $B$. 
This number is most likely a conservative estimate and it is estimated by comparing the systematic uncertainties of ATLAS and CMS analyses with similar final states, especially high-mass $\Zgam$ searches \cite{Aaboud:2017uhw,Sirunyan:2017hsb} and high mass $\gamgam$ searches \cite{Aaboud:2018ewm,Aaboud:2018ftw,Sirunyan:2018wnk}.

\subsection{Model interpretation}
\label{sec:modelinterpretation}

Recall that the main focus is the study of models where the top partner has 100\% BSM \BR $\tprime\to \St$ and $S$ decays into EW gauge bosons. 
Even within this limited framework, we still need to discuss the relative strengths of the various $S$ decay channels, controlled by the couplings in \eq{eq:LBSM}.

\begin{figure}[t]
\centering
\includegraphics[width=0.49\textwidth]{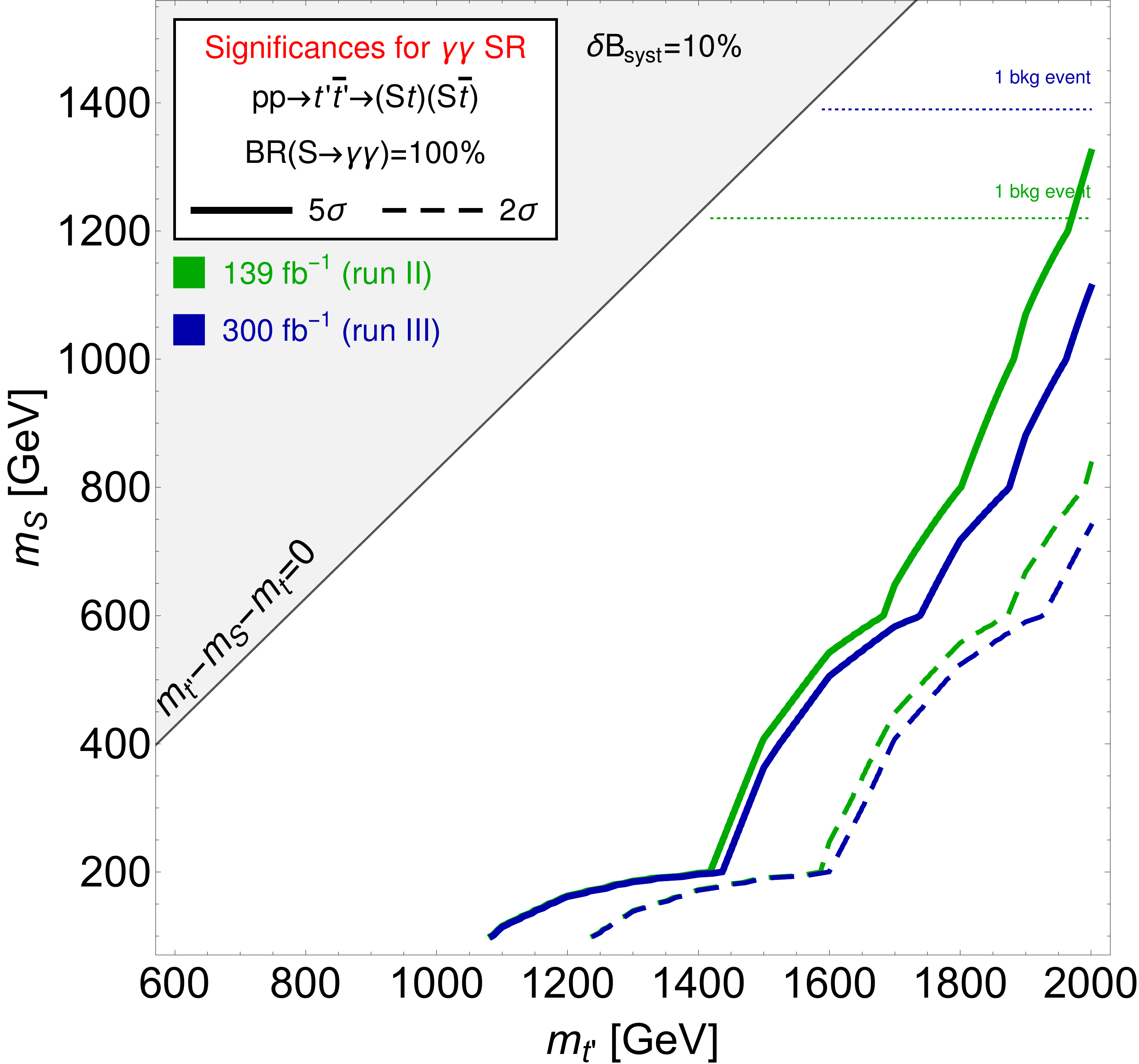}
\includegraphics[width=0.49\textwidth]{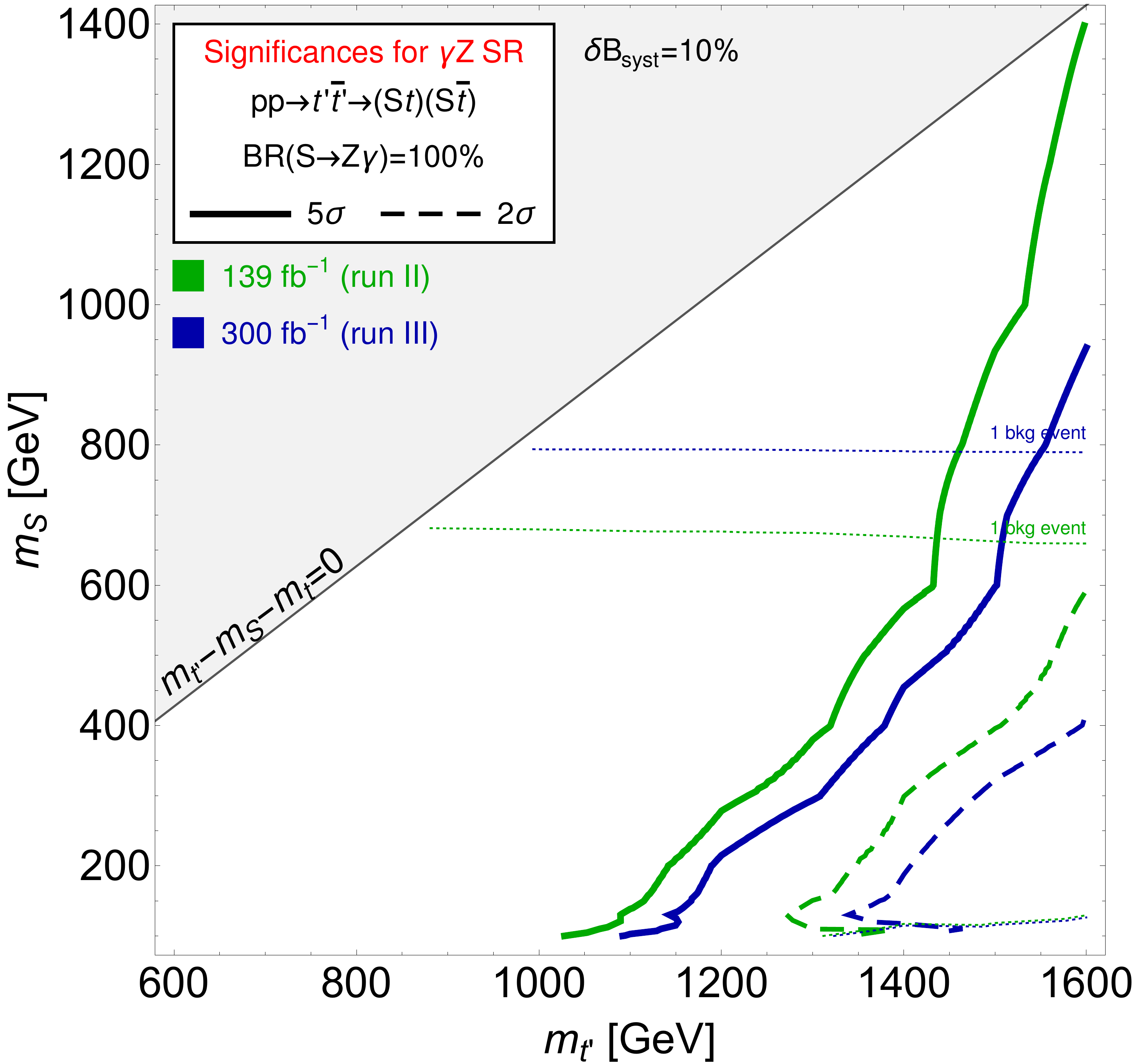}
\caption{\label{fig:optimalreach} LHC optimal reach for different LHC luminosities for the $\gamgam$ SR (left) and $\Zgam$ SR (right). The solid lines correspond to the 5$\sigma$ discovery reach, while the dashed lines correspond to the 2$\sigma$ exclusion reach. The dotted lines identify the region with 1 irreducible background event, where the contribution of fake rates can become relevant.}
\end{figure}

We start by considering the optimal reaches for the two SR considered in this analysis, corresponding to scenarios where $S$ decays fully either into $\gamgam$ or $\Zgam$. 
Such scenarios are likely non-physical, but they allow to determine the maximum potential of the selections. 
The LHC reaches for this simplified scenario are presented in \fig{fig:optimalreach} for two different LHC luminosities, corresponding to the final luminosity at the end of \runtwo and the nominal final luminosity of \runthree. 
It can be noticed that the sensitivity of the search diminishes for increasing $m_{\tprime}$ due to the reduction of production cross section, but it improves with increasing $m_S$ because of the reduction of the background yields (see \tab{tab:BGZgamma}).

We now move on to more theoretically motivated scenarios.
We first  consider the benchmark motivated by partial compositeness, where only the anomaly induced pseudoscalar couplings $\tilde \kappa_B$ and  $\tilde \kappa_W$ are non-zero. 

\begin{figure}[htbp]
\centering
\includegraphics[width=0.49\textwidth]{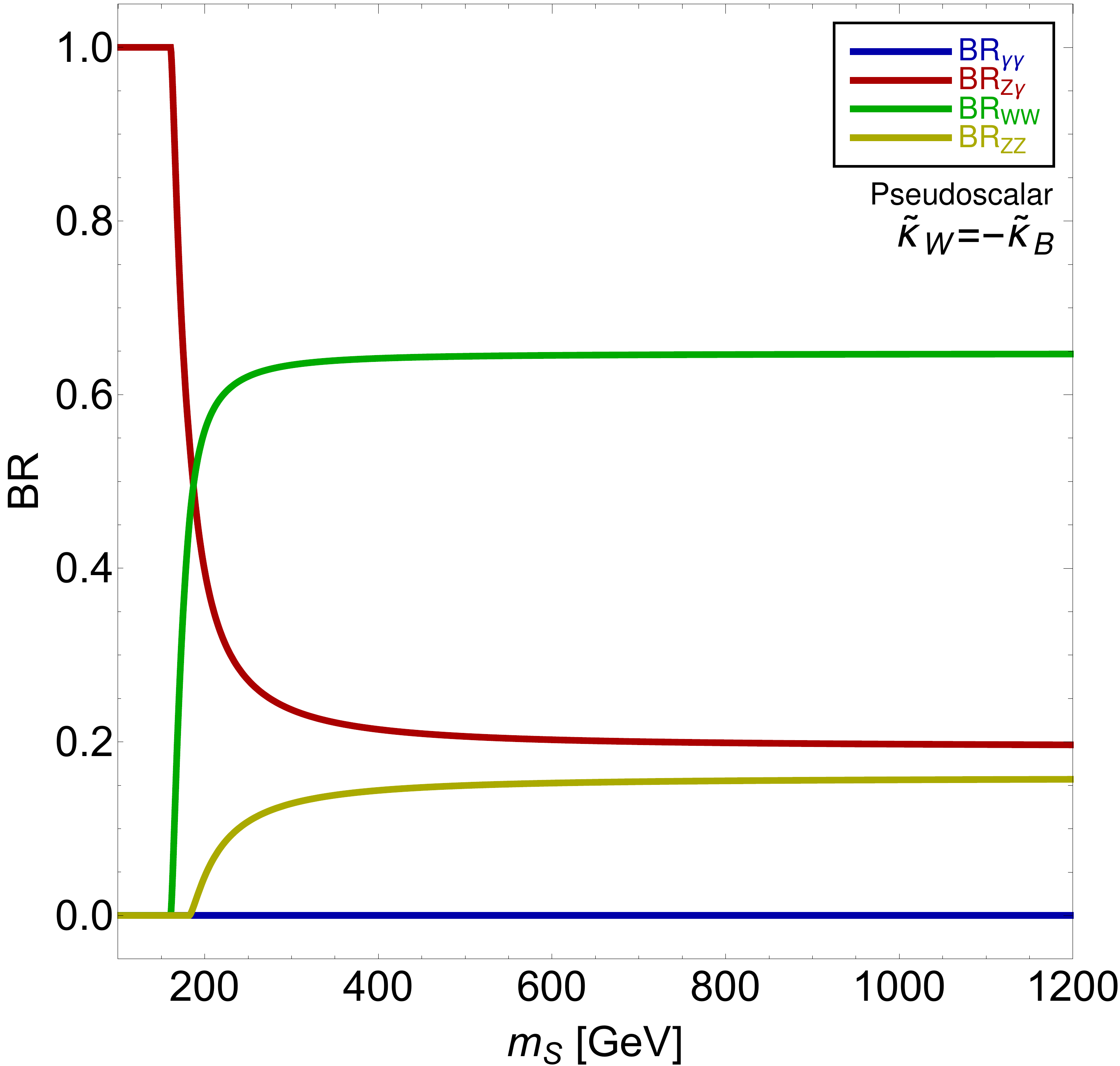}
\includegraphics[width=0.49\textwidth]{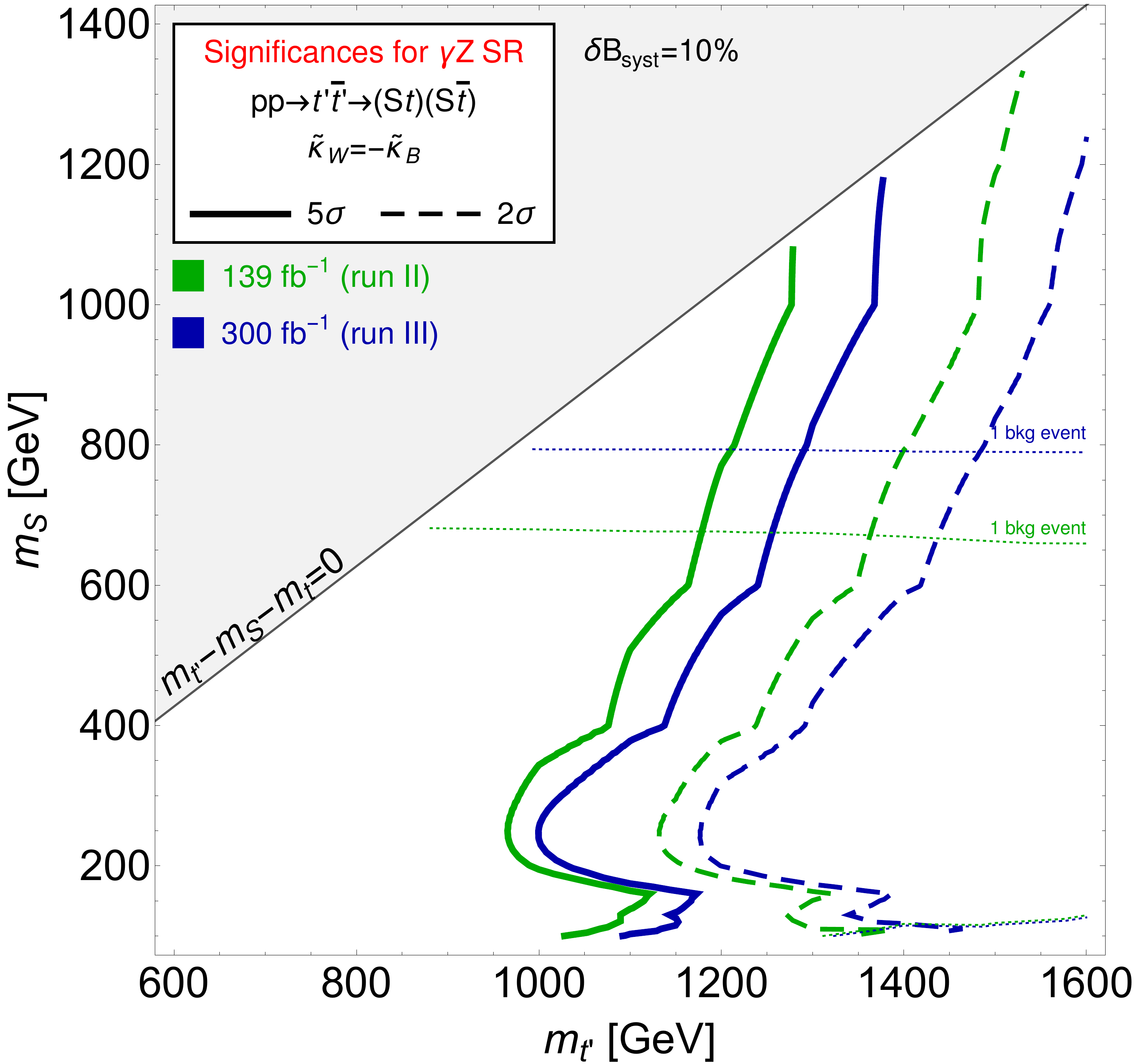}
\caption{\label{fig:noGammaGamma} Left panel: \BRs of $S$ resonance into EW bosons for the pseudoscalar case ($\kappa_B = \kappa_W = 0$) in the photophobic $S$ case ($\tilde \kappa_B=-\tilde \kappa_W$). Right panel: LHC reach for different LHC luminosities; the meaning of contours is the same as in \fig{fig:optimalreach}.}
\end{figure}

In this case, the structure of the anomaly coefficients~\cite{Bizot:2018tds} in all explicit realizations gives $\tilde\kappa_B+\tilde\kappa_W = 0$, thus suppressing the $S\to\gamgam$ decay. This leads to a 100\% $\BR({S\to \Zgam})$ below the $WW$ threshold and still an acceptably large value above it, as displayed in~\fig{fig:noGammaGamma}~(left).
The LHC reaches for this scenario are presented in \fig{fig:noGammaGamma}~(right) for two different LHC luminosities, corresponding to the final luminosity at the end of \runtwo and the nominal final luminosity of \runthree. 
Here,  we consider only the $\Zgam$ SR because of the negligible sensitivity of the $\gamgam$ SR.  

Different effects are present in the reach of \fig{fig:noGammaGamma}. For  $m_S\lesssim 2m_W$ the sensitivity is optimal due to a 100\% decay rate of both $S$ into \Zgam ($S\to \Zgam,\, S\to \Zgam$) and a high efficiency (\fig{fig:efficiencies} (right)). Above threshold the $S\to VV,\, S\to VV$ ($V=W,Z$) decay channels kick in with $\approx 64\%$ rate and negligibible efficiency, while the $S\to \Zgam,\, S\to \Zgam$ rate reduces to $\approx 4\%$. The mixed decay $S\to VV, S\to \Zgam$ takes 16\% of the branching ratio and have an efficiency approximately constant and near 40\% compared to the pure $\Zgam$ case (\fig{fig:eff_azSR_azchannels}). This depletion in the signal explains the kink of sensitivity lost near the $m_S\approx 2m_W$ threshold. 
In both regions the sensitivity improves with increasing values of $m_S$ due to a rapid decrease of the background, as noticed in \fig{fig:optimalreach}. 

The interpretation for the composite Higgs model described in \sec{sec:interpretation_composite} is straightforward. 
The $S$ is photophobic and we can read the bounds directly from \fig{fig:noGammaGamma}. 
It is encouraging to see that even for not optimised cuts this channel could be competitive with the search for the $+5/3$ charged partner~\cite{Sirunyan:2018yun}. 
Some more details for this model are given in \app{sec:appendixmodels}.

\begin{figure}[tbp]
\centering
\includegraphics[width=0.43\textwidth]{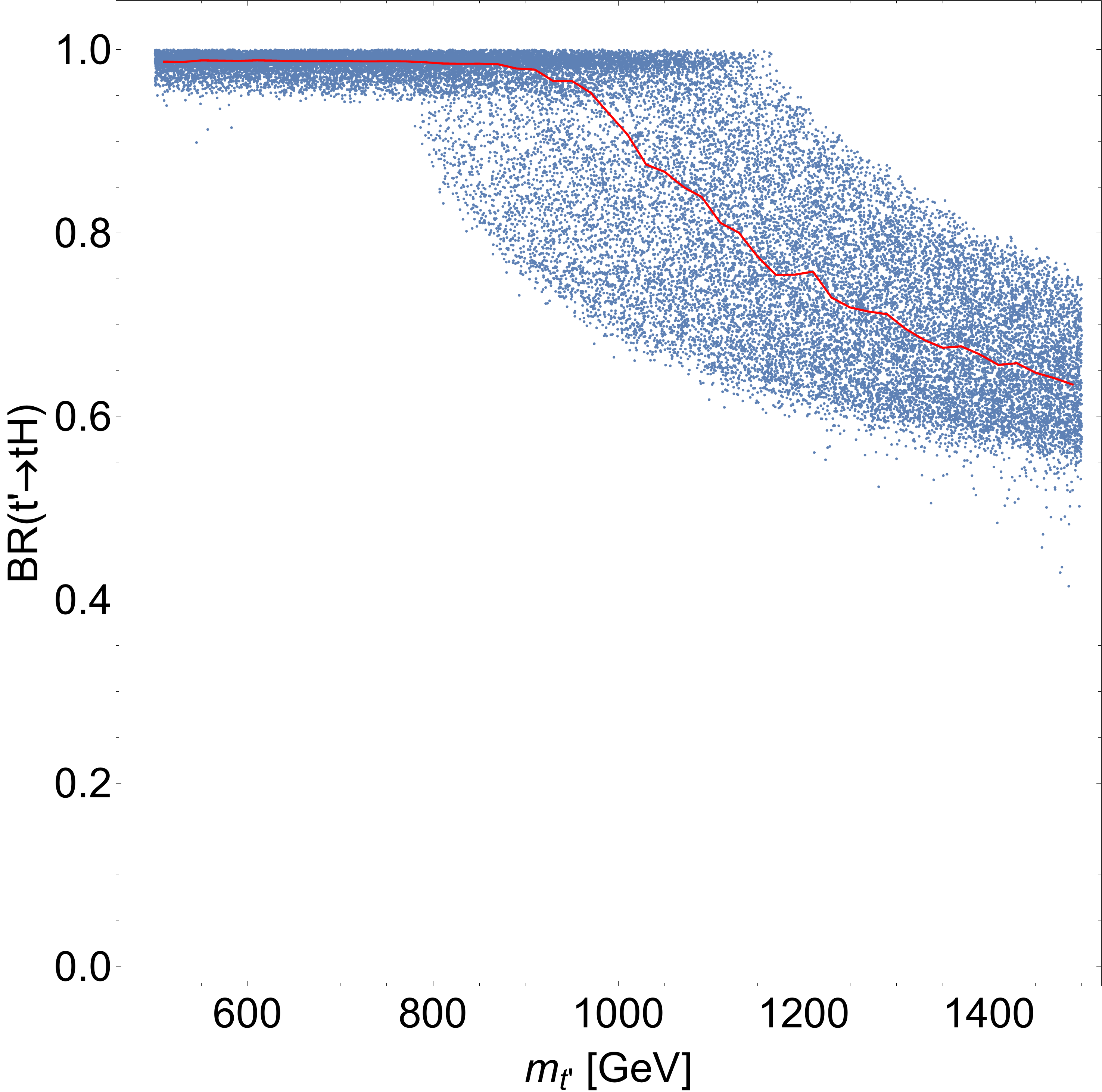}\hspace*{0.5truecm}
\includegraphics[width=0.45\textwidth]{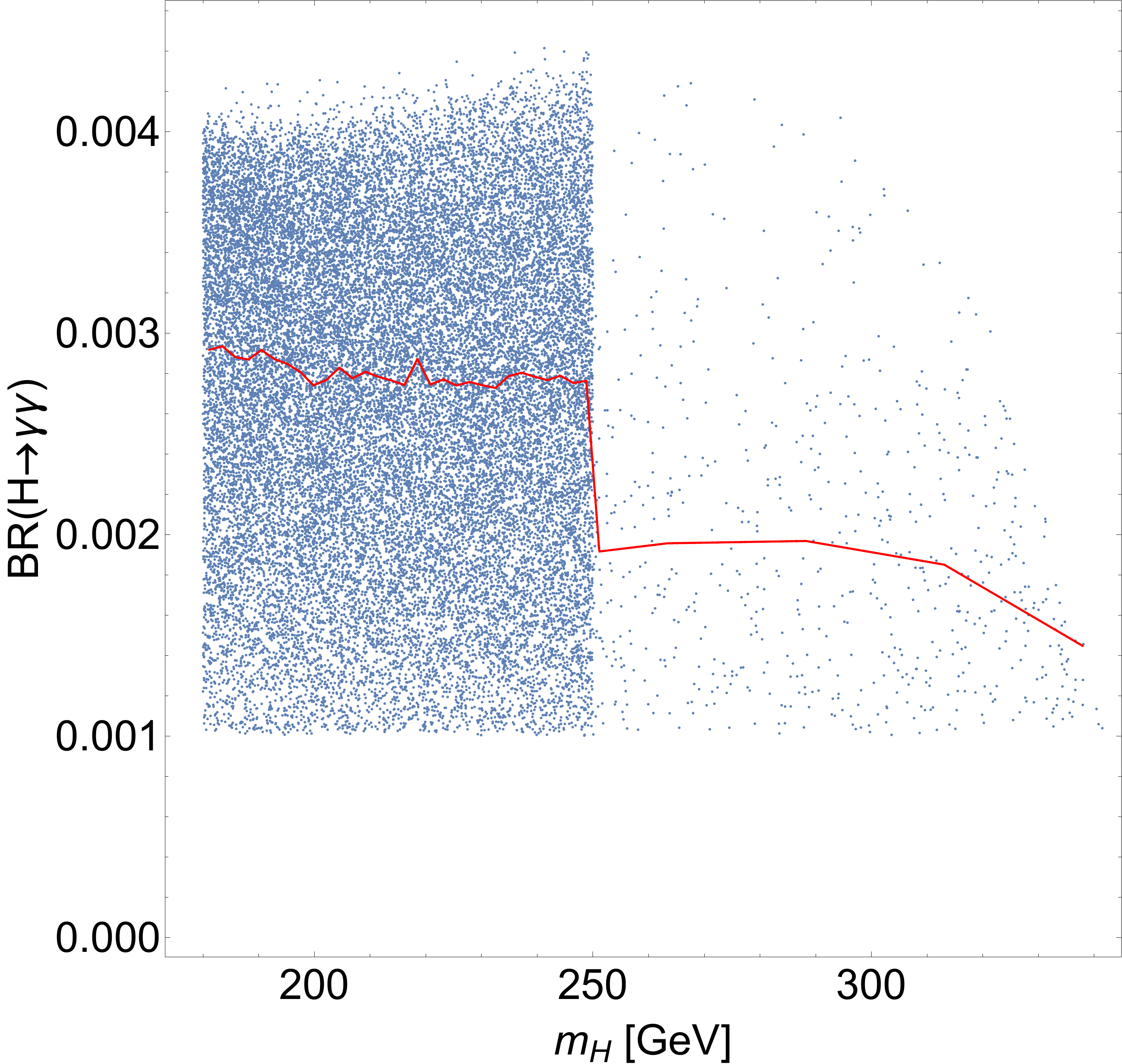}
\caption{Scatter plots (in blue) of the BRs of $\tprime\to Ht$ and $H\to\gamgam$ as a function of the respective masses and (in red) the median value on the binned samples. The binning for the $\tprime$ BR is $20\, \GeV$ while for the $H$ BR it is $2.3\, \GeV$ for $180\, \GeV <m_H< 250\, \GeV$ and $25\, \GeV$ for $m_H>250\, \GeV$.\label{fig:BRsH} }
\end{figure}

For the 2HDM+VLQ case, the interpretation is somewhat more complicated because of the more numerous parameters, richer particle spectrum and, hence, the decay patterns of \tprime and $S$.
A scan has been performed by varying the 2HDM input parameters described in appendix~\ref{sec:appendix2HDM} to obtain benchmark points characterised by the highest BRs of \tprime and $S$ into the final states considered in this analysis in order to maximise the sensitivity. Such points are simply representative of the 2HDM spectrum, as we ignored the fact that {\sc HiggsBounds} excludes the  majority of them. In fact, the scope of this selection is to illustrate the potential of the model independent analysis developed in this paper rather than to constrain specific theoretical models.
We first restricted the scan by enforcing an almost exclusive decay of the \tprime into the CP-even scalar $H$ by setting the masses of the CP-odd $A$ and charged $H^\pm$ states to high values  and by restricting the 2HDM input parameters in such a way that SM decays of the \tprime are also suppressed. We then computed the BRs of $\tprime$ and $H$ as a binned function of their masses by considering the median of the sample for each bin. This procedure approximates the BRs neglecting any correlation point-by-point and is reasonably accurate given the size of the sample (approximately 30,000 points). In fact, we have verified that the sum of the BR functions obtained with this procedure is approximately 1 for all $\tprime$ and $H$ masses. Examples of the distribution of scanned points and of the median BRs are provided in figure~\ref{fig:BRsH} for $\tprime\to Ht$ and $H\to\gamgam$. (The procedure is identical for the case of $\tprime \to At$ and $A\to\gamgam$, though the point distributions and median values are obviously different.)

The decay of $H$ into $\gamgam$ is around 0.3\% below the $hh$ threshold, while its decay into \Zgam is $\sim$0.05\%. The generically dominant decay of $H$ is into $gg$, which is on average around 70\%, followed by $WW$ ($\sim$20\%) and $ZZ$ (5\% to 10\%), while the BRs into $b\bar b$ and $c\bar c$ are 1\% or less. Above the $2m_h\approx250\, \GeV$ threshold, $H\to hh$  dominates and all other BRs drop significantly, until the (on-shell) $t\bar t$ channel opens and becomes dominant. Then, we do a second scan with the role of $H$ and $A$ interchanged (approximately 80,000 points) and compute the BRs as described above. Here, there cannot be $WW$ and $ZZ$ decays of the $A$ state, so that $gg$ and $b\bar b$ decays share the majority oft the decay rate (about 90\% of it, with the remainder saturated by $\tau^+\tau^-$ and $Zh$, which we then neglect in the MC generation) till the (off-shell) $t\bar t^*$ channel opens (coincidentally enough,  around $m_t+m_b+m_W\approx 260\, \GeV$), with the $\gamgam$ and \Zgam rates being generally  lower than in the previous case.  
Given the low BR into \Zgam, in addition to the subleading  BR of the $Z$ into leptons, no significant sensitivity is expected in the $Z(\to \ell^+\ell^-))\gamma$ final state and, therefore, we will focus only on the $\gamgam$ SR for the 2HDM+VLQ  in  the case of both a light $H$ and $A$. 

The efficiencies for $\{\gamgam,WW\}$ and $\{\gamgam,ZZ\}$ are provided in appendix~\ref{sec:appendixefficiencies}. Given the high BR into $gg$, the efficiencies have been computed for the $\{\gamgam,gg\}$ final state as well. This has been done only in the region of parameter space where high sensitivity is obtained, {i.e.,} for \mtp less than 1~\TeV, and are in average around 20\%. The efficiencies for the $\{\gamgam,hh\}$ channel have also been calculated above the $H\to 2h$ threshold, in the region of high sensitivity, and are found to be around 30\%. Given the illustrative nature of this example, we assumed the efficiencies for the $\{\gamgam,t\bar t^*\}$ in the case of a light $A$ channel to be flat and around 30\%. 
% Given the illustrative nature of this example, we assumed the efficiencies for the $\{\gamgam,hh\}$ (in the case of a light $H$) and $\{\gamgam,t\bar t^*\}$ (in the case of a light $A$)  channels to be flat and around 30\%. 

\begin{figure}[tbp]
\centering
\includegraphics[width=0.45\textwidth]{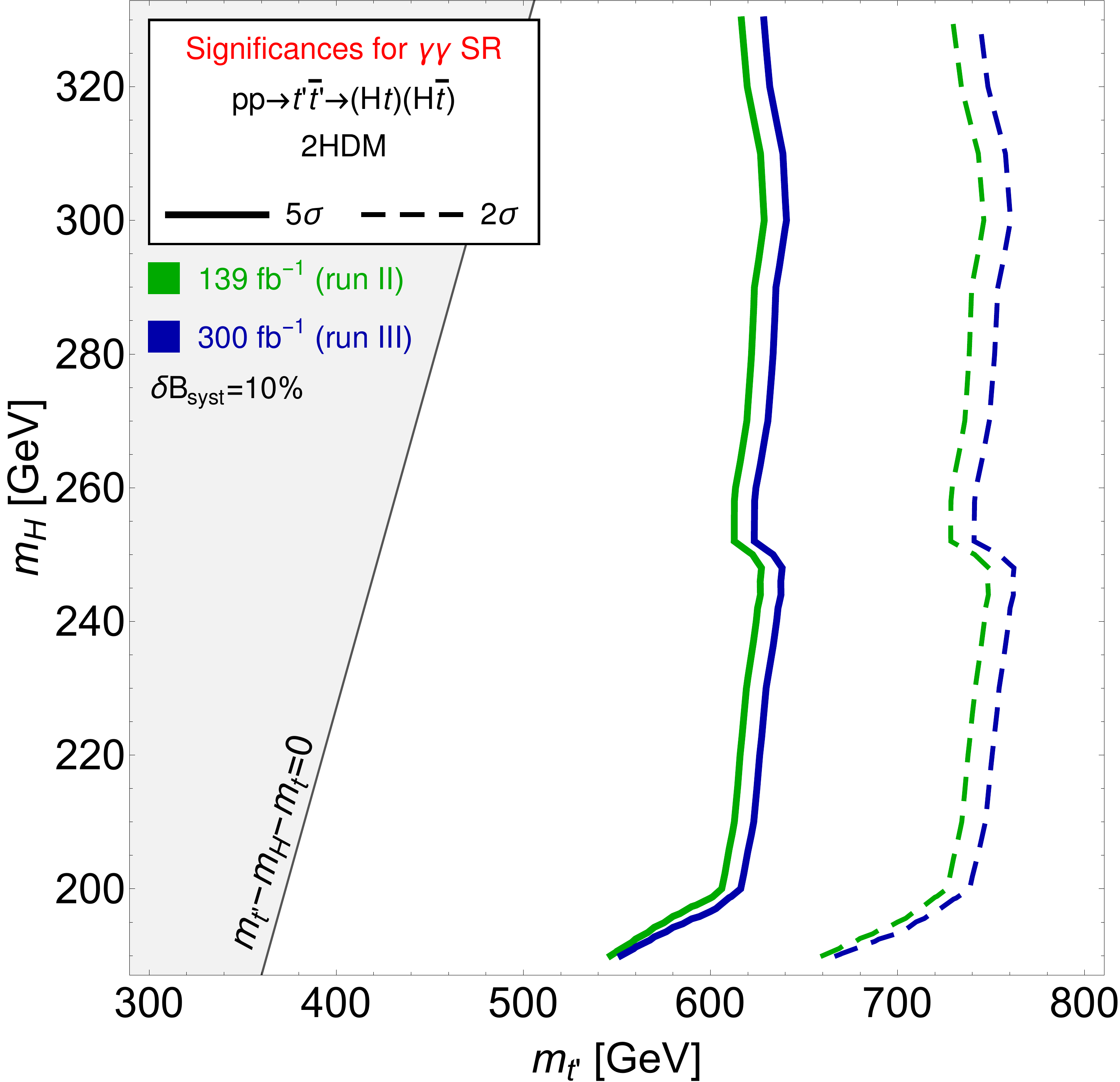}\hspace*{0.5truecm}
\includegraphics[width=0.45\textwidth]{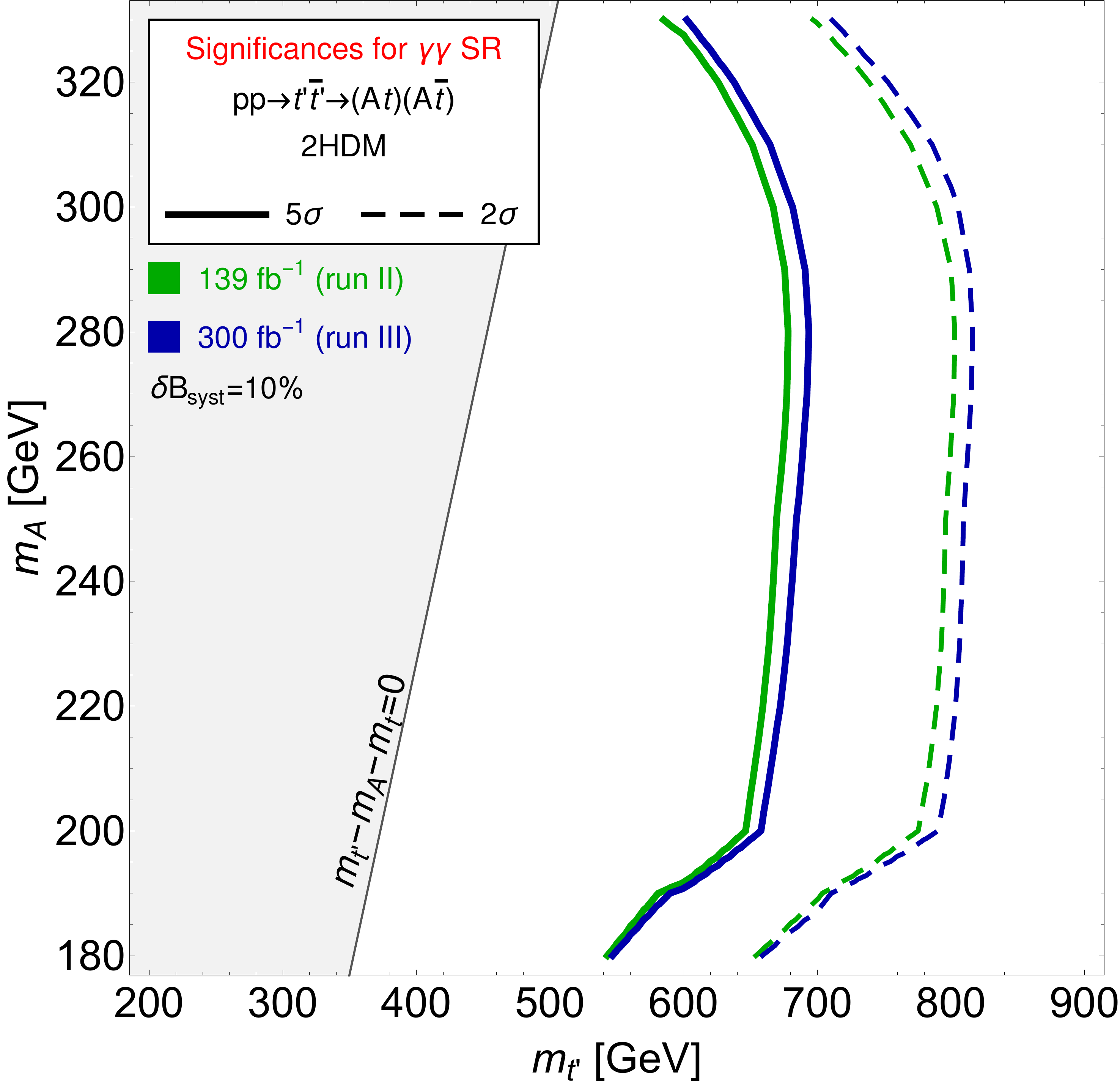}
\caption{Significances in the 2HDM+VLQ for points with large product of BRs of $\tprime\to H t$ and $H\to\gamgam$ (left)  or
$A\to\gamgam$ (right). The meaning of the contours is the same as in \fig{fig:optimalreach}. \label{fig:2hdm} }
\end{figure}

The  results for the 2HDM+VLQ are shown in figure~\ref{fig:2hdm}. For the case of a light $H$ state, some discovery reach has been found for $\mtp$ around 600~\GeV and exclusion is possible up to $\mtp$ around 700~\GeV, almost independently of $m_S$ below the (on-shell) $t\bar t$ threshold. For the case of a light $A$ state, the reach in $m_{\tprime}$ for both discovery and exclusion is somewhat deeper than in the previous case, by some 50~\GeV. In contrast, the one in $m_S$ is very similar, as it again collapses at approximately $2m_t$.\footnote{Note that, while one may want to consider the case of both $H$ and $A$ being light (and possibly degenerate in mass) in order to benefit from an increased $S$ signal rate, this is impractical because, on the one hand, the $H^\pm$ boson also ought to be light to preserve EWPT compliance (thereby increasing the $\tprime \to H^\pm b$ BR) and, on the other hand, the impact of the restrictions from \higgsbounds\  increases substantially.}

%% file: conclusions.tex
While the case for VLQs, especially those of top flavour, has already been well established from the theoretical side, the experimental pursuit of their signatures at the LHC has been somewhat limited, as ATLAS and CMS analyses have primarily been carried out under the assumption that such new states of matter decay into SM particles only, i.e., via $\tprime \to \Wb, \Zt$ and $\higgst$. 
This approach clearly enables one to make the most in terms of optimising the signal-to-background ratio in an analysis, chiefly because one can attempt reconstructing the measured $W^+$, $Z$ and $h$ masses. 
However, if one considers VLQ models with additional particles this is overly restrictive since the VLQ may decay via exotic channels involving scalars or pseudoscalars. 
While the kinematic handles available to enhance these exotic channels may be apparently limited in comparison (as the exotic scalar or pseudoscalar states may have not been discovered already and/or their mass not measured), the size of the associated \BRs could be large enough so as to nonetheless enable sensitivity to these channels. 
Furthermore, if the companion Higgs states are heavier than the   $W^+$, $Z$ and $h$ objects of the SM, the signal would anyhow be present in a region of space where the background contamination is minimised.  
Based on this reasoning, in this paper, we have set out to assess the scope of the LHC to test \tprime decays into neutral (pseudo)scalar states, whose nature could be either fundamental or composite. 
As an example of spin-0 fundamental states, we have assumed here a Higgs sector comprised of the SM state supplemented by a scalar boson as well as a 2HDM (Type-II) containing both a scalar and pseudoscalar state (which we have taken light one at a time). 
As an example of spin-0 composite states, we have looked at a CHM where an additional pseudoscalar state emerges as a pNGB of the underlying new strong dynamics. 
In fact, we have also shown how all such models can conveniently be parametrised in the form of a simplified model onto which they can be mapped. 

Of the various possible decay modes of this additional neutral (pseudo)scalar bosons, which we have collectively labelled as $S$, we have considered here two of the cleanest probes possible at the LHC, i.e., $S\to \gamgam$ and $\Zgam$ (with the $Z$ decaying into electron/muon pairs). 
In doing so, we have performed a dedicated signal-to-background analysis exploiting parton level event generation, QCD shower and hadronisation effects as well as detector emulation aimed at establishing the sensitivity of the LHC experiments to such decays, where the $S$ state emerges from a companion top decay, $\tprime\to \St$, following $\tprime \antitprime$ production (with the $\antitprime$ decay treated inclusively). 
In the case of both $S$ signatures, we have not attempted any reconstruction of the SM top quark entering the $\tprime$ decay chain although, on a trial-and-error basis, we have assumed knowledge of the $S$ mass, to be able to exploit both the cleanliness of the two $S$ decay channels and the ability of a standard LHC detector in sampling $\gamgam$ and $Z(\to\ell^+\ell^-)\gamma$ invariant masses with high resolution. 
Indeed, this approach  also enables us to compare on a more equal footing the scope of $\tprime\to S\, t$ signatures with that of $\tprime\to \Wb,\, \Zt$ and $\higgst$ ones, where a mass reconstruction is normally imposed on the $W^\pm$, $Z$ and $h$ decay products. 

As a result of this approach, we have found that  the $\tprime\to \St$ signatures give a level of sensitivity not dissimilar from that obtained through studies of $\tprime\to \Wb,\, \Zt$ and $\higgst$. 
For specific regions of the parameter space of VLQ models with exotic Higgs states, which have survived all available constraints from both direct and indirect \tprime and $S$ searches (including those obtained by ourselves from recasting experimental studies for other sectors), we have found the following exclusion and discovery reaches. 
For a simplified model maximising both the $\tprime$ and $S$ \BRs, \mtp can be probed in both the $\gamgam$ and $\Zgam$ channels up to approximately \unit{2}{\TeV} for $S$ masses  well into the \TeV region. 
In the CHM scenario considered, coverage is not dissimilar for the $\gamgam$ case but for the $\Zgam$ the \tprime reach is limited to \unit{1.6}{\TeV}. 
Finally, in the 2HDM+VLQ, it is possible to exclude $\mtp$ up to around 700(750)~\GeV and discover $\mtp$ up to around 600(650)~\GeV\ almost  independently of $m_S$ when $S$ represents the CP-even(odd) $H(A)$ state and below the (on-shell) $t \bar t$ threshold for the decay of $S$. This is limited to the $\gamgam$ case, though, as $\Zgam$ gives no sensitivity at both \runtwo and III.

Hence, in  connection to all of the above, we can confidently conclude to have surpassed the state-of-the-art in VLQ searches in two respects: firstly, by testing the scope of non-SM decays of the $\tprime$ state and, secondly, by deploying a selection procedure which is model independent yet enables one to interpret its results in a variety of  theoretical scenarios. Furthermore, it should be noted that, while restricting ourselves to the case of $\gamma\gamma$ and $Z\gamma$ signatures of the (pseudo)scalar states emerging from the described VLQ decays, there is no reason why our procedure cannot be applied to other $S$ decays. Indeed, it can also be further improved (e.g., by reconstructing top-quark decays). 

In summary, we believe that there is significant margin for improving the sensitivity of the LHC to models with a heavy top partner, through the exploitation of  its decay channels into exotic (i.e., non-SM-like) neutral (pseudo)scalar states, which are ubiquitous in BSM constructs containing such a new fermion. 
In fact, over sizeable regions of the parameter space of the realistic VLQ models considered here, we have found that sensitivity to both the \tprime and $S$ mass can extend well into the  \TeV region, thereby being competitive with the currently studied SM channels. 
While in this paper we have limited ourselves to illustrating this through a few benchmarks examples, in a forthcoming paper, we shall quantify the regions of parameter space of our models where such a phenomenology can be realised, including tensioning the scope of standard and exotic $\tprime$ decays against each other.

\clearpage

%% file: appendix_models.tex
In this appendix, additional details are given of the models; the 2HDM+VLQ model in \app{sec:appendix2HDM} and the composite Higgs model in \app{sec:appendixcomp}. 

\subsection{The 2HDM with an additional VLQ}
\label{sec:appendix2HDM}
The scalar potential of the model includes two identical scalar doublets ($\Phi_1, \Phi_2)$ and a discrete symmetry $\Phi_i\to (-1)^i\Phi_i$ ($i=1,2$), which is only violated softly by dimension-two terms \cite{Branco:2011iw},
\begin{align}
V  &= m_{11}^2 \Phi_1^\dag \Phi_1 + m_{22}^2 \Phi_2^\dag \Phi_2 - m_{12}^2 \left( \Phi_1^\dag \Phi_2 + \Phi_2^\dag \Phi_1 \right) 
+ \frac{\lambda_1}{2} \left( \Phi_1^\dag \Phi_1 \right)^2 + \frac{\lambda_2}{2} \left( \Phi_2^\dag \Phi_2 \right)^2 \nn \\
&+ \lambda_3 \left( \Phi_1^\dag \Phi_1 \right) \left( \Phi_2^\dag \Phi_2 \right)  
+ \lambda_4 \left( \Phi_1^\dag \Phi_2 \right) \left( \Phi_2^\dag \Phi_1 \right) + \frac{\lambda_5}{2} \left[ \left( \Phi_1^\dag \Phi_2 \right)^2 + \left( \Phi_2^\dag \Phi_1 \right)^2 \right] .
\label{thdmV}
\end{align}
We take all parameters in the above potential to be real (although $m_{12}^2$ and $\lambda_5$ could in principle be complex). 
The two complex scalar doublets may be rotated into a basis where only one doublet acquires a VEV, the \emph{Higgs basis},
\begin{eqnarray}
 H_1= \frac{1}{\sqrt{2}} \left(
   \begin{array}{c}
     \sqrt{2} \, G^+ \\
     v+\varphi^0_1+iG^0 \\
   \end{array}
 \right),\quad  H_2= \frac{1}{\sqrt{2}} \left(
   \begin{array}{c}
     \sqrt{2}\, H^+ \\
     \varphi^0_2+iA \\
   \end{array}
 \right),
\end{eqnarray}
where $G^0$ and $G^\pm$ are the would-be Goldstone bosons and $H^\pm$ are a pair of charged Higgs bosons. 
$A$ is the CP odd pseudoscalar, which does not mix with the other neutral states. 
The Goldstone bosons are aligned with the VEV in Higgs flavor space, while the $A$ is orthogonal.
The physical CP even scalars $h$ and $H$ are mixtures of $\varphi^0_{1,2}$ and the scalar mixing is parametrized as
\begin{eqnarray}
\left(
 \begin{array}{c}
 h\\
         H \\
   \end{array}
 \right)= \left(
          \begin{array}{cc}
            s_{\beta-\alpha} & c_{\beta-\alpha} \\
            c_{\beta-\alpha} & -s_{\beta-\alpha} \\
          \end{array}
        \right)\left(
   \begin{array}{c}
 \varphi^0_1\\
 \varphi^0_2 \\
   \end{array}
 \right),
\end{eqnarray}
where $\tan\beta=v_1/v_2$ is the angle used to rotate $\Phi_{1,2}$ to the Higgs basis fields
$H_{1,2}$, $\alpha$ is the additional mixing angle needed to diagonalize
the mass matrix of the CP-even scalars, and $s_{\beta-\alpha}=\sin(\beta-\alpha)$, $c_{\beta-\alpha}=\cos(\beta-\alpha)$.
The most general renormalisable interaction and mass terms involving the VLQ can be described by the following Lagrangian (where we only include the third generation SM quarks),
\begin{equation}
-\mathcal{L}_Y \supset
y_T\overline{Q}_L \widetilde{H}_2 T_R +
\xi_T\overline{Q}_L \widetilde{H}_1T_R+M \overline{T}_L T_R \ ,
\label{eq:appLyuk}
\end{equation}
where $\widetilde H_i \equiv i \sigma_2 H^*_i$ ($i=1,2$), 
$Q_{L}$ is the SM quark doublet and $M$ is a bare mass term for the VLQ, which is unrelated to the Higgs mechanism of EWSB.
Note that often the Yukawa couplings 
of the 2HDM are written in terms of the fields $\Phi_1, \Phi_2$. In \eq{eq:appLyuk} we use the Higgs basis fields, so the Yukawa couplings $y_T,\xi_T$ must be defined accordingly. In a Type II-model, as we are considering in this paper, the up-type quarks only couple to the doublet $\Phi_2$, while down-type quarks only couple to $\Phi_1$.
Additional mixing terms of the form
$\overline{T}_L t_R$ 
 can always be rotated away and reabsorbed into the definitions of the Yukawa couplings. 
In the weak eigenstate basis $(\widetilde t,T)$, where $\widetilde{t}$ is the SM top quark, the top quark and VLQ mass matrix is
\begin{eqnarray}
{\mathcal M}=\left(
\begin{array}{cc}
    \frac{y_t v}{\sqrt{2}} & \frac{\xi_T v}{\sqrt{2}} \\
    0 & M \\
  \end{array}
        \right) ,
\label{eq:massmat}
\end{eqnarray} 
where $y_t$ is the Yukawa coupling of the top quark.
It is clear from the above mass matrix that the physical mass of the heavy top, $\mtp$, is different from $M$ due to the $t$--$T$ mixing. 
The mass matrix ${\mathcal M}$ can be diagonalised by a bi-unitary transformation in the same way as in \sec{sec:simpPC} to obtain the physical states $(t_{L,R} , \tprime_{L,R})$ in terms of the gauge eigenstates $(\widetilde t_{L,R},T_{L,R})$,
\begin{eqnarray}\label{eq:ULUR}
\left(
\begin{array}{c}
 t_{L,R} \\
 \tprime_{L,R}  \\
    \end{array}
\right)=\left( \begin{array}{cc}
c_{L,R} &  -s_{L,R} \\
s_{L,R} &  c_{L,R} \\
\end{array}
\right)\left(
\begin{array}{c}
\widetilde t_{L,R} \\
T_{L,R}  \\
\end{array}
\right) = U_{L,R} \left(
\begin{array}{c}
\widetilde t_{L,R} \\
T_{L,R}  \\
\end{array}
\right)
\end{eqnarray}
The mixing angles $\theta_L$ and $\theta_R$ are not independent parameters. 
From the bi-unitary transformations we can derive the relations 
\begin{equation}
\tan(2\theta_L) = \frac{\sqrt{2} M v \xi_T}{M^2 - \frac{y^2_t v^2}{2} - \frac{\xi^2_T v^2}{2}}, \quad\quad 
\tan(2\theta_R) = \frac{y_t \xi_T v^2}{M^2 - \frac{y^2_t v^2}{2} + \frac{\xi^2_T v^2}{2}} ,
\label{eq:appmix2hdm}
\end{equation}
and by using the traces and determinants
\begin{eqnarray}
\Tr\left(U_L {\mathcal M} {\mathcal M}^\dagger U^\dagger_L\right) &=& \mtp^2 + m^2_t \\
\det\left(U_L {\mathcal M} {\mathcal M}^\dagger U^\dagger_L \right) &=& m^2_t \mtp^2
\end{eqnarray}
we end up with the relations
\begin{eqnarray}
\mtp^2 + m^2_t &=& M^2 + \frac{y^2_t v^2}{2} + \frac{\xi^2_T v^2}{2} \\
\frac{y^2_t v^2 M^2}{2} &=& m^2_t \mtp^2 \\
M^2 &=& m^2_t \sin^2\theta_L + \mtp^2\cos^2\theta_L \ ,
\end{eqnarray}
and a relationship between $\theta_L$ and $\theta_R$ and the Yukawa couplings,
\begin{equation}
\tan\theta_L = \frac{\mtp}{m_t} \tan\theta_R  , \quad\quad  
\quad\quad \frac{\xi_T}{y_t} = s_L c_L \frac{\mtp^2 - m^2_t}{m_t \mtp}.
\end{equation}
The $\tprime$--$t$ interaction can thus be described by three independent physical 
parameters:\ two quark masses $m_t, \mtp$ and a mixing angle $s_L=\sin\theta_{L}$.

After rotating the weak eigenstates $(\widetilde t_L,T_L)$ into the mass eigenstates, 
the Yukawa Lagrangian takes the following form~\cite{Arhrib:2016rlj}:
\begin{eqnarray}
-\mathcal{L}_Y &\supset& \frac{1}{\sqrt{2}} (\tbar_L,\antitprime_L)U_L
\left[\varphi^0_1\left(
 \begin{array}{cc}
 {y_t} & {\xi_T} \\
  0 & 0 \\
 \end{array}
 \right)+\varphi^0_2\left(
\begin{array}{cc}
  y_t \cot\beta & y_T \\
         0 & 0 \\
 \end{array}
 \right)\right]U_R^\dag \left(
 \begin{array}{c}
    t_R \\  \tprime_R  \\
 \end{array}
 \right)\nonumber\\
   &-& i (\tbar_L,\antitprime_L) U_L A \left(
  \begin{array}{cc}
  {y_t}\cot\beta & {y_T} \\
                     0 & 0 \\
   \end{array}
  \right)U_R^\dag \left(
  \begin{array}{c}  t_R \\  \tprime_R \\
 \end{array}
 \right),
\label{equa:intr}
\end{eqnarray}
where $U_{L,R}$ are the matrices appearing in \eq{eq:ULUR}.  
The neutral Higgs couplings to top ($t$) and top partner ($\tprime$) pairs are in the notation of \eq{eq:LBSM} given by (with $S=H$ or $A$)
\begin{align}
\kappa^H_{t} &=  \left(\cba-\frac{\sba}{\tan\beta}\right) c_L c_R
- \left(\frac{\xi_T}{y_t}\cba -\frac{y_T}{y_t}\sba \right)  c_L s_R \nonumber \\
\kappa^H_{\tprime} &=  \left(\cba-\frac{\sba}{\tan\beta}\right) s_L s_R
+ \left(\frac{\xi_T}{y_t}\cba -\frac{y_T}{y_t}\sba \right)  s_L c_R \nonumber \\
\kappa^H_{L}&=  \left(\cba-\frac{\sba}{\tan\beta}\right) c_L s_R
+ \left(\frac{\xi_T}{y_t}\cba -\frac{y_T}{y_t}\sba \right)  c_L c_R \nonumber \\
\kappa^H_{R} &=  \left(\cba-\frac{\sba}{\tan\beta}\right) s_L c_R
- \left(\frac{\xi_T}{y_t}\cba -\frac{y_T}{y_t}\sba \right)  s_L s_R 
\label{eq:rulesHA} \\
\widetilde\kappa^A_{t} &= \frac{1}{\tan\beta}\,c_L c_R -\frac{y_T}{y_t}\,c_L s_R \nonumber \\
\widetilde\kappa^A_{\tprime} &= \frac{1}{\tan\beta}\, s_L s_R +\frac{y_T}{y_t}\, s_Lc_R\nonumber \\
i\kappa^A_{L} &= -\frac{1}{\tan\beta} \, c_L s_R -\frac{y_T}{y_t} \, c_Lc_R \nonumber \\
i\kappa^A_{R}&= -\frac{1}{\tan\beta}\, s_L c_R + \frac{y_T}{y_t}\, s_Ls_R  \nonumber .
\end{align}
The couplings here are normalised to $y_t/\sqrt{2}$, which is what the $H_{\rm SM}t\bar{t}$ coupling would be in the case of no mixing between the $\widetilde t$ and $T$ and additionally, in the alignment limit of the 2HDM, $\sin(\beta -\alpha)\to 1$ where the lightest neutral scalar $h$ is the SM-like Higgs boson.
Note that in \eq{eq:LBSM} the terms with diagonal couplings $S\tprime\antitprime$ of the top partner to the scalars are not included, since they are not phenomenologically relevant in this paper. We include them in \eq{eq:rulesHA} for completeness, however. Note also that the combination $(\cba-\sba \cot\beta)$ that occurs in \eq{eq:rulesHA} is proportional to the 2HDM Type II Yukawa coupling of the heavier Higgs boson $H$.

In our analysis we have used a modified version of the public code 2HDMC~\cite{Eriksson:2010zzb} with a VLQ added according to the description above. We have scanned over the parameter space of the model, which is constrained by Higgs data from the LHC that can be evaluated using the public code \higgsbounds~\cite{Bechtle:2013wla}. In addition, 2HDMC can evaluate oblique parameters and theoretical constraints on unitarity, perturbativity and positivity of the potential. However, since our aim here is rather to demonstrate the use of the method developed in this paper, we have not made a comprehensive scan to satisfy these bounds, but instead we have considered parameter points that provide large BRs of $\tprime\to St $ and $S\to\gamma\gamma$ for $S=H$ or $A$. We have therefore chosen to make the Higgs boson that does \textit{not} play the role of $S$ as well as the charged Higgs boson heavy. We perform random scans over the parameters and generate $10^5$ points for each of the scenarios with $S=H,A$. We then keep those points where the product $\BR({\tprime\to \St})\times\BR({S\to\gamma\gamma})>10^{-3}$. The scalar $S$ is taken in the range $180 \, \GeV < m_S < 350 \, \GeV$, while for $S=H$ the other heavy scalar is taken in the range $600 \, \GeV < m_A < 1000 \, \GeV$. For $S=A$, instead we choose $m_H=1\, \TeV$. The charged Higgs mass is always $m_{H^\pm}=1\, \TeV$. The remaining Higgs sector parameters are in the ranges $0.99 < |\sba| < 1$, $0.1 < \tan\beta < 1$ and we take $m_{12}^2=m_A^2\sin\beta\cos\beta$. Finally, the VLQ couplings are taken in the ranges
$500\, \GeV < \mtp < 1500\, \GeV$, $-0.15 < s_L < 0.15$ and $10 < y_T < 15$.

\subsection{The composite Higgs model}
\label{sec:appendixcomp}

As mentioned in the main text, the SM Higgs $\mathcal{H}$ field in this model is a bi-doublet of $\su{2}_L\times \su{2}_R$, which together with a singlet $S$ forms the five dimensional anti-symmetric irrep of $\sp{4}$,
\begin{equation}
\mathcal{H} \oplus S \equiv \begin{pmatrix} H^{0*}&H^+\\ -H^{+*}&H^0\end{pmatrix} \oplus S \in (\mathbf{2},\mathbf{2}) \oplus (\mathbf{1},\mathbf{1}) = \mathbf{5}.
\end{equation}
The fermionic sector also consists of a bi-doublet and a singlet in the $\mathbf{5}$ of $\sp{4}$,
\begin{equation}
\Psi \equiv  \begin{pmatrix} T& X\\ B& T'\end{pmatrix} \oplus \widetilde T \in (\mathbf{2},\mathbf{2}) \oplus  (\mathbf{1},\mathbf{1}) = \mathbf{5} .
\end{equation}
The new fermions mix with the third family quarks of the SM. 
The mixing is obtained by choosing to embed both the left-handed $Q_L= ( \widetilde t_L, b_L)^T$ and the right-handed $\widetilde t_R$ as spurions into the $\mathbf{6}$ of $\su{4}$. 
The non-zero components of $Q_L$ fit into the bi-doublet of the $\su{2}_L\times \su{2}_R$ subgroup, while $\widetilde t_R$ is in the singlet of the $\mathbf{6}\to  \mathbf{5}+\mathbf{1}$ decomposition of $\su{4}\to \sp{4}$.  
The choice for $Q_L$ is essentially dictated by the need to preserve the custodial symmetry of~\cite{Agashe:2006at}. 

The construction of the interaction Lagrangian from the general formalism has been addressed in many papers and will not be reviewed here. 
Suffice it to say that we combine the five pNGBs into a $4\times 4$ matrix $\Pi$ and exponentiate it to obtain
\begin{equation}
\Sigma = \exp\left(\frac{i\sqrt{2}}{f} \Pi\right), \mbox{~transforming as:~ } \Sigma\to g \Sigma h^{-1}, \mbox{~for~} g\in \su{4},~h\in \sp{4},
\end{equation}
and use it to ``dress'' the fermionic field $\Psi$, written as a $4\times 4$ anti-symmetric matrix.
In this notation, the Lagrangian becomes
\begin{equation}
\mathcal{L} = y_L f \tr\left(\bar Q_L \Sigma \Psi_R \Sigma^T \right) + y_R f  \tr\left(\Sigma^* \bar \Psi_L \Sigma^\dagger \widetilde t_R\right) -  M  \tr\left(\bar \Psi_L \Psi_R \right)  + \mbox{ h.c.}\label{eq:CHlagr}
\end{equation}
where we indicated the dressing explicitly. 
(Note that $Q_L\to g Q_L g^T$ and $\Psi\to h \Psi h^T$.)

We allow only the Higgs field to acquire a VEV, and we denote the mixing angle by $\sin\theta = v/f$, where $v=246\, \GeV$. 
Generically $f>800\, \GeV$ from EWPT, although one can envisage mechanisms that would allow to lower that bound~\cite{BuarqueFranzosi:2018eaj}.

Computing \eq{eq:CHlagr} to all orders in $\theta$ and retaining only terms linear in $h$ and $S$, $h$ being the canonically normalised physical Higgs with VEV shifted to zero, we can write the part of \eq{eq:CHlagr} concerning top partners as (see also~\cite{Bizot:2018tds})
\begin{equation}
\mathcal{L}_{\mathrm{tops}} = -  \begin{pmatrix} \bar{\widetilde t}_L &\bar T_L&\bar{T'}_L&\bar{\widetilde{T}}_L\end{pmatrix}\bigg(  \mathcal{M}  + h  \mathcal{I}_h + S \mathcal{I}_S \bigg) \begin{pmatrix} \widetilde t_R\\ T_R\\T'_R\\ \widetilde{T}_R \end{pmatrix}  + \mbox{ h.c.}
\end{equation}
where the mass and Yukawa matrices are given by
\begin{align}
\mathcal{M} &= \begin{pmatrix} 0 &  y_L f \cos^2\left(\frac{\theta}{2}\right)  & -y_L f \sin^2\left(\frac{\theta}{2}\right) &0\\
\frac{y_R f}{\sqrt{2}} \sin\theta & M & 0 & 0\\
\frac{y_R f}{\sqrt{2}} \sin\theta & 0 & M  & 0\\
0 & 0 & 0 & M\\
\end{pmatrix}  \nonumber \\
\nonumber \\
\mathcal{I}_h &= \begin{pmatrix} 0 &  -\frac{1}{2} y_L  \sin\theta  & -\frac{1}{2} y_L  \sin\theta &0\\
\frac{y_R}{\sqrt{2}} \cos\theta & 0 & 0 & 0\\
\frac{y_R}{\sqrt{2}} \cos\theta & 0 & 0  & 0\\
0 & 0 & 0 & 0\\
\end{pmatrix}   \\
\nonumber \\
\mathcal{I}_S &= \begin{pmatrix} 0 & 0 & 0 & \frac{i y_L}{\sqrt{2}} \sin\theta \\
0 & 0 & 0 & 0\\
0  & 0 & 0  & 0\\
i y_R \cos\theta & 0 & 0 & 0\\
\end{pmatrix}  .  \nonumber
\end{align}
The singular value decomposition of $\mathcal{M}$ is unwieldy, but can be performed numerically or perturbatively to order $\theta \approx v/f$.  For the four top quark mass eigenstates  $t, \tprime, \tpp, \tppp$, the perturbative expressions for the masses are
\begin{equation}
m_t = \frac{y_L y_R f v}{\sqrt{2}\Mhat}+{\mathcal{O}}\left(\frac{v^2}{f^2}\right), \quad \mtp = M, \quad m_{\tpp} = M + {\mathcal{O}}\left(\frac{v^2}{f^2}\right),\quad m_{\tppp} = \Mhat + {\mathcal{O}}\left(\frac{v^2}{f^2}\right).
\end{equation}
The mass of the bottom partner (mostly aligned with $B$) turns out to be of the same order as that of the heaviest top partner $m_{\tppp}$, while $X$ has mass equal to $M\equiv \mtp$ since it does not mix with anything. 
For the top quarks, the conversion from gauge to mass eigenbasis reads, to ${\mathcal{O}}\left(v/f\right)$,
\begin{align}
& \widetilde t_L = -\frac{M}{\Mhat} t_{L} + \frac{y_L f}{\Mhat} \tppp_{L},\quad
T_L = \frac{y_L  f}{\Mhat} t_{L} +\frac{M}{\Mhat} \tppp_{L},\quad
T'_L = \tpp_{L},\quad  \widetilde T_L = \tprime_{L} \\
& \widetilde t_R = t_{R} + \frac{y_R v }{\sqrt{2} M} \tpp_{R} +   \frac{y_R v M }{\sqrt{2} \Mhat^2} \tppp_{R}, \quad
T_R = \tppp_{R} -  \frac{y_R v M }{\sqrt{2} \Mhat^2} t_{R},\quad
T'_R = \tpp_{R} - \frac{y_R v }{\sqrt{2} M} t_{R},\quad
\widetilde T_R = \tprime_{R} .\nonumber
\end{align}
This spectrum justifies that choice of simplified model in the text where we neglect all the top partners other than the lightest one.

\begin{figure}[t]
\begin{center}
\includegraphics[width=0.7\textwidth]{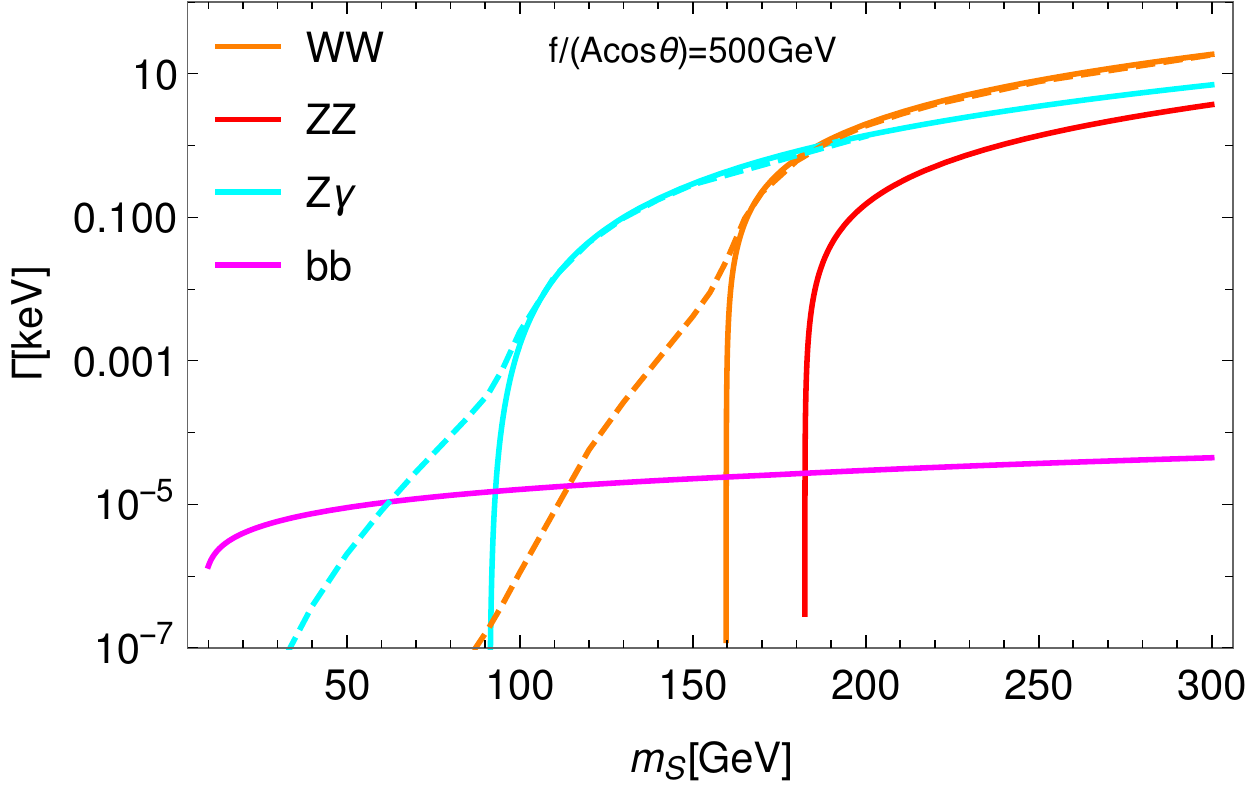}
\caption{Partial width of $S$ in the dominant decay channels for the composite Higgs model benchmark scenario discussed in \sec{sec:interpretation_composite}. The dashed lines denote the contribution with at least one off-shell weak boson.  \label{etadecays}}
\end{center}
\end{figure}

Regarding decays of the pseudoscalar in this model,
in \fig{etadecays} we show the partial widths of $S$ as a function of its mass, including the dominant loop induced  fermionic channel $S\to \bar b\, b$ relevant below the $Z\, \gamma$ threshold.
We use $f/(A\cos\theta)=500\, \GeV$ but all curves rescale by $\left(500\,\GeV\,A\cos\theta/f\right)^2$. We see that for all interesting regions of parameters the width is always very narrow, but still prompt.

The most promising  parameter region for this class of  models is $m_S\lesssim 160\, \GeV$, where the $S$ decays dominantly to $\Zgam$. 
This region is motivated from the model building perspective since it is expected $m_S<m_h$. 
From the experimental point of view it offers a clear benchmark of a $\Zgam$ channel. 
Above $2m_W$ the $WW$ channel overcomes, and for $m_S\lesssim 80\, \GeV$ the $b\bar{b}$ channel dominates, both of which are less clean channels experimentally.

\clearpage

%% file: appendix_rangeofvalidity.tex
\label{sec:appendixnwa}

\begin{figure}[tbp]
\centering
\includegraphics[width=.45\textwidth]{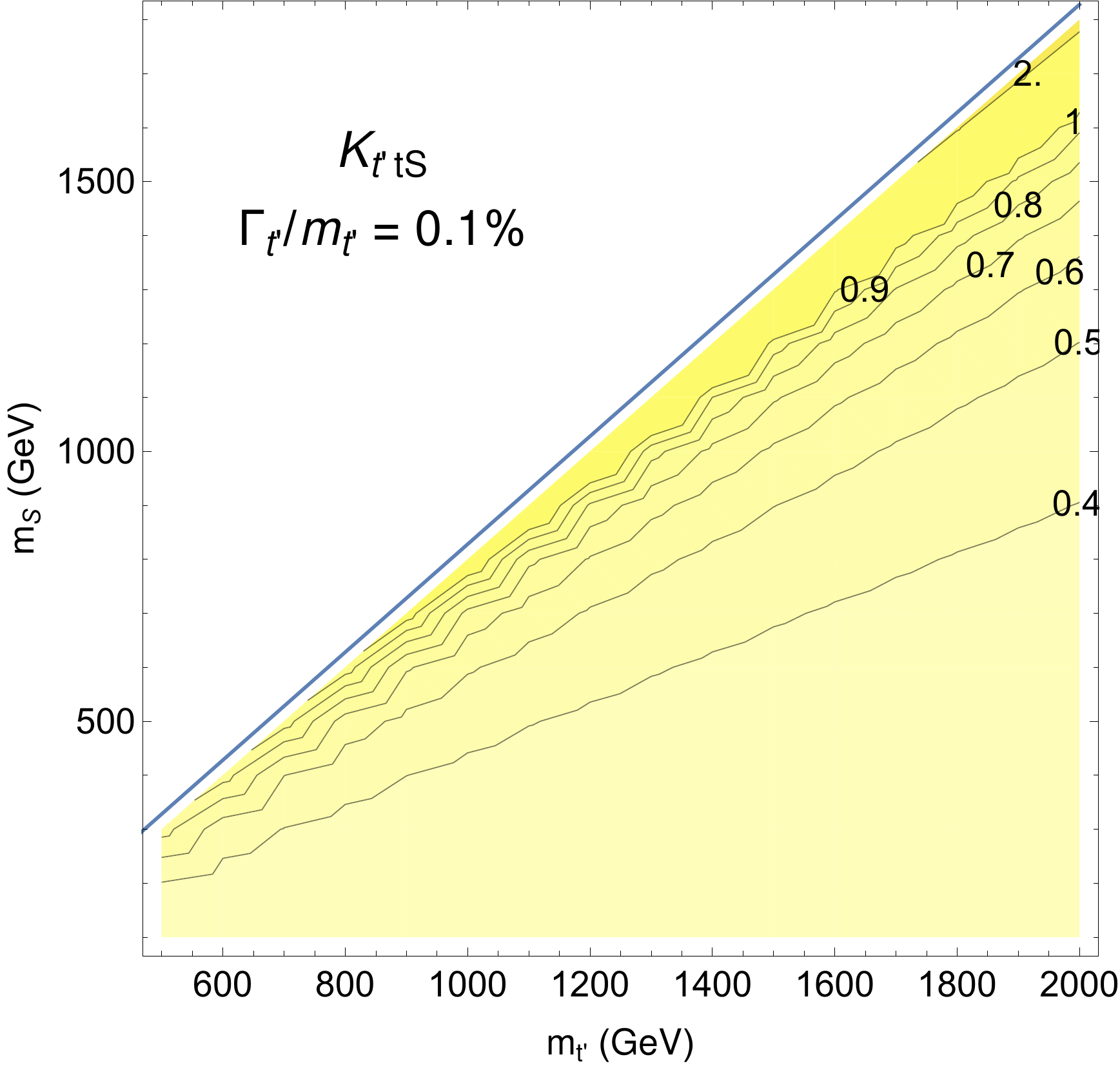}
\includegraphics[width=.45\textwidth]{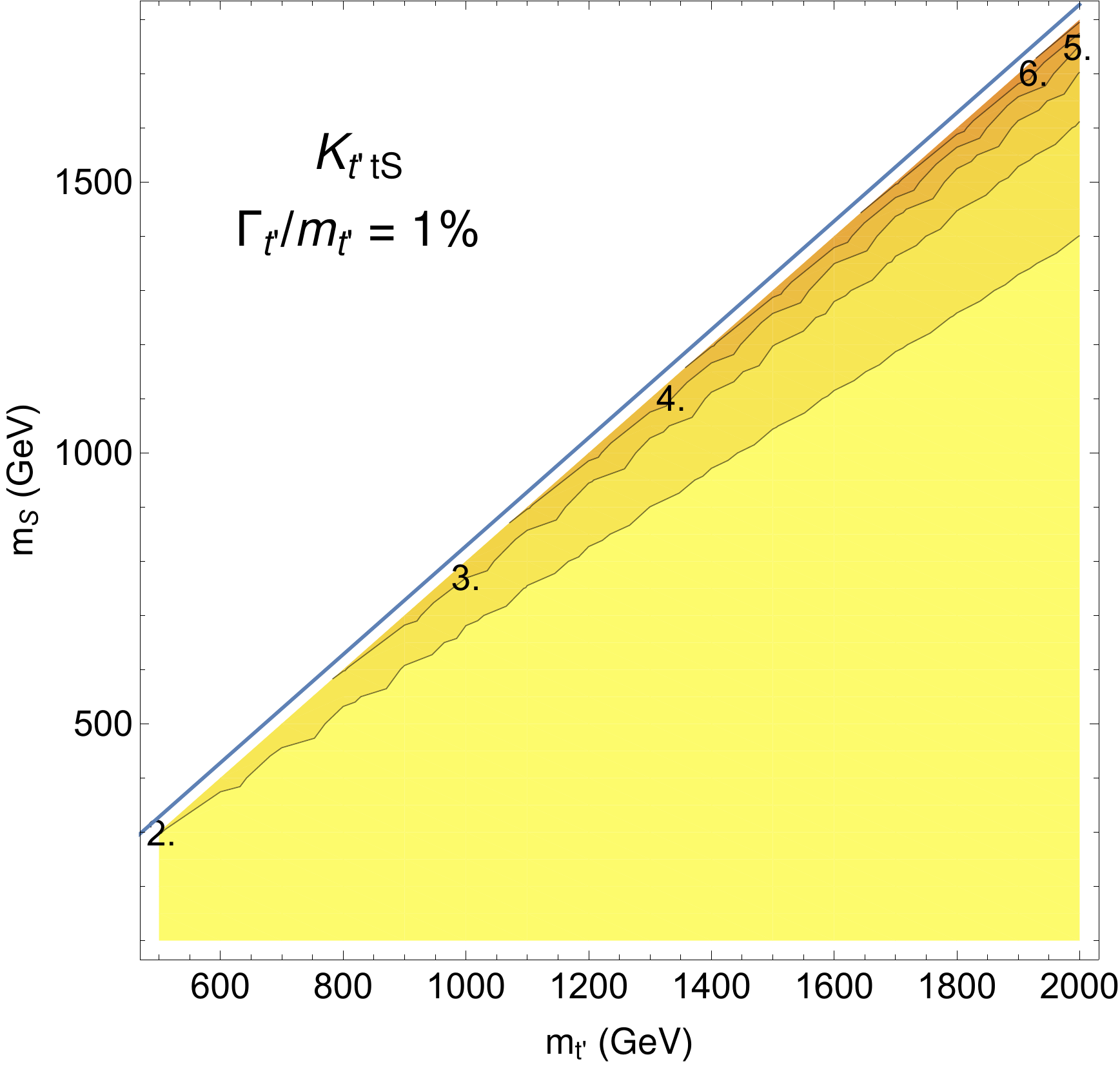} \\ 
\includegraphics[width=.45\textwidth]{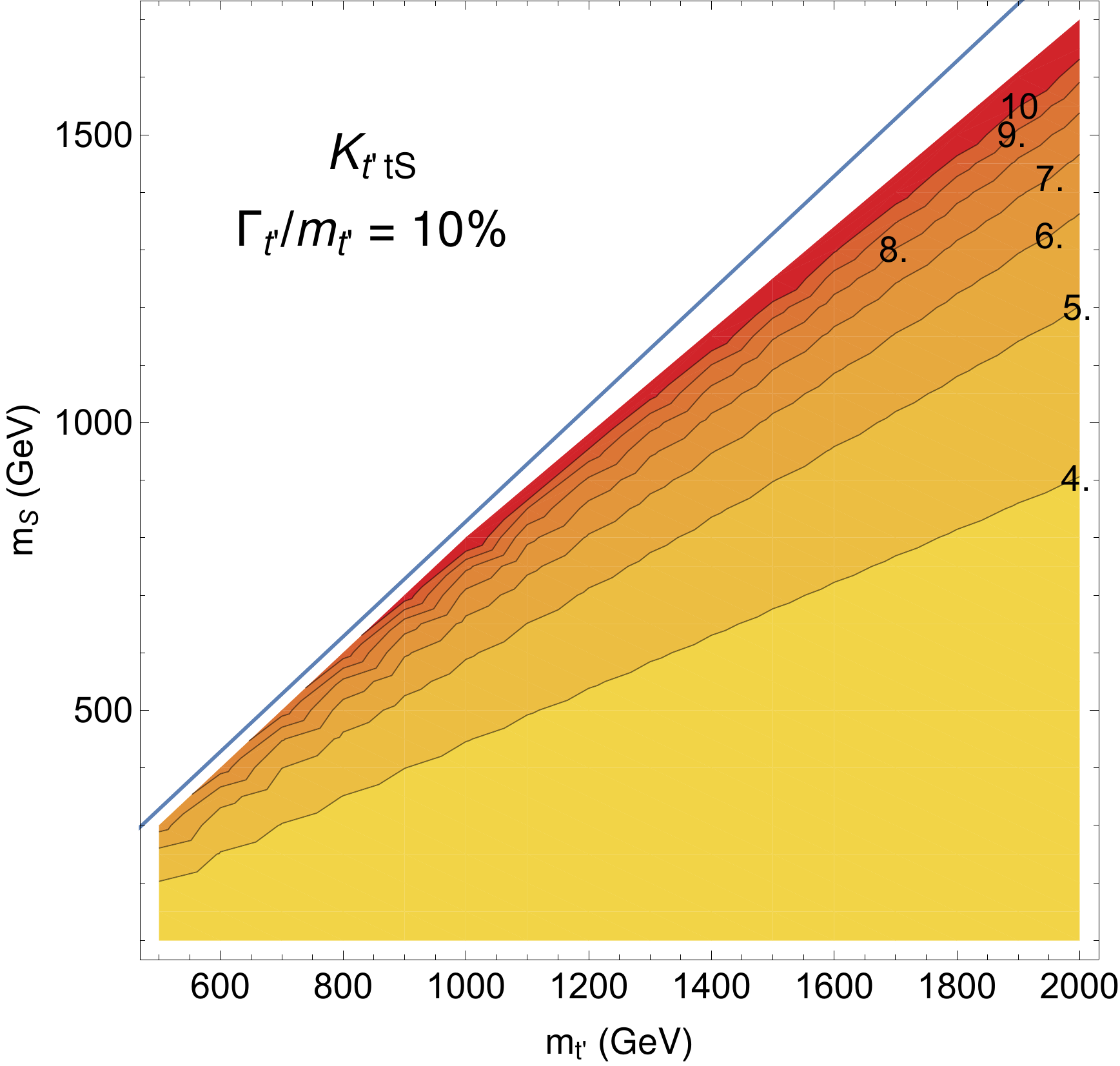}
\caption{\label{fig:TtetacouplingfixedWMratio} Values of the $\kappa^{S }_{L,R}$ coupling corresponding to fixed $\Gamma_{\tprime}/\mtp$ ratios (0.1\%, 1\% and 10\%) in the $\{\mtp, \ms \}$ plane. The blue contour corresponds to the kinematic limit $\mtp - m_t - \ms =0$. The maximum value for the coupling to be in the perturbative region has been limited to $4\pi$.}
\end{figure}

\begin{figure}[tbp]
\centering
\includegraphics[width=.9\textwidth]{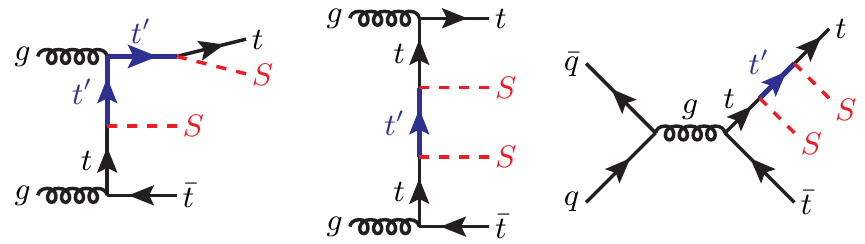}
\caption{\label{fig:topologiesLW} Examples of topologies containing at least one $\tprime$ propagator and leading to the same final state for the process $pp \to t \bar t  S   S $.}
\end{figure}

In the processes under consideration both $\tprime$ and $S$ are assumed to be in the narrow-width approximation (NWA), in order to factorise the production of the top partner from its decay chain. 
Such assumption, however, implies that the coupling $\tprime t S $ cannot exceed specific values which depend on the masses of $\tprime$ and $ S $ according to the relation in \eq{GamTtotS.EQ}.
Considering as a simplifying and extreme assumption that the only available decay channel for $\tprime$ is into the SM top and $S$ and that one chirality of the couplings is dominant with respect to the other, such that either $\kappa^{S}_R \ll \kappa^{S}_L$ or vice versa, the values of the coupling corresponding to different $\Gamma_{\tprime}/\mtp$ ratios is shown in \fig{fig:TtetacouplingfixedWMratio}.
For a specific $\{\mtp, \ms \}$ configuration, values of the coupling larger than those in the contours of \fig{fig:TtetacouplingfixedWMratio} would produce a larger width. 
The determination of the validity of the NWA approximation is important to understand the reliability of the results. 
If the $\tprime$ width is not narrow, off-shellness effects in the process of pair production and the contribution from topologies which are neglected in the NWA, represented by the examples of \fig{fig:topologiesLW}, can become more and more relevant.

\begin{figure}[tbp]
\centering
\includegraphics[width=.45\textwidth]{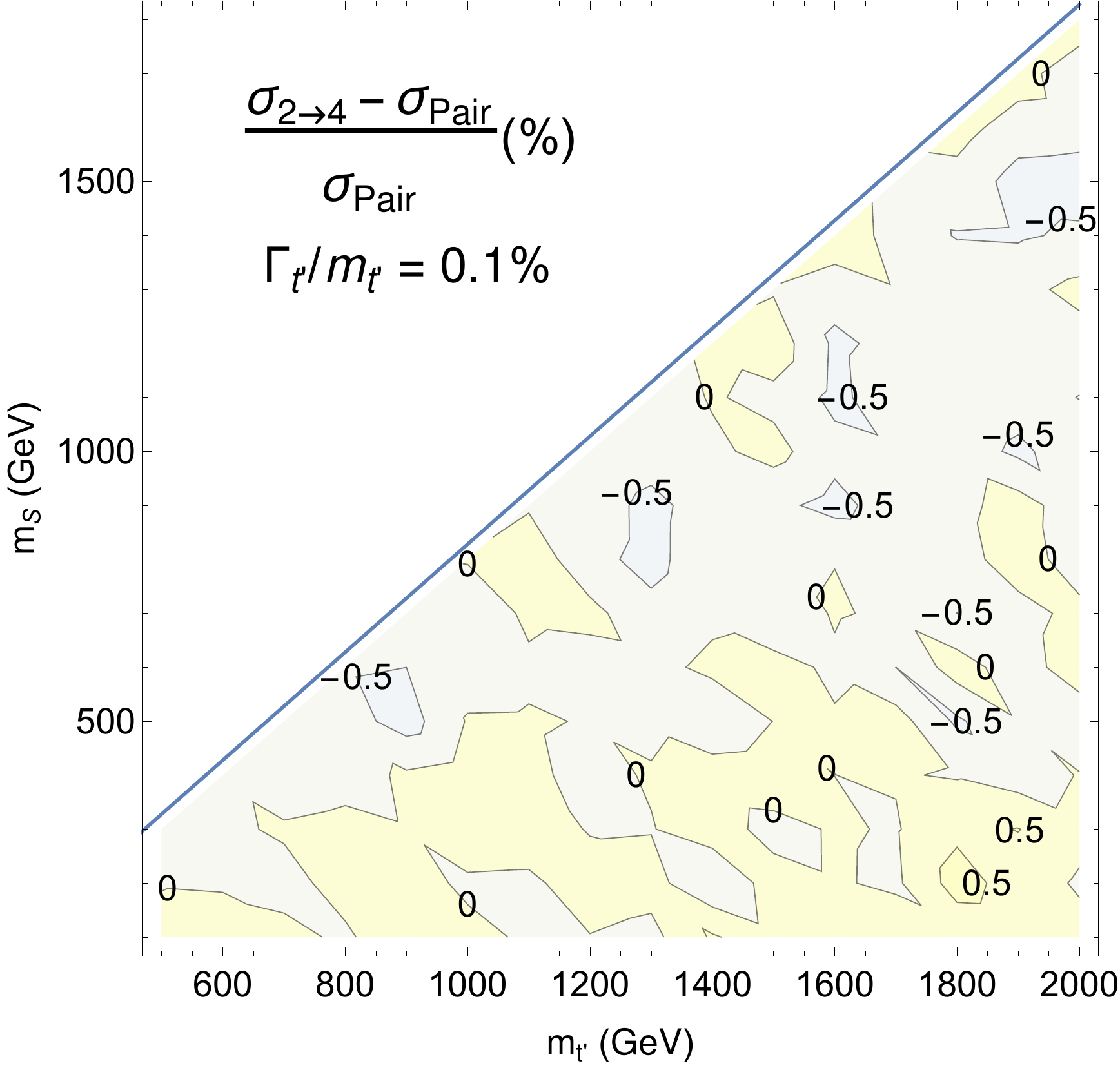}
\includegraphics[width=.45\textwidth]{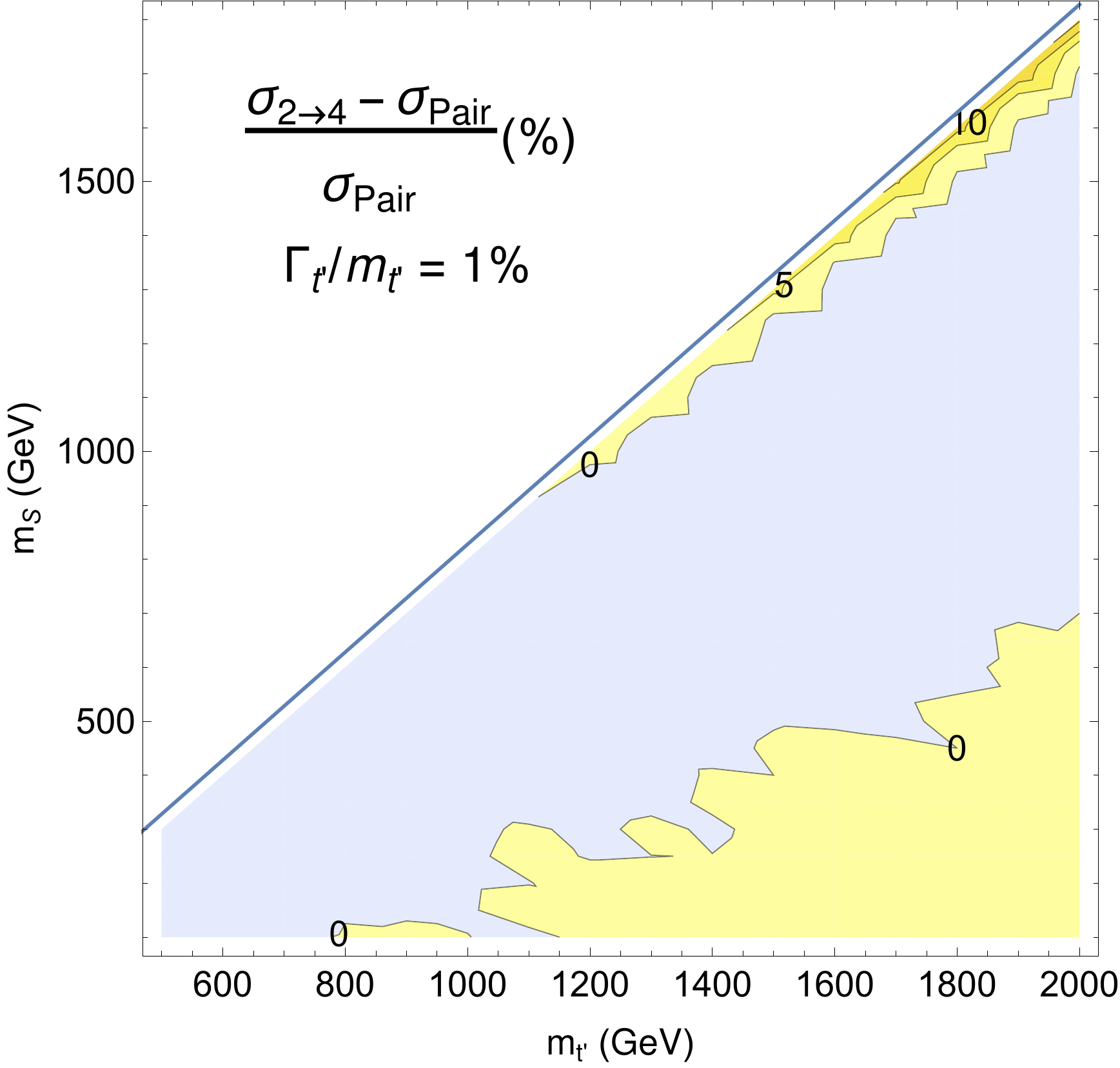}\\
\includegraphics[width=.45\textwidth]{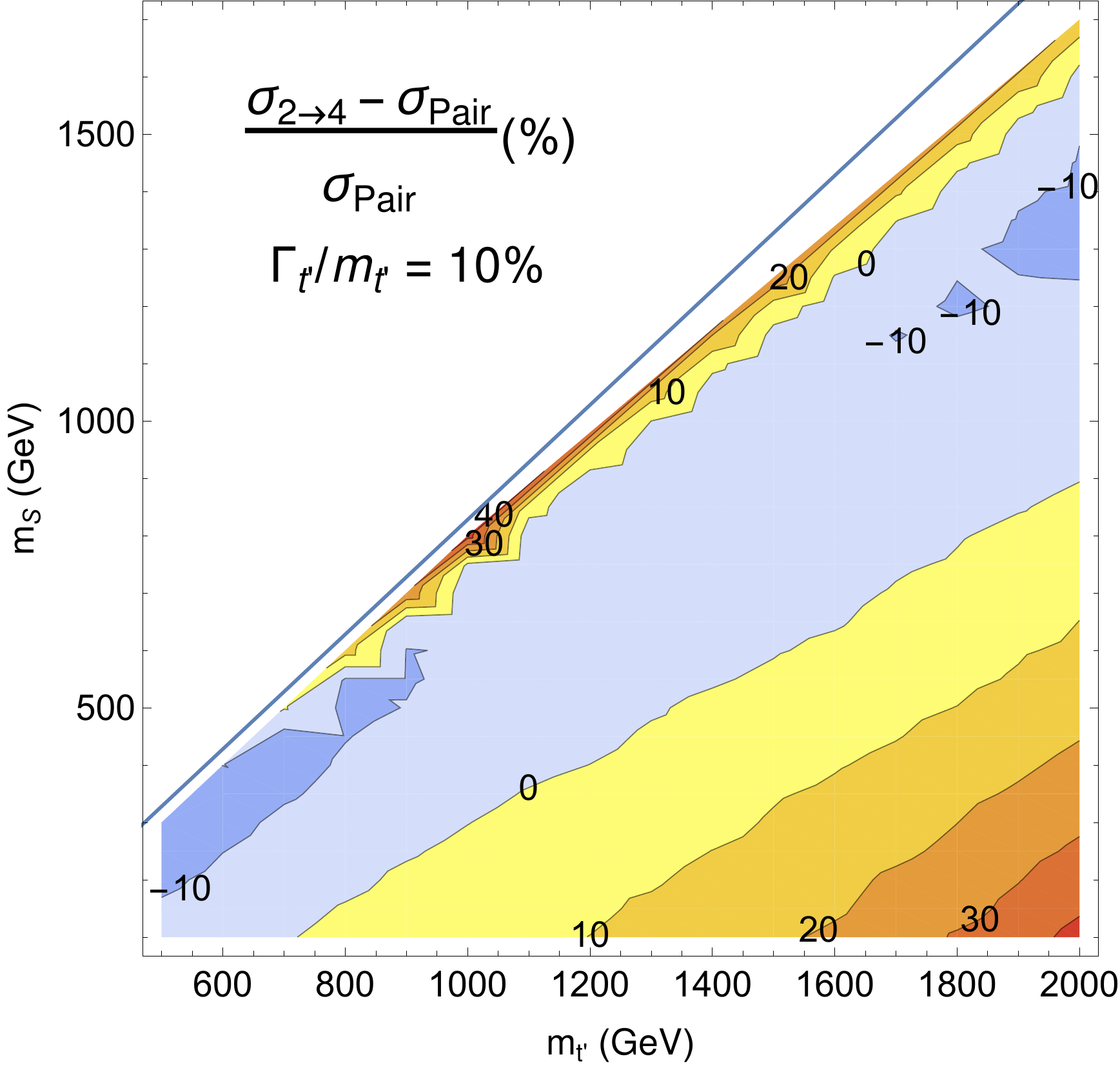}
\caption{\label{fig:TPratio} Relative ratio of the cross-sections for the full process $p p\to t \bar t  S   S $ ($\sigma_{2\to4}$) and for the pair production process $p p \to \tprime  \antitprime  \to (\St ) (S\, \bar t )$ where the $\tprime$ production and decay are factorised in the NWA approximation ($\sigma_{\rm Pair}$). The ratio is shown for different values of the $\Gamma_{\tprime}/\mtp$ ratios (0.1\%, 1\% and 10\%), and the couplings $\kappa^{S}_{L,R}$ are not allowed to exceed the perturbative limit $4\pi$.}
\end{figure}

To assess how the width of $\tprime$ affects the determination of the cross-section, the full $2\to4$ process $p p \to t \bar t  S   S $ has been evaluated by imposing the presence of at least one $\tprime$ propagator in the topologies, in order to obtain the signal under the assumption of negligible $S t t$ coupling. 
With such process, the off-shellness effects and contribution of topologies such as those in \fig{fig:topologiesLW} are fully taken into account. 
Still under the assumption that $\tprime$ can only decay to $\St$ and therefore that the only way to increase the total width of $\tprime$ for a given $\{\mtp, \ms \}$ configuration is by increasing $\kappa^{S}_{L,R}$, the ratio between the cross-sections of the full process and of the pair-production process in the NWA is shown in \fig{fig:TPratio}.

The effect of a large width is already noticeable when the $\Gamma_{\tprime}/\mtp$ ratio reaches 1\%, when the interference between the resonant channels and all the other contributions is negative and of the order of few percents in a region where $m_S + m_t$ is around 80\% of $\mtp$. 
If the $\Gamma_{\tprime}/\mtp$ ratio is below 1\% the relative ratio between cross-sections is dominated by the statistical fluctuations of the simulation. 
For this reason, the numerical results in the following sections assume $\Gamma_{\tprime}/\mtp$ to be of order 0.1\%.

\clearpage

%% file: appendix_objects.tex
In the following, more details for the definition and selection of objects at reconstructed level are presented, elaborating on the brief description in \sec{sec:object_definition}, in order to facilitate reproducibility and as a guide for possible future searches at colliders. 

For all objects, the default ATLAS \delphes\ card~\cite{delphes} is used, with minor modifications in a few cases, as explained below. 
Objects that partially fall in the calorimeter transition region $1.37 < |\eta| < 1.53$ are excluded, if they are reconstructed in the calorimeter, where $\eta$ is the pseudorapidity. 
Relative angular distances in the detector are typically expressed as \dR in the $\eta$-$\phi$ plane where $\phi$ is the azimuthal angle around the beampipe. 
A particle's transverse momentum \pt is the momentum component in the plane transverse to the beam axis. 

Isolation and overlap removal are needed to distinguish the objects from each other in the detector simulation,\footnote{In detectors at colliders, the same energy deposits can be associated with different objects, e.g., an electron can also be identified as a jet. In order to make sure each energy deposit is counted only once, every object has to be energetically isolated and the objects are not allowed to overlap in the detector.}
\label{sec:isol}
which is done in the \delphes\ card, unless otherwise specified. 
This is achieved by creating the containers for the objects in mind: jets, photons, electrons and muons. 
In \delphes\ all objects passing their respective efficiency cut are first reconstructed as the respective object and as a jet. 
The object will then be put into the jet container and the container corresponding to the reconstructed object. 
By passing an isolation criterion the object is removed from the jet container and only kept in the container corresponding to the correct reconstruction. 
The criterion is met when an isolation variable $I$ is within a certain constraint. 
The variable is defined by summing the \pt of all objects, not including the candidate, within a cone of \dR\ around the candidate and dividing by the candidate \pt. 
That is, 
\begin{equation}
I = \frac{\sum_{i\neq \text{candidate}} \pt(i)}{\pt(\text{candidate})},
\end{equation}
where the sum runs over all the objects $i$ around the candidate within the \dR cone.

The objects used in the analysis are defined below. 

\textbf{Photons}, $\gamma$, are reconstructed by considering energy deposits in the electromagnetic calorimeter (ECAL) and no tracks in the inner detector. 
Objects successfully reconstructed as photons are required to have a $\pt > 30\,\GeV$ and $ |\eta|  < 2.37$. 
Photons in the transition region are not taken into account. 
Overlap removals are done in the modified \delphes\ card as described above, where the photon candidate is identified and put in the correct container by passing the photon efficiency cut corresponding to the ATLAS tight quality efficiency cuts~\cite{Aaboud:2018yqu}. 
Isolation of the photon is done after the simulation and it is considered isolated when the isolation variable $I < 0.008$, where $I$ is defined as described above.

\textbf{Leptons}, $\ell$, are in the following understood to mean {electrons} or {muons} only, and not $\tau$-leptons.
Electrons are reconstructed by looking at both energy deposit in the ECAL and having a track in the inner tracking system. 
For the following, simulation in \delphes\ reconstruction of the electron is done by combining the reconstruction efficiency of the two subsystems and parametrise it as a function of energy and pseudorapidity. 
Muons pass the calorimeters and are reconstructed by combining the information from the inner tracker and the muon spectrometer. 
In \delphes, the user specifies the efficiency of the muons such that a muon is only reconstructed with a certain probability~\cite{delphes}. 
Leptons are required to pass an isolation criterion for which $I < 0.12$ within the cone $\dR < 0.2$ for electrons and $\dR < 0.3$ for muons. 
Furthermore, leptons are required to have $\pt > 25\, \GeV$ and be in the region of $| \eta | < 2.47$, excluding the transition region in the case of electrons. 
Further overlap removals of leptons are done in \delphes\ where the lepton candidate is identified and put into the correct container by passing the given lepton efficiency. 
For electrons, the efficiencies correspond to the ATLAS tight quality efficiency cut~\cite{ATLAS-CONF-2016-024}. 
For muons, the default \delphes\ values are used. 

\textbf{$Z$ bosons}, $Z$, are identified as two leptons with same flavour and opposite signs, whose invariant mass fall within the window $|\mll - \mz|<10\, \GeV$ where \mll is the invariant mass of the reconstructed leptons. 

\textbf{Jets}, $j$, are reconstructed by using the \fastjet~\cite{fastjet} package together with \delphes.
Here the anti-$k_t$ algorithm~\cite{Cacciari:2008gp} with a $R$ parameter of $R=0.4$ is in use for jet reconstruction. 
Jets are required to pass $\pt > 25\, \GeV$ and $| \eta | < 2.47$, excluding the transition region. 

\textbf{$B$-jets}, $j_b$, are jets which originate from the hadronisation of a $b$-quark. In \delphes\ this means a jet which contains a truth $b$-quark. 
The efficiency and misidentification rate is parametrised in \delphes\ based on estimates from ATLAS~\cite{delphes,ATL-PHYS-PUB-2015-022}.

\textbf{Missing transverse energy}, \MET, is computed in \delphes\ by taking the negative scalar sum of the transverse component of the momenta of all calorimeter towers (i.e., energy deposits in the calorimeter),  $\vec{E}_{\text{T}}^{\text{miss}} = - \sum_i \vec{\pt}(i)$~\cite{delphes}. 

\textbf{The scalar transverse energy}, \Ht, is computed by taking the scalar sum of the \pt of all reconstructed basic objects used in the analysis, in this case: jets, muons, electrons and photons.  
All these objects which enter the \Ht definition are required to pass the stated analysis \pt and $\eta$ cuts.

%% file: appendix_efficiencies.tex
\begin{figure}[tbp]
\centering
\includegraphics[width=0.49\textwidth]{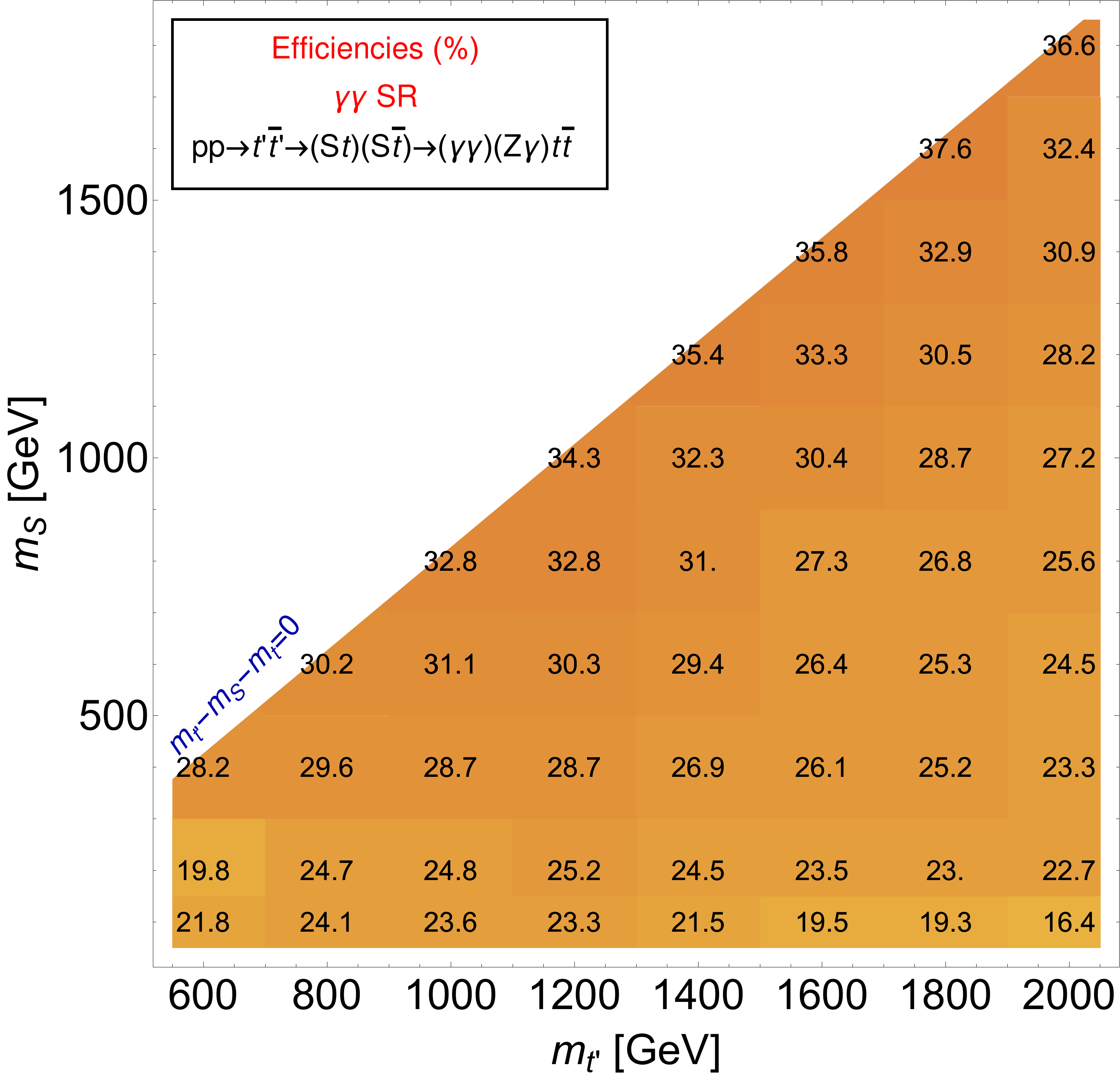}\\
\includegraphics[width=0.49\textwidth]{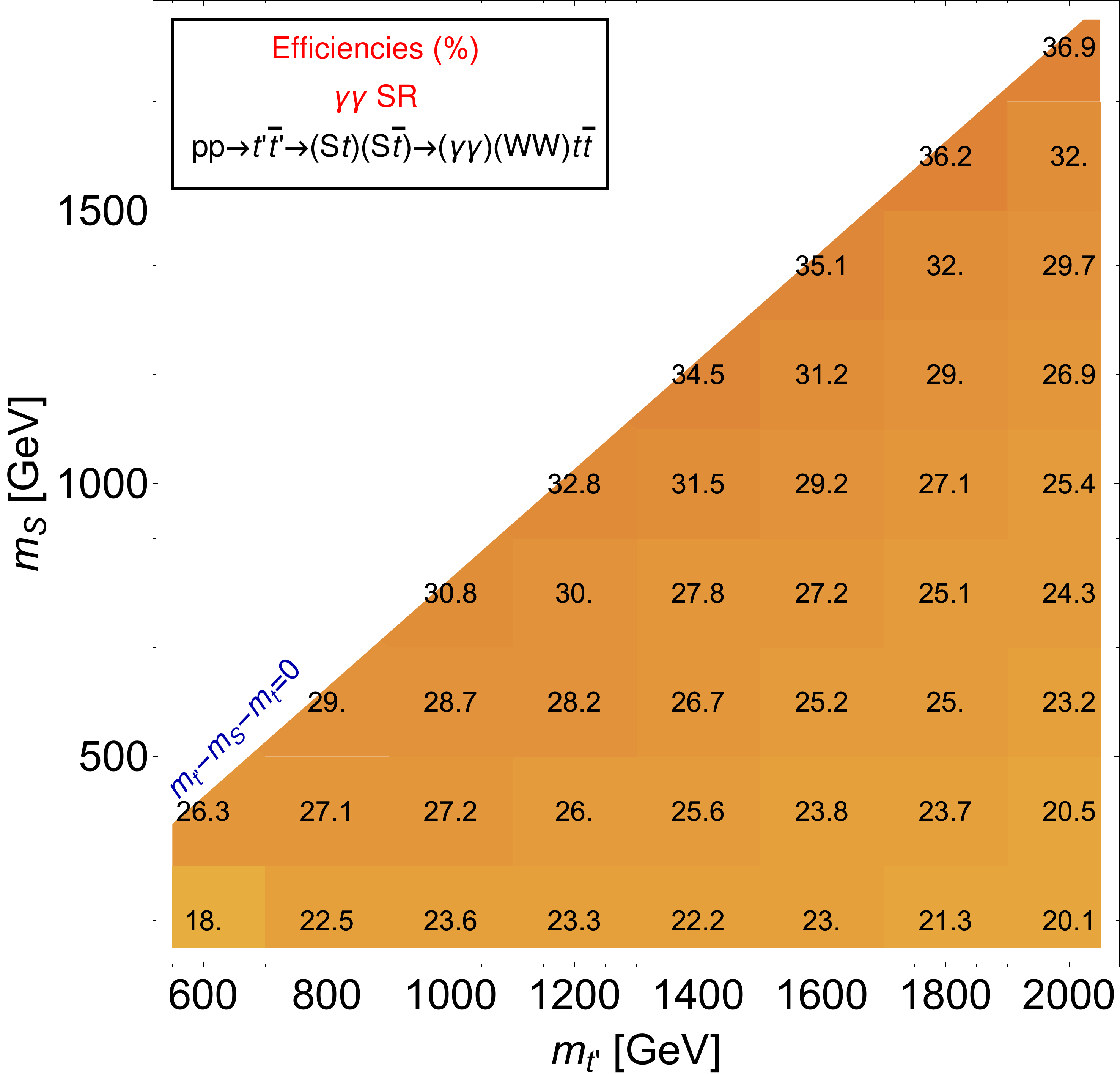}
\includegraphics[width=0.49\textwidth]{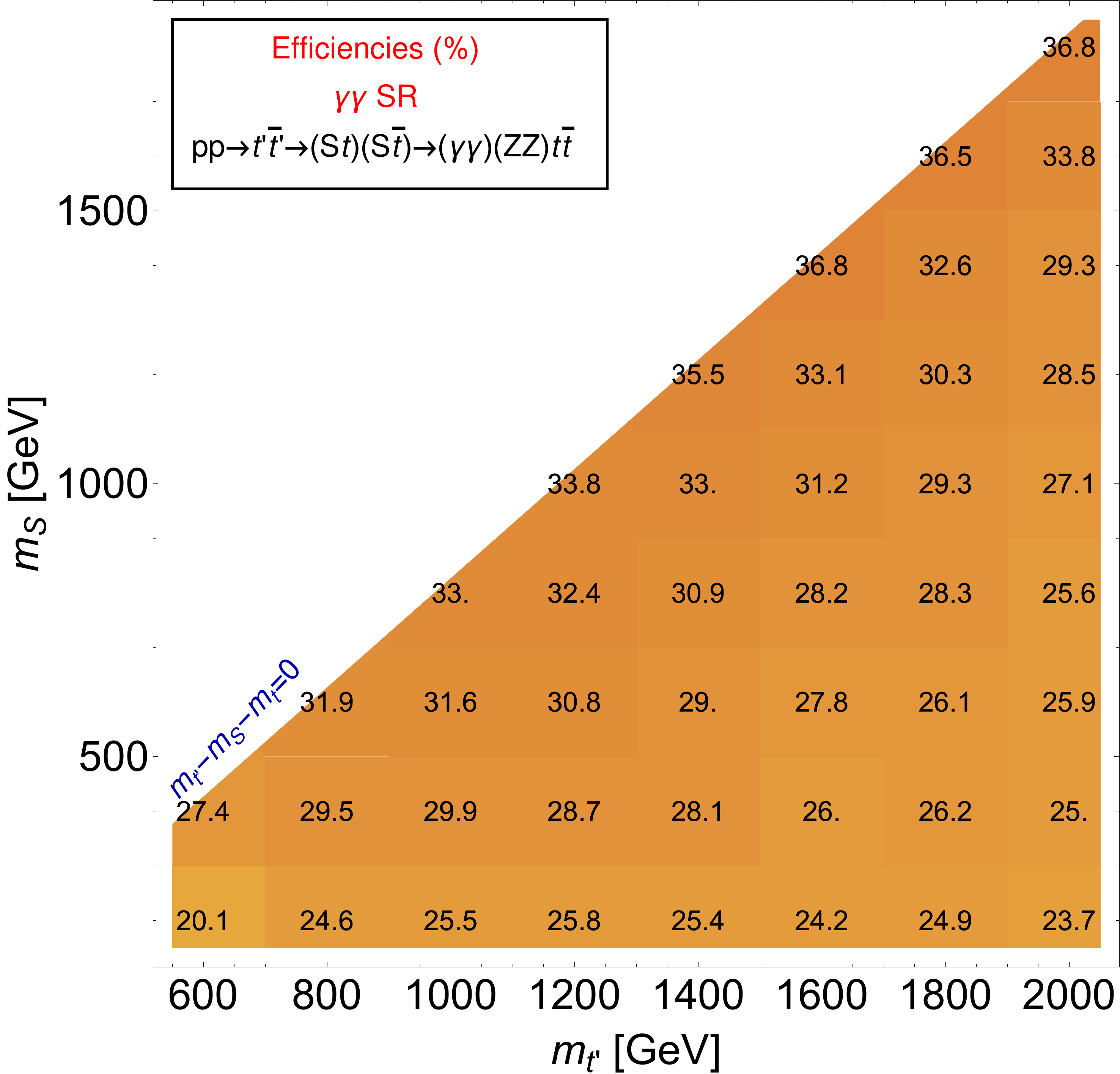}\\
\caption{\label{fig:eff_aaSR_aachannels} Efficiencies for the $\gamgam$ SR and for channels where at least one of the two $S$ decays into $\gamgam$.}
\end{figure}
\begin{figure}[tbp]
\centering
\includegraphics[width=0.49\textwidth]{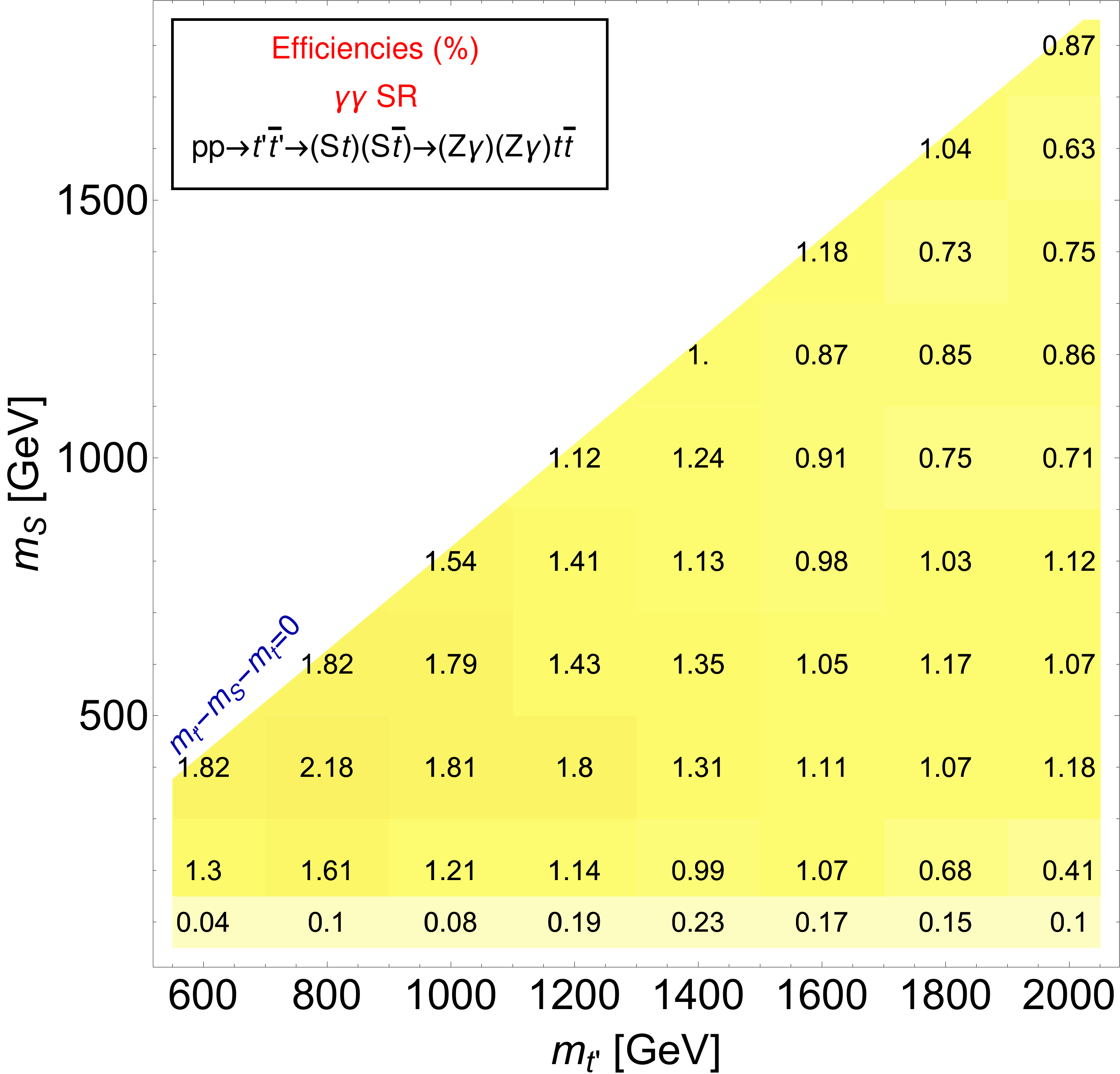}
\includegraphics[width=0.49\textwidth]{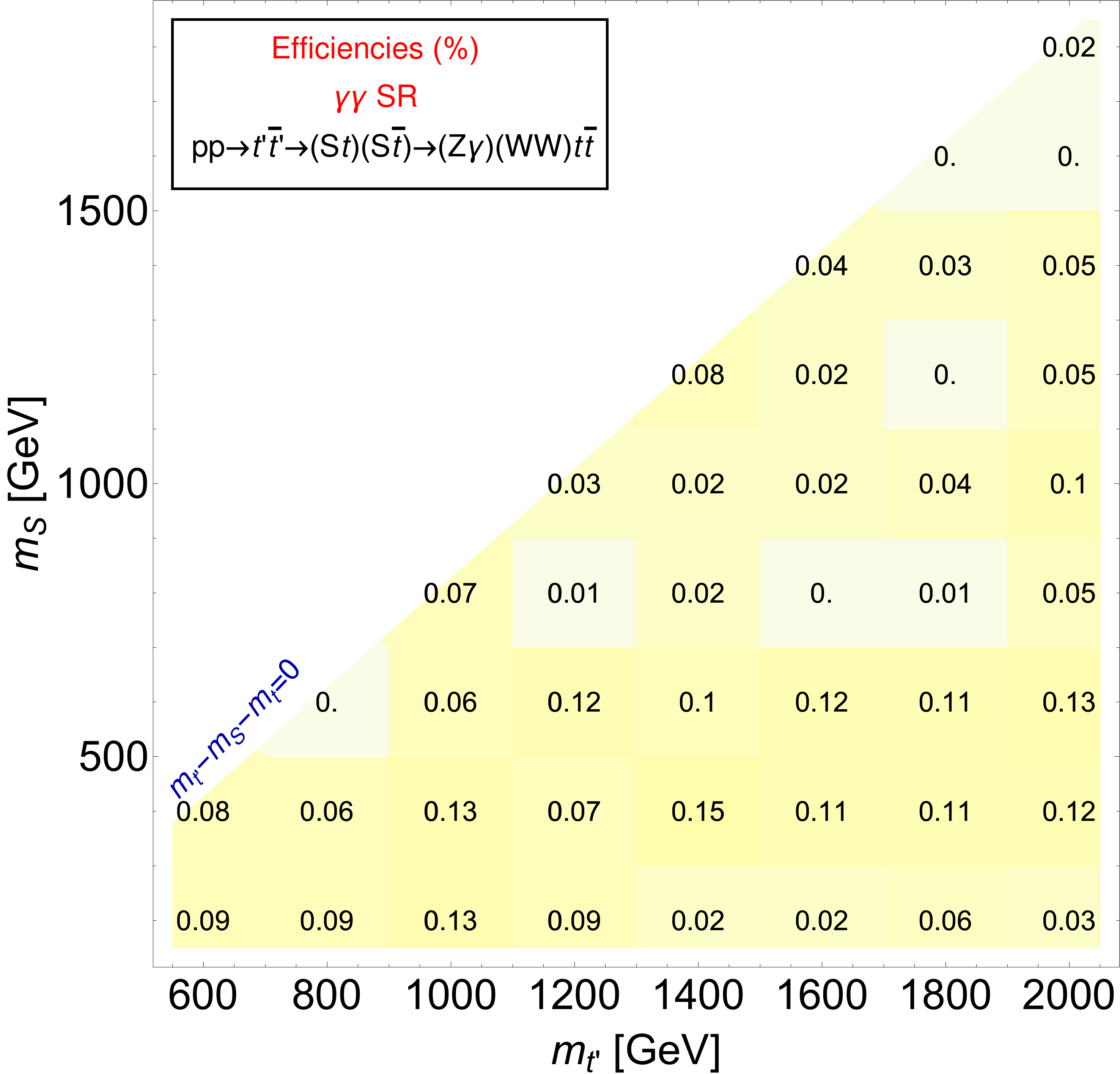}\\
\includegraphics[width=0.49\textwidth]{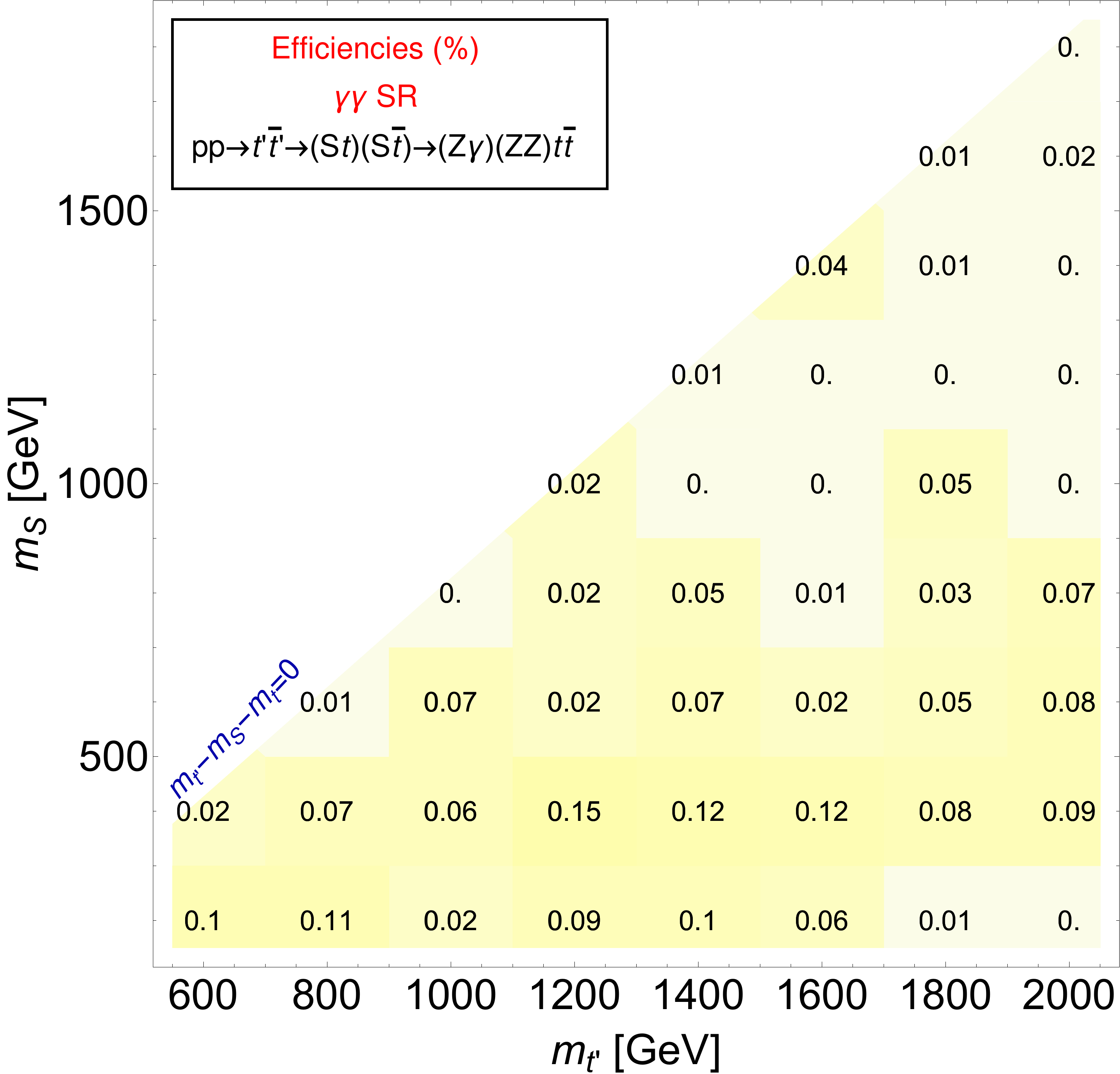}
\caption{\label{fig:eff_aaSR_azchannels} Efficiencies for the $\gamgam$ SR and for channels where at least one of the two $S$ decays into $\Zgam$.}
\end{figure}
\begin{figure}[tbp]
\centering
\includegraphics[width=0.49\textwidth]{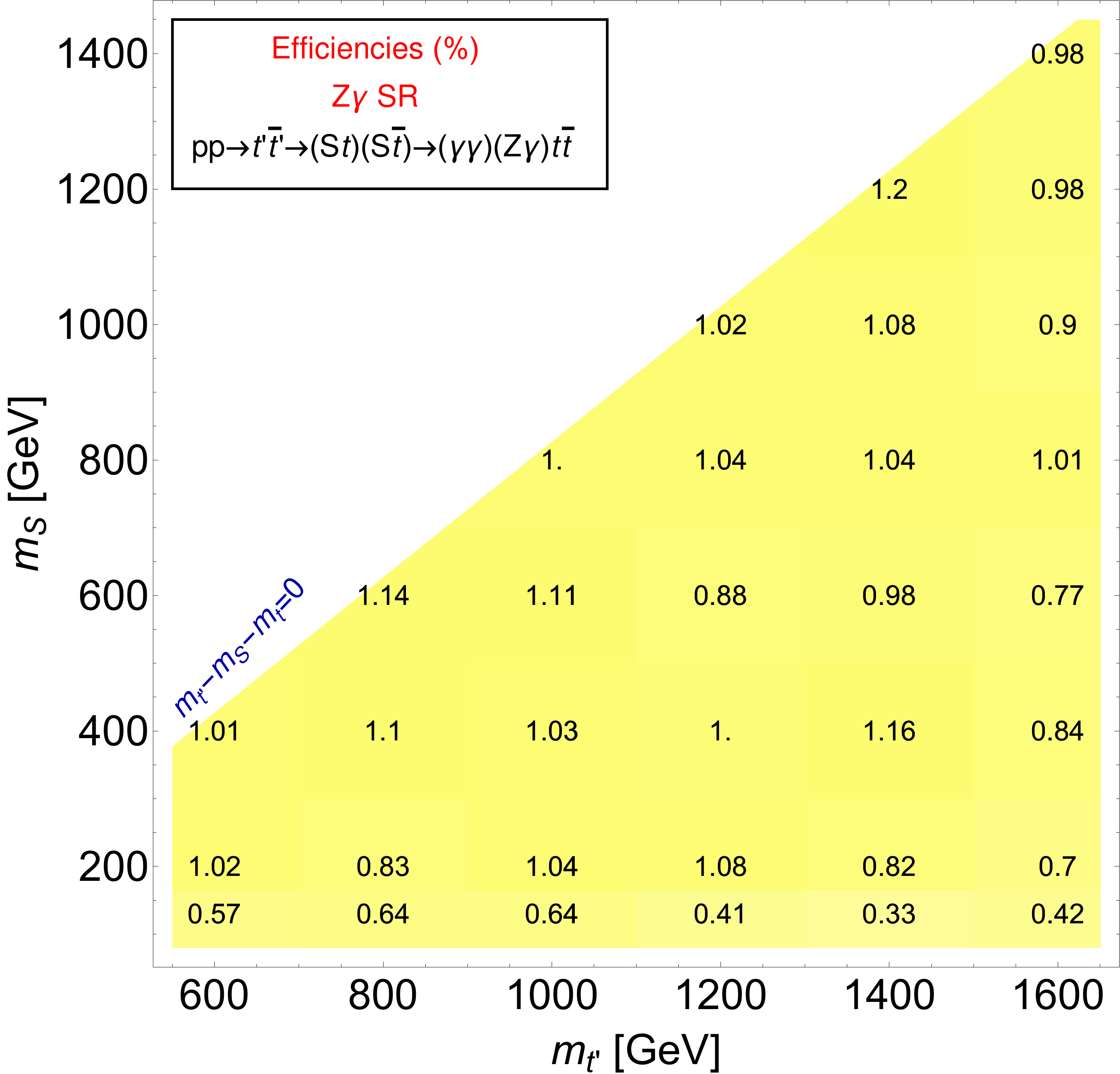}
\includegraphics[width=0.49\textwidth]{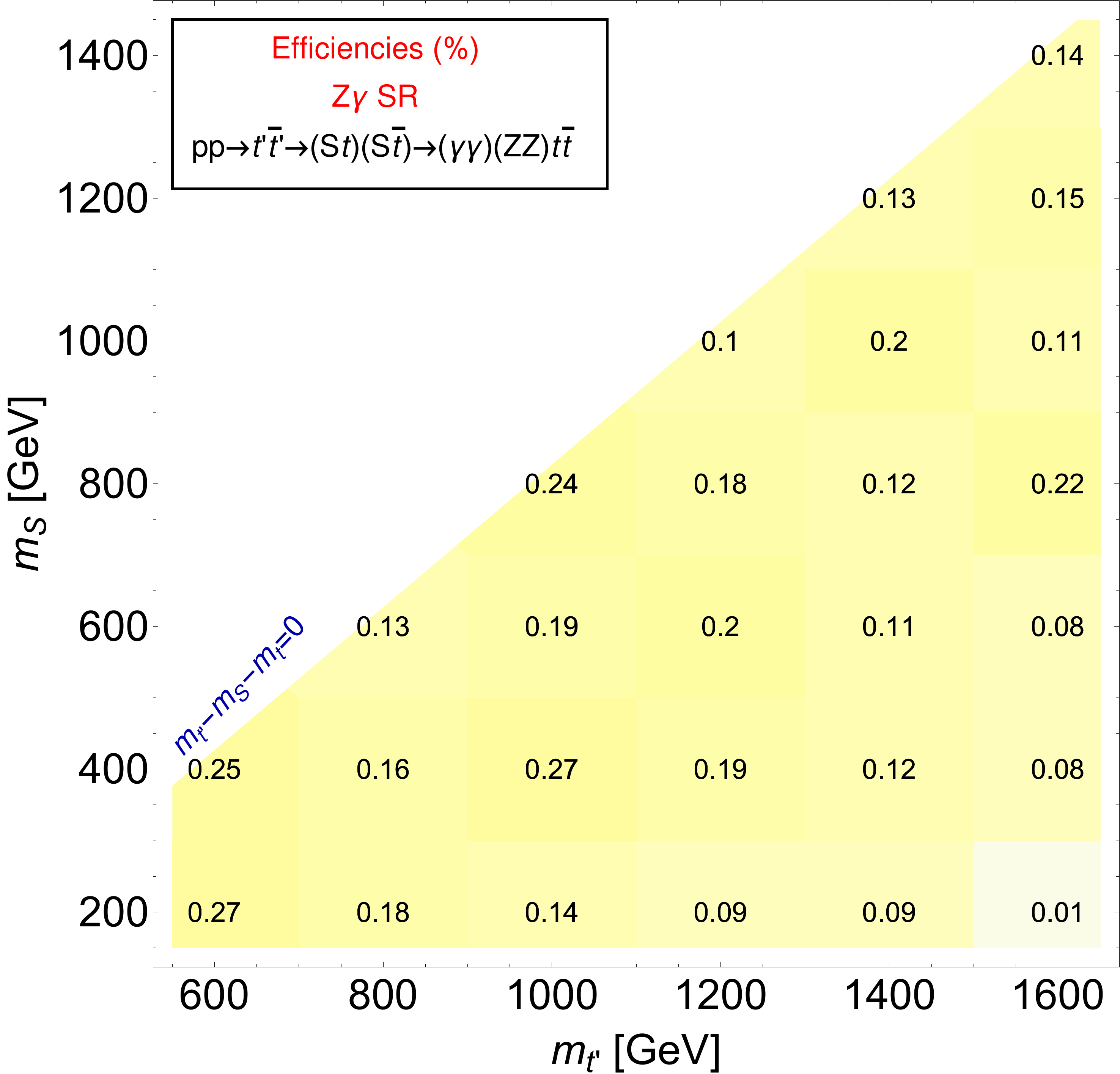}
\caption{\label{fig:eff_azSR_aachannels} Efficiencies for the $\Zgam$ SR and for channels where at least one of the two $S$ decays into $\gamgam$.}
\end{figure}
\begin{figure}[tbp]
\centering
\includegraphics[width=0.49\textwidth]{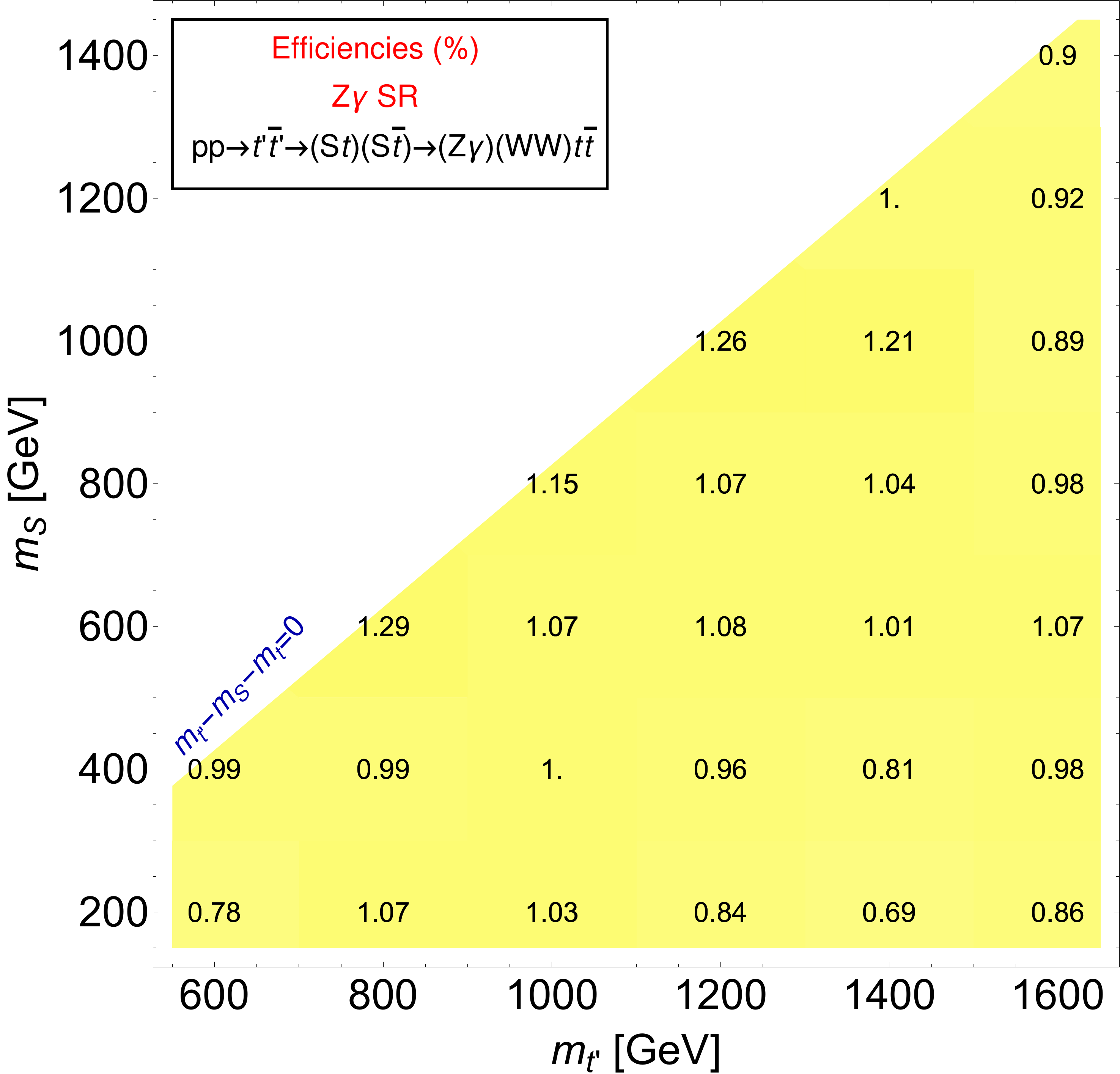}
\includegraphics[width=0.49\textwidth]{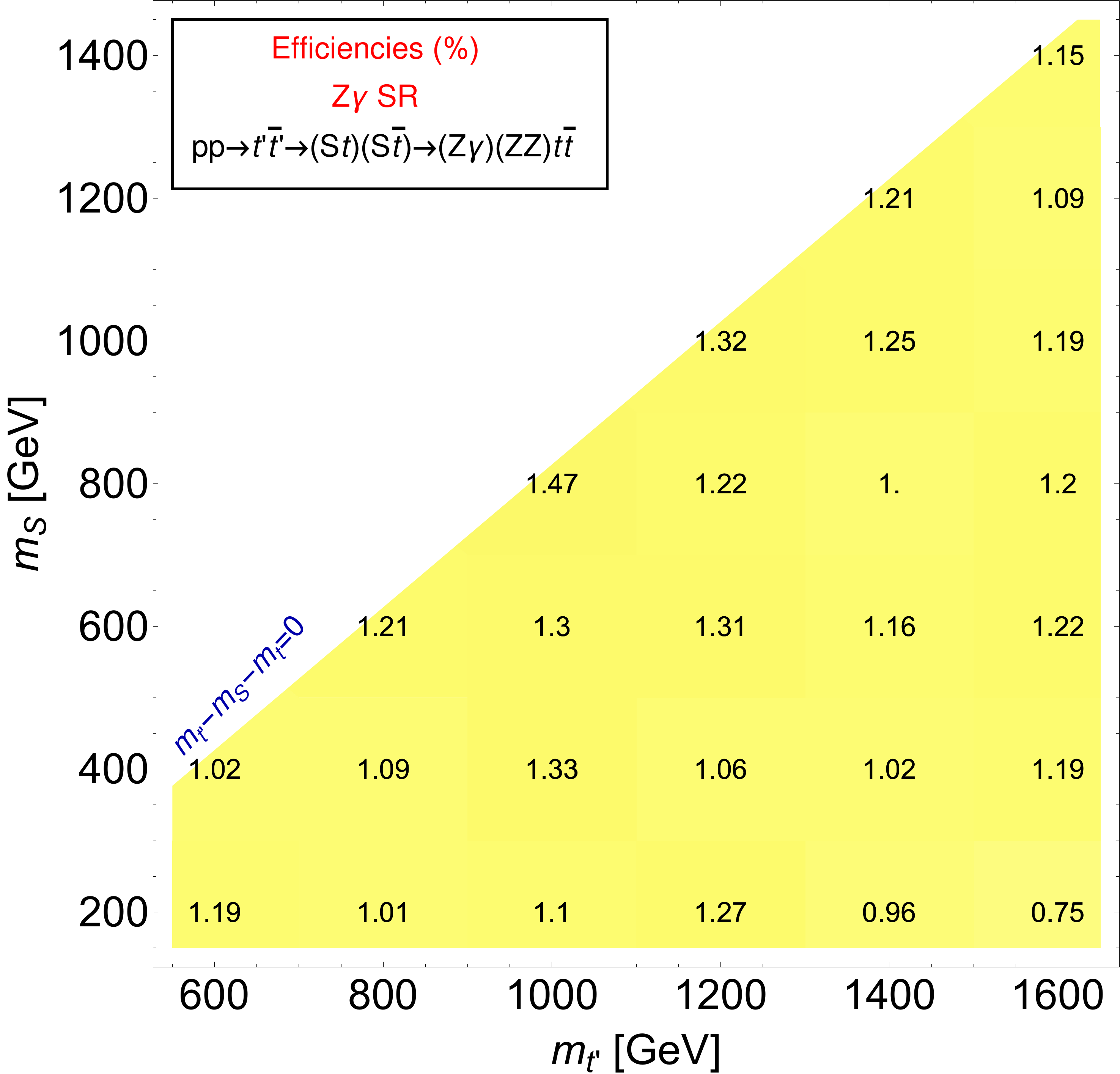}
\caption{\label{fig:eff_azSR_azchannels} Efficiencies for the $\Zgam$ SR and for channels where at least one of the two $S$ decays into $\Zgam$.}
\end{figure}

In this appendix we present the signal efficiencies for each channel and mass point considered in the analysis, except those already shown in \fig{fig:efficiencies}.
Figures~\ref{fig:eff_aaSR_aachannels}, \ref{fig:eff_aaSR_azchannels}, \ref{fig:eff_azSR_aachannels} and \ref{fig:eff_azSR_azchannels} show, respectively, the efficiencies for the $\gamgam$ SR and for channels where at least one of the two $S$ decays into $\gamgam$, $\gamgam$ SR and at least one $S$ decaying to $\Zgam$, $\Zgam$ SR and at least one $S$ decaying to $\gamma \gamma$ and $\Zgam$ SR and at least one $S$ decaying to $\Zgam$.

\clearpage